\documentclass[sn-mathphys]{sn-jnl}

\jyear{2022}%

\theoremstyle{thmstyleone}%
\theoremstyle{thmstyletwo}%

\theoremstyle{thmstylethree}%

\usepackage{natbib}

\usepackage{subcaption}
\usepackage{longtable,tabularx}
\usepackage{booktabs}
\usepackage{amsmath}
\usepackage{verbatim}
\usepackage{bm}
\usepackage{mathtools}

\newcommand{\mat}[1]{{\mathbf{#1}}}

\newcommand{\mA}{\mat{A}}

\newcommand{\mI}{\mat{I}}

\newcommand{\mQ}{\mat{Q}}

\newcommand{\mU}{\mat{U}}
\newcommand{\mV}{\mat{V}}

\newcommand{\mSigma}{\mat{\Sigma}}

\newcommand{\bbf}{{ \bm{f}}}

\newcommand{\bu}{{ \mathbf{u}}}

\newcommand{\bv}{{ \bm{v}}}

\newcommand{\ba}{{ \mathbf {a}}}
\newcommand{\bq}{{ \mathbf {q}}}

\newcommand{\defeq}{\vcentcolon=}

\graphicspath{{Figures/}} 

\newcommand{\di}{i}

\begin{document}

\title[Galerkin spectral estimation]{Galerkin spectral estimation of vortex-dominated wake flows}

\author*[1,2]{\fnm{Katherine J.} \sur{Asztalos}}\email{kasztalos@anl.gov}

\author[2]{\fnm{Abdulrahman} \sur{Almashjary}}\email{aalmashjary@hawk.iit.edu}

\author[2]{\fnm{Scott T. M.} \sur{Dawson}}\email{scott.dawson@iit.edu}

\affil*[1]{\orgdiv{Transportation and Power Systems}, \orgname{Argonne National Laboratory}, \orgaddress{\street{S. Cass Ave.}, \city{Lemont}, \postcode{60439}, \state{IL}, \country{USA}}}

\affil[2]{\orgdiv{Mechanical, Materials, and Aerospace Engineering Department}, \orgname{Illinois Institute of Technology}, \orgaddress{\street{10 W. 32$^{\text{nd}}$ St.}, \city{Chicago}, \postcode{60616}, \state{IL}, \country{USA}}}


\abstract{We propose a technique for performing spectral (in time) analysis of spatially-resolved flowfield data, without needing any temporal resolution or information. This is achieved by combining projection-based reduced-order modeling with spectral proper orthogonal decomposition.  In this method, space-only proper orthogonal decomposition is first performed on velocity data to identify a subspace onto which the known equations of motion are projected, following standard Galerkin projection techniques. The resulting reduced-order model is then utilized to generate time-resolved trajectories of data. Spectral proper orthogonal decomposition (SPOD) is then applied to this model-generated data to obtain a prediction of the spectral content of the system, while predicted SPOD modes can be obtained by lifting back to the original velocity field domain. This method is first demonstrated on a forced, randomly generated linear system, before being applied to study and reconstruct the spectral content of two-dimensional flow over two collinear flat plates perpendicular to an oncoming flow. At the range of Reynolds numbers considered, this configuration features an unsteady wake characterized by the formation and interaction of vortical structures in the wake. Depending on the Reynolds number, the wake can be periodic or feature broadband behavior, making it an insightful test case to assess the performance of the proposed method. In particular, we show that this method can accurately recover the spectral content of periodic, quasi-periodic, and broadband flows without utilizing any temporal information in the original data. To emphasize that temporal resolution is not required, we show that the predictive accuracy of the proposed method is robust to using temporally-subsampled data.

}


\maketitle

\section{Introduction}\label{sec:intro}

Understanding the formation, evolution, and interaction of coherent vortical structures in the wakes of bluff body and aerodynamic flows is important for the prediction of aerodynamic loads and potential fluid-structure interactions.
There are a number of approaches for developing models for such systems, which may utilize some combination of the known/assumed equations of motion, and data collected from 
simulations or experiments. 
Perhaps the most ubiquitous method to identify coherent structures within data is the (spatial) proper orthogonal decomposition (POD) \cite{Lumley1967,lumley1970stochastic,sirovich1987turbulence,holmes2012pod,rempfer2000low}, which identifies an ordered set of spatial modes that are optimal at capturing the energy of the data.  
As well as identifying coherent structures, the modes identified from POD can be utilized to find a reduced-order model (ROM) of the system dynamics. 
This can be achieved via Galerkin projection, where the equations of motion are projected onto a subspace spanned by a set of POD modes, hereafter referred to as a POD-Galerkin model. 
This approach has been applied across a wide variety of applications, including wall-bounded turbulent flows \cite{aubry1988turbbl,rempfer1994bl,moehlis2002couette,smith2005pod,borggaard2008galerkinpipe,podvin2009galerkin}, the wake of a circular cylinder \cite{deane1991galerkin,noack1994galerkin,noack:03cyl}, cavity flows \cite{rowley2001cavity,rowley2006cavity}, and mixing layers \cite{rajaee1994mixing,ukeiley2001mixing,balajewicz2013jfm}. 
However, there are well-known limitations in the accuracy of Galerkin projection models, in part due to the fact that they are often unable to fully capture the dissipative dynamics of small scales that may not be captured by leading POD modes.  

A variety of methods have been proposed to mitigate this and other shortcomings of Galerkin-POD models, including the addition of eddy viscosity \cite{aubry1988turbbl,rempfer1994bl,podvin2009galerkin,osth2014podvisc}, LES-based closure models \cite{wang2012podclosure}, finite-time thermodynamic \cite{noack2008finite} and other multiscale statistical closure models \cite{callaham2022multiscale}, or directly constraining the model to be energy preserving \cite{balajewicz2013jfm,balajewicz2015minimal}. Cordier et al.~\cite{cordier2013identification} compare several of these and additional enhancements to the standard POD-Galerkin procedure to improve accuracy. Improvements to the accuracy of such models can also be achieved through error-minimization in a discrete-time setting \cite{carlberg2011efficient,carlberg2013gnat}, and in changing the subspace and direction of projection to optimally preserve the observability and controllability of the full system \cite{willcox:2002,rowley:05pod}. 

In this work, for reasons that will become apparent with the description of our methodology, we will not be concerned with the long-time stability of POD-Galerkin models, but will rather only need them to produce short time-series of approximate data that can be used as a surrogate for the full system.  In this way, we leverage the properties of POD-Galerkin models that are typically quite faithful to the full-order system.

More recently, the increasing availability of sufficient quantities of time-resolved data has lead to the development and application of methods for modeling and analysis of fluids problems that (in contrast to the POD method described above) directly utilize this temporal information within the data. If data is time-resolved, the original formulation of POD given in Lumley \cite{lumley1970stochastic} allows for the identification of spatiotemporal modes, which converge to Fourier modes in time for statistically-stationary systems. This space-time POD formulation was applied sparingly in the  decades following Lumley's formulation (e.g.~\cite{glauser1987coherent,picard2000pressure}), but has recently become popular due (aside from data availability) to the emergence of practical algorithms for its computation described in Towne et al.~\cite{towne2018spectral}, who also illuminate connections between this SPOD and other spectral analysis methods. Schmidt \& Colonius \cite{schmidt2020guide} further discuss and summarize the implementation of SPOD. 
The standard SPOD algorithm utilizes Welch's method that compiles Fourier transforms of many temporal windows of data, and thus requires a large total number of time-resolved snapshots. 

An alternative method to decompose time-resolved data to identify spatial modes corresponding to a single (and possibly complex) frequency is the dynamic mode decomposition \cite{schmid2008,schmid2010dynamic,rowley2009spectral}, which considers the eigendecomposition of a linear operator that approximates the time-evolution of the data.  

While advances in both experimental techniques and computational power and storage capacity make time-resolved data more readily available, there remain cost and technological limitations that can hamper the acquisition of adequate quantities of sufficiently time-resolved data to enable the application of the aforementioned methods. Accordingly, a number of methods have been proposed to enable the application of methods such as SPOD and DMD even when full time-resolved data is not available. 
For example, DMD can be modified to enable the use of data that is
under-resolved and/or non-uniformly-sampled in time by fitting temporal eigenfunctions to the full time series of data rather than only considering the discrete-time map between consecutive snapshots  \cite{chen2012variants,gueniat2015dynamic,leroux2016dynamic,askham2018variable}. 
Compressive sensing methods \cite{donoho2006compressed} can also be utilized to reconstruct DMD on data that is underresolved and randomly sampled in time \cite{tu2014spectral}, in doing so bypassing the restrictions of the Nyquist-Shannon sampling criterion \cite{nyquist1928certain,shannon1949communication}.
It is also possible to combine time-resolvent measurements of a small number of quantities with underresolved (in time) full state measurements (e.g. of a velocity field measured with particle image velocimetry), using stochastic estimation techniques. To this end, Tu et al.~\cite{tu2013integration} applies linear stochastic estimation and a Kalman smoother for flowfield reconstruction, while Zhang et al.~\cite{zhang2020spectral} utilize spectral stochastic estimation \cite{tinney2006spectral} to recover the SPOD of a velocity field using time-resolved pressure and non-time-resolved velocity field measurements. In addition to the above data-driven methods, it is also possible to apply physics-based methods to interpolate data that is under-resolved in time \cite{krishna2020reconstructing,wang2021model}.

The methodology for recovering spectral information proposed in the present work differs from the above methods in that it will not utilize any temporal information in the data. Instead, the temporal information will come from POD-Galerkin models, which will be used to generate the temporal windows of data required to perform SPOD.
 This is also distinct from promising recent work that directly uses SPOD modes in Galerkin projection models \cite{chu2021stochastic}, or that projects the governing equations in both space and time onto SPOD or resolvent modes \cite{towne2021space}.

Our proposed method is applicable in cases where spatially-resolved data is available, as required for computing these models. In particular, it should be well suited for flows where the dynamics are dominated by the interactions of large coherent structures that are well captured by the chosen POD subspace. Vortex-dominated wake flows are a particularly well-suited example, which we focus on here. In particular, we consider the two-dimensional wake behind two collinear plates, at a range of (relatively low) Reynolds numbers.

Flow over arrays of bluff bodies have a range of academic and industrial applications, including in the modeling and design of heat exchangers and high-rise buildings \cite{sohankar2006flow,gao2008airborne}. While flow over a single  bluff body is often characterized by periodic von K\'arm\'an vortex shedding, the presence of additional bodies can add considerable complexity, even for two-dimensional flows at relatively low Reynolds numbers.

Vortex shedding from adjacent bluff bodies  can feature vortex shedding that is either in-phase or anti-phase, with in-phase interactions forming a large-scale wake and anti-phase interactions forming parallel vortex streets \cite{williamson1985evolution}. There have been numerous studies examining the dependence upon Reynolds number and gap spacing between two or more adjacent bodies to quantify the flow patterns observed in the wake \cite{supradeepananalysis,bai2016flip,alam2003aerodynamic,kang2003characteristics}. At smaller gaps, ``flip-flopping" behavior is observed, while at higher gaps independent vortex streets start to form. Intermediate gaps can contain a rich array of interesting dynamics, with transitory behavior exhibiting, for example, random flip-flopping patterns. Common adjacent geometrical bodies of interest include circular cylinders; see \cite{zhou2016wake} for a detailed review of the interacting wake between two adjacent cylinders in various configurations. One of the problems of interest for this work is the flow over two adjacent plates, where the dynamics of the vortex-vortex interactions are dominated by the Reynolds number and gap spacing, as described in Refs.~\cite{guillaume2000investigation,miau1996flopping}. 
 Flow over a three-cylinder ``fluidic pinball" arrangement is a related configuration that has been the subject of  a number of recent modeling \cite{deng2020low,deng2022cluster} and control \cite{maceda2021stabilization} studies.  
 The bluff body array wake behaviour observed in these works are all characterized by the manner by which vortices form and shed from either side of the bodies, and subsequently evolve and interact in the wake. By varying the Reynolds number, we will use such a configuration to demonstrate the applicability of our proposed methodology to recover the spectral content of vortex-dominated wake flows featuring a single dominant frequency, multiple such frequencies, and broadband behavior.

The paper proceeds as follows. In section \ref{sec:methods} we describe the POD, SPOD, and Galerkin projection methods that underpin our proposed methodology for recovering spectral content from non-time-resolved data, which is presented in section \ref{sec:newmethod}. This method is subsequently applied and validated on a randomly-generated forced linear system in section \ref{results_toy}, and applied to flow over two collinear plates in section \ref{results_twoplates}.

\section{Modal analysis and projection-based modeling}\label{sec:methods}

This section describes the various existing methods that are utilized and combined in the present work. We first introduce the space-only proper orthogonal decomposition (POD) in section~\ref{sec:methods_POD}, before describing spectral POD (SPOD) in section \ref{sec:methods_SPOD}, and Galerkin projection of governing equations onto a spatial POD basis in section \ref{sec:methods_GP}.

\subsection{Proper orthogonal decomposition}\label{sec:methods_POD} Proper orthogonal decomposition (POD) is a technique that can be utilized to obtain a set of  spatial modes that optimally represent the data from an energetic perspective. Known variously in other fields as principal component analysis, the singular value decomposition of appropriately-formed matrices, and Karhunen-Loev\`{e} transformation in signal processing, the discussion of POD here follows much of its introduction to the field of fluid dynamics by Lumley \cite{Lumley1967,lumley1970stochastic} and utilizes the popular snapshot-method for discrete datasets \cite{sirovich1987turbulence}, which reduces the computational burden for cases where the number of measurements made at each timestep exceeds the total number of timesteps considered.

A vector field $\bu(x,t)$ (in our context, a velocity field) defined on a domain $\Omega_x$ can be decomposed into a superposition of time-varying coefficients with spatial basis functions such that 
\begin{equation}
     \bu(x,t) = \bu_{0}(x) + \sum_{i = 1}^{\infty} a_{i}(t)\bu_{i}(x), \label{POD}
\end{equation}
where $\bu_{0}(x)$ represents the mean of the data, $a_{i}(t)$ are the time-dependent coefficients, and $\{ \bu_i (x) \}_{i=1}^{\infty}$ represent a set of orthonormal basis functions, which are POD modes. Here space and time are denoted by $x$ and $t$ respectively. The POD modes are those which capture the most energy of the vector field; if velocity data is utilized, the POD modes capture the kinetic energy of the velocity field. Mathematically, this can be expressed as the POD modes being eigenfunctions satisfying
\begin{equation}
\label{eq:PODcont}
\int_{\Omega_x} C(x,x') \bu_i(x') dx' = \sigma_i^2 \bu_i(x),
\end{equation}
where $C(x,x') = \mathbb{E}[ \left(\bu(x,t)-\bu_0(x)\right)\left(\bu(x',t)-\bu_0(x')\right)]$ is the spatial correlation function, and the eigenvalues $\sigma_i^2$ are ordered such that $\sigma_i\geq\sigma_{i+1}\geq 0$. 
This property allows for the truncation of the infinite sum in  Eq.~\ref{POD} with minimal loss in accuracy, so that 
\begin{equation}
     \bu(x,t) \approx \bu_{0}(x) + \sum_{i = 1}^{r} a_{i}(t)\bu_{i}(x), \label{POD2}
\end{equation}
for some finite truncation index $r$. 
As in Eqs.~\ref{POD}-\ref{POD2}, POD is typically performed after first subtracting the mean of the given dataset (or in some cases, an equilibrium point). The set of POD modes $\bu_i$ then satisfy the homogeneous boundary conditions, with the mean $\bu_{0}(x)$ satisfying any nonhomogeneous boundary conditions of the system.  
This decomposition enables any vector field to be approximated as a linear superposition of the mean with a finite number of POD modes, which automatically satisfies the boundary conditions contained within the data.

Data can be collected from any system of interest and arranged to compute the POD, amounting to a discrete, finite-data approximation to Eq.~\ref{eq:PODcont}. 
Letting $\bq_j$ be the mean-subtracted measurements taken at time $t_j$, 
we form the snapshot matrix
\begin{equation}
    \mQ = 
    \left[ 
            \begin{array}{cccc}
            \vrule & \vrule &        & \vrule \\
            {\bq}_{1}   & {\bq}_{2}    & \cdots & \bq_{N_t}    \\
            \vrule & \vrule &        & \vrule 
  \end{array}
    \right].
\end{equation}
 The discrete analogue of Eq.~\ref{eq:PODcont} is then given by 
\begin{equation}
\mQ\mQ^T \bu\defeq \boldsymbol{C} \bu_i = \sigma_i^2 \bu_i,
\end{equation}
where here and throughout we suppress quadrature weights associated with the numerical approximation of the integral. 
In our case, each snapshot $\bq^{i}$ will include the components of a two-dimensional velocity field. Rather than computing the discrete spatial correlation matrix $\boldsymbol{C}$ directly, we can instead use the (typically smaller) time-correlation matrix, with the equivalent eigenproblem
\begin{equation}
\mQ^T\mQ \ba_i = \sigma_i^2 \ba_i, 
\end{equation}
where $\ba_i$ are the discrete temporal coefficients of the POD modes. 
Equivalently, the POD modes and coefficients can be found from the left and right (respectively) singular vectors in the singular value decomposition
\begin{equation}
    \mQ = \mU\mSigma\mV^{T} \approx \sum_{i=1}^{r}\bu_{i}\left(\sigma_{i}\bv_{i}^{T} \right) = \sum_{i = 1}^{r} \bu_{i}(x)a_{i}(t), \label{svd}
\end{equation} 
where the right equality shows  the explicit relationship between the truncated SVD and the finite-dimensional vector form of the POD expansion given in Eq.~\ref{POD2}. Here the POD modes $\bu_i$ are the columns of $\mU$, $\bv_i$ are the columns of $\mV$, and $\sigma_i^2$ denotes the energy content of the $i$-th mode. 
Note that POD produces an optimal  approximation of the data in the sense of minimizing the $l_2$-norm of the error between the original data and the low-rank approximation using a subspace of POD modes.

\subsection{Spectral proper orthogonal decomposition}\label{sec:methods_SPOD} In this section we provide an overview of the methodology for spectral proper orthogonal decomposition (SPOD); additional details can be found in Refs.~\cite{towne2018spectral,schmidt2020guide,schmidt2019efficient}. 
Unlike space-only POD described in section \ref{sec:methods_POD}, SPOD decomposes data into spatio-temporal functions. Theoretically, it can be formulated from a similar eigenproblem to Eq.~\ref{eq:PODcont}, 
\begin{equation}
\label{eq:SPODcont}
\int_{\Omega_{x},\Omega_t} C(x,x',t,t') \bu_i(x',t') dx'dt' = \sigma_i^2 \bu_i(x,t),
\end{equation}
where $C(x,x',t,t')$ is the spatio-temporal correlation function, and the integration is performed over both the spatial ($\Omega_x$) and temporal ($\Omega_t$) domains. For systems that are statistically-stationary in time, it can be shown that the temporal eigenfunctions are Fourier modes \cite{lumley1970stochastic}. This property means that the practical computation of these spatio-temporal eigenfunctions from data first requires a discrete Fourier transform (DFT) in time. 
Following the methods described in Refs.~\cite{towne2018spectral,schmidt2020guide}, this is achieved in practice by utilizing techniques such as Welch's method \cite{welch1967use}, where the power spectral density (PSD) of vector-valued data is computed from the DFT of a number of (possibly overlapping) windows of data.

In this method, time series of data consisting of $N_t$ equally-spaced snapshots can be segmented into a series of $N_{\text{blk}}$ blocks which cover a time period $T$, where each segment contains $N_{\text{FFT}}$ snapshots. 
 In contrast to the space-only formulation of POD described in section \ref{sec:methods_POD}, we emphasize that SPOD requires that data be resolvent in time (and with a constant timestep). 
In our case, we will also consider the case where windows of data come not from a single time series, but instead from running models from different initial conditions.

To estimate the spectrum, Welch's method utilizes the discrete Fourier transform of each block of data (indexed by $k$), with this Fourier transform and its inverse given by
\begin{align}
    \hat{\bq}^{\left(k \right)}_m = \sum_{j=0}^{N_{\text{FFT}}-1} \bq^{\left(k \right)} \left(t_{j+1} \right) \exp\left(-\dfrac{\di 2 \pi j m}{N_{\text{FFT}}}\right), \\ \nonumber m=-\frac{N_{\text{FFT}}}{2}+1, \dots , \frac{N_{\text{FFT}}}{2},
\end{align}
and
\begin{align}
    \bq^{\left(k \right)}\left( t_{j+1} \right) = \dfrac{1}{N_{\text{FFT}}} \sum_{m=-N_{\text{FFT}}/2+1}^{N_{\text{FFT}}/2} \hat{\bq}^{\left(k \right)} _m \exp\left({\dfrac{\di 2 \pi j m}{N_{\text{FFT}}}}\right), \\ \nonumber j=0, \dots , N_{\text{FFT}}-1 . 
\end{align}

After this transformation, the data is represented at a set of discrete frequencies given by $f_m = {m}/T$, where $m=-N_{FFT}/2+1, \dots, N_{FFT}/2$ is the frequency index. The 
 Fourier-transformed data at each frequency is then selected from each block, and arranged into a frequency-specific 
 data matrix 
\begin{equation}
    {\hat{\mQ}_m} = 
    \left[ 
            \begin{array}{cccc}
            \vrule & \vrule &        & \vrule \\
            \hat{\bq}_{m}^{\left(1 \right)}   & \hat{\bq}_{m}^{\left(2 \right)}    & \cdots & \hat{\bq}^{\left(N_{\text{blk}} \right)}_{m}    \\
            \vrule & \vrule &        & \vrule 
  \end{array}
    \right], 
\end{equation}

In analogy with Eq.~\ref{svd}, the SVD $\hat\mQ_m= \hat{\mU}_m\hat{\bf\Sigma}_m\hat\mV_m^{*} 
$ yields left singular vectors $\hat\bu_{m,i}$ that are the SPOD modes corresponding to frequency index $m$, with the energy of the modes given by the corresponding singular values (squared). Note that using a $\hat{\cdot}$ to indicate SPOD modes is a slight abuse of notation, since these do not directly correspond to the Fourier transform of the space-only POD modes. 
 Unlike the case for space-only POD, the right singular vectors do not have any physical meaning in the time domain, and instead represent coefficients relating the Fourier transformed data for a given window and frequency to the SPOD modes at that frequency.

\subsection{Galerkin Projection}\label{sec:methods_GP} 
Galerkin projection is a method that can be used to  approximate a differential equation by projecting onto a finite-dimensional basis. 
Suppose that we have a differential equation that can be represented by
\begin{equation}
\label{eq:DE}
    \dot\bu'=\bbf(\bu'),
\end{equation}
where $\dot\bu$ is the time derivative of $\bu$, and $\bu'(x,t) = \bu(x,t)-\bu_0(x)$ is the system state relative to some fixed mean or equilibrium state. 

We seek an approximate solution to Eq.~\ref{eq:DE} of the form 
\begin{equation}
    \bu'(x,t) =  \sum_{i=1}^{r} a_{i}(t) \bu_{i}(x), \label{approxsoln}
\end{equation}
where the $\bu_{i}$'s are the trial functions and the $a_{i}$'s their coefficients. That is to say, we look for a solution that is in the span of the set of functions $\{ \bu_i\}_{i=1}^r$. In matrix form, we can express Eq.~\ref{approxsoln} as
\begin{equation}
\bu' =  \mU_r \ba, \label{approxsoln2}
\end{equation}
where $\mU_r$ is a matrix with columns $\bu_i$, and at a given time $\ba$ is a column vector containing the coefficients $a_i$. Substituting Eq.~\ref{approxsoln2} into Eq.~\ref{eq:DE}, we obtain 
\begin{equation}
\label{eq:DE2}
    \mU_r\dot\ba=\bbf(\mU_r\ba).
\end{equation}
If the trial functions $\bu_i$ are orthonormal, then $\mU_r^T\mU_r = \mI$, so we can premultiply both sides of Eq.~\ref{eq:DE2} by $\mU_r^T$ to obtain
\begin{equation}
\label{eq:GalerkinModel}
    \dot\ba= \mU_r^T\bbf(\mU_r\ba),
\end{equation}
which is the general form of an $r$-th order Galerkin projection model for the original differential equation.

To give an alternate description to this analysis, substituting the trial function expansion Eq.~\ref{approxsoln} into the original differential produces (in general) a non-zero residual, which we can set to zero once projected onto a set of test functions. Here, the test functions and trial functions are both chosen to be the functions $\bu_i$. Geometrically, we are projecting the equations onto a subspace spanned by the $\bu_i$'s in a direction orthogonal to this subspace. In more general Petrov-Galerkin methods, the test and trial functions do not need to be the same (and equivalently the projection does not need to be orthogonal to the subspace).

As suggested by the notation and the similarity between Eqs.~\ref{POD} and \ref{approxsoln}, we can choose the test and trial functions to be the first $r$ POD modes identified from data obtained from simulations of Eq.~\ref{eq:DE}, giving the methodology to obtain POD-Galerkin ROMs.

For a linear system with $\bbf(\bu') = \mA \bu'$ for a $n\times n$ square matrix $\mA$, the $r$-th order Galerkin projection model is
\begin{equation}
    \dot\ba = \left(\mU_r^T\mA\mU_r\right)\ba\defeq \mA_r \ba,
    \label{eq:gplin}
\end{equation}
where $\mA_r$ is an $r\times r$ matrix.

We now describe the application of this Galerkin projection procedure to the incompressible Navier--Stokes equations, 
\begin{equation}
  \frac{\partial \bu}{\partial t} = -(\bu \cdot \nabla)\bu - \dfrac{1}{\mathrm{Re}} \Delta\bu +\nabla p, \label{NS} 
\end{equation}
 where $\bu$ and $p$ are the velocity and pressure fields.
  Upon substituting the POD approximation 
  \begin{equation}
    \bu(x,t) \approx \bu_{0}(x) + \sum_{i=1}^{r} a_{i}(t)\bu_{i}(x), \label{veldecomp}
\end{equation}
into these equations and taking an inner product with each of the first $r$ POD modes, we obtain a ROM of the form
\begin{align}
    \nonumber \dfrac{d}{dt}a_{i} = \dfrac{1}{\text{Re}} \sum_{j=0}^{r} \textbf{l}_{ij}  a_{j} + \sum_{j=0}^{r} \sum_{k=0}^{r} \textbf{q}_{ijk}  a_{j}  a_{k} + f_{i}^{p} \\
    \quad \text{for} \quad i = 1, ..., N, \label{dadt_NS}
\end{align}
where the coefficients are given by $\textbf{l}_{ij} = \langle \bu_{i}, \Delta \bu_{j} \rangle_{\Omega_x}$ and $\textbf{q}_{ijk} = -\langle \bu_{i}, \bu_k \langle\bu_{j} \cdot \nabla \rangle \rangle_{\Omega_x}$, and with the unknown pressure term given by $f_{i}^{p} = \langle \bu_{i}, -\nabla p \rangle_{\Omega_x}$. In this work the pressure term has been neglected, as the homogeneous boundary conditions result in negligible pressure terms. Note also that the accurate computation of these inner products (and the terms within them) requires spatially resolved data. 
For more details concerning the formulation and properties of POD-Galerkin models, see Refs.~\cite{berkooz1993pod,holmes2012pod,rempfer2000low,noack:03cyl}, for example.

\section{Galerkin spectral estimation}
\label{sec:newmethod} 
This section introduces a methodology, which we term Galerkin spectral estimation (GSE), that can utilize POD-Galerkin models, identified from the procedure described in section \ref{sec:methods_GP}, to estimate the spectral (in time) behavior of a system, as characterized by SPOD as described in section \ref{sec:methods_SPOD}, without needing the original data to be resolved in time. 
For clarity and convenience, the full algorithm is summarized as Algorithm 1 below.

\smallskip \noindent \textbf{Algorithm 1}: Galerkin Spectral Estimation 

\begin{enumerate}
    \item Collect $N_t$  velocity field snapshots, compute and subtract the temporal mean at each spatial location, and assemble the mean-subtracted snapshots into the data matrix
    \begin{equation*}
    \mQ = 
    \left[ 
            \begin{array}{cccc}
            \vrule & \vrule &        & \vrule \\
            {\bq}^{1}   & {\bq}^{2}    & \cdots & \bq^{N_t}    \\
            \vrule & \vrule &        & \vrule 
  \end{array}
    \right].
\end{equation*}
    \item Compute the POD of the data matrix $\mQ$, from the SVD
    \begin{equation*}
    \mQ=\mU\mSigma\mV^{T} \approx \sum_{i=1}^{r}\bu_{i}\left(\sigma_{i}\bv_{i}^{T} \right) = \sum_{i = 1}^{r} \bu_{i}(x)a_{i}(t)
    \end{equation*}
    \item Compute the Galerkin projection model for the time-evolution of the POD coefficients from
    \begin{equation*}
 \dot\ba= \mU_r^T\bbf(\mU_r\ba).
    \end{equation*}
    \item Evolve the Galerkin projection model from a chosen set of initial conditions to obtain time-resolved data (in terms of the POD coefficients, $\ba(t)$).
    \item Perform SPOD on the predicted POD coefficient trajectories, where each trajectory is used for one or more (possibly overlapping) DFT windows.
    \item The resulting SPOD mode energies should approximate those for the full system. The SPOD modes computed for the POD-Galerkin models $\hat\psi_{m,i}$ can be mapped back into physical space using the (space-only) POD modes:
    \begin{equation*}
        \hat{\bu}_{m,i} = \mU_r\hat\psi_{m,i}.
    \end{equation*}
\end{enumerate}
There are several choices to be made when implementing this procedure. In step 2, a decision must be made for the truncation level, $r$. 
Note that in this method the computed SPOD modes are limited to the subspace spanned by the first $r$ POD modes, which could make it difficult to accurately estimate the results of SPOD at frequencies for which the system has relatively low energy. The choice of $r$ can also affect the accuracy of the resulting POD-Galerkin model computed in step 3. Note, however, that these models do not strictly need to be asymptotically stable, since only finite-time windows of data are required for the application of SPOD. 

In step 4, the POD-Galerkin model is used to generate data to be used to estimate the SPOD of the full system. If we started with a single trajectory of time-resolved data in step 1, then one could most simply try to reconstruct this trajectory from the first snapshot. However, the ROM is not limited to simply reconstructing the initial data. To start with, the comparative low computational cost of evaluating (and low storage requirements for storing states from) a ROM likely makes it feasible to generate many more snapshots than were present in the initial dataset. This could include both running the ROM for a longer total time, and saving snapshots at shorter time intervals, to enable the analysis of higher frequencies in SPOD. Indeed, the initial data used for POD does not need to be sampled uniformly in time, or be time-resolved at all. Rather than using a single trajectory of data, the ROM could also be evaluated along multiple trajectories, each with a different initial condition. This could be advisable if the ROM loses accuracy over long time horizons (e.g. with non-physical growth or decay of total energy). These initial conditions could be chosen from selected snapshots of the initial data, or could be sampled more broadly. 
Particularly if choosing initial conditions that are not directly taken from  the initial data, it could be advisable to remove an initial section of the data before further analysis to eliminate any rapid transients associated with a nonphysical initial condition.
The minimum length of each trajectory is the window size used for each block in the SPOD algorithm; a longer trajectory could be used to generate multiple windows. If concatenating data from several trajectories, it is important to ensure that each window is only associated with a single trajectory.

In step 6, the SPOD of the ROM is mapped back into physical space. This mapping from POD coefficients to velocity fields could also be performed prior to applying SPOD in step 5 with no effect on the results, though this would make the SPOD computationally expensive due to the larger state size.

The following sections demonstrate the use of this proposed GSE method in several contexts. First, in section~\ref{results_toy}, we apply this method to a randomly-generated linear system that is forced with Gaussian white noise. Following this, in section~\ref{results_twoplates} we apply this method to two dimensional flow past two collinear flat plates, a system governed by the incompressible nonlinear Navier--Stokes equations.  In this section example, we vary both the manner in which the data is collected (in terms of time resolution), and the complexity of the dynamics through variation of the Reynolds number.

\section{Galerkin spectral estimation of a forced random linear system}\label{results_toy}

To test and validate the proposed method, we first apply it to data obtained from a randomly-generated linear system that is continually forced with white noise.  
The random system (which is generated using the rss command in MATLAB) takes the form
\begin{align}
    \dot{\textbf{x}}(t) = \textbf{A}\textbf{x}(t) + \textbf{B}u(t) 
    \label{eq:ss}
\end{align} 
Here the $n$-dimensional state vector is $\textbf{x}(t)$, 
and the input (or control) vector is $u(t)$. A single input channel is used, with $\textbf{B}$ being a column vector of ones, so that the random input excites each state in the same manner.

\begin{figure}[!htb]
  \centering
\begin{subfigure}{0.9\textwidth}
  \centering
    \includegraphics[trim=4cm 14cm 4cm 15.5cm,clip,width=1.0\linewidth]{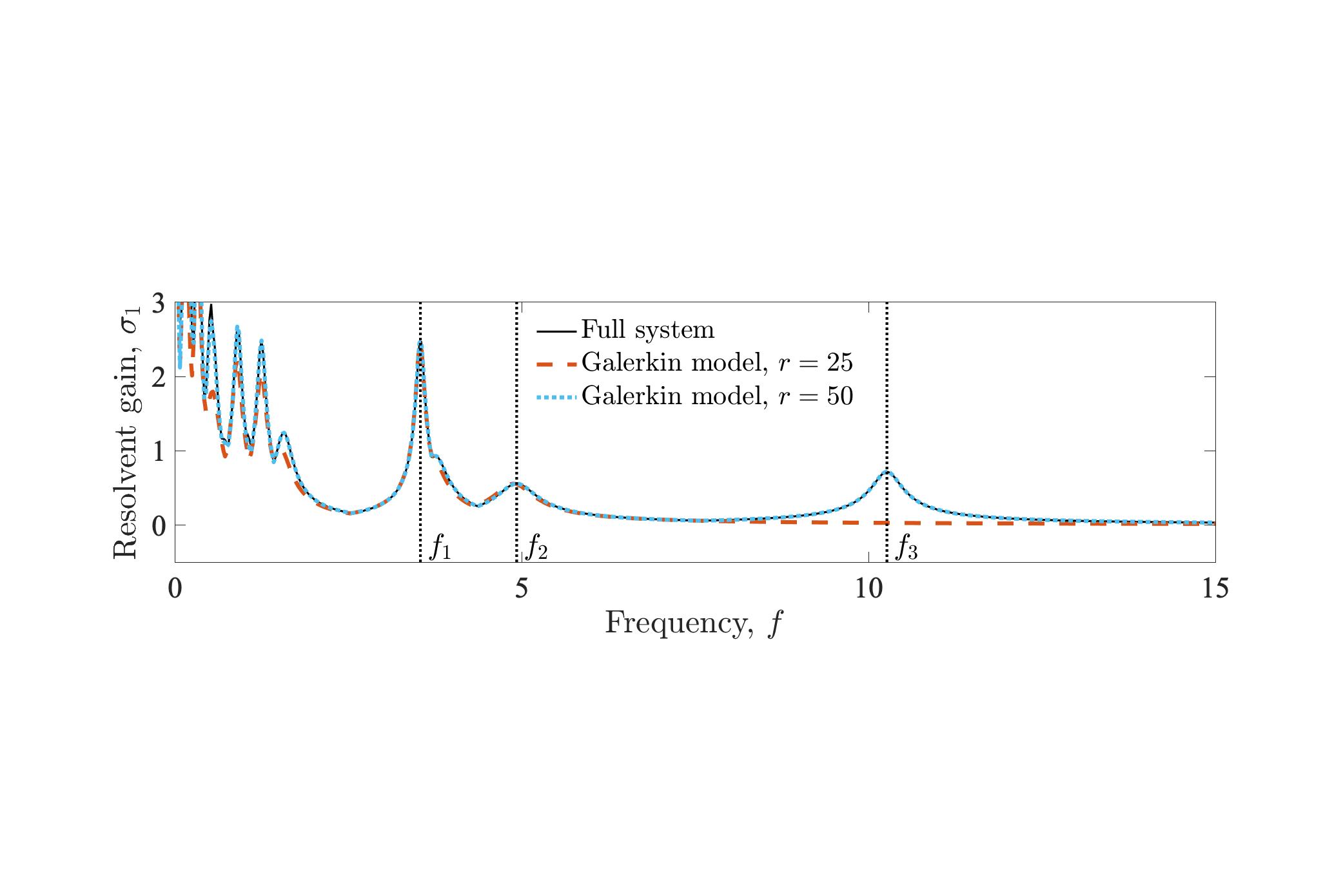}
  \caption{}
  \label{fig:pseudo_tp}
\end{subfigure}
\begin{subfigure}{0.9\textwidth}
  \centering
\includegraphics[trim=4cm 14cm 4cm 15.5cm,clip,width=1.0\linewidth]{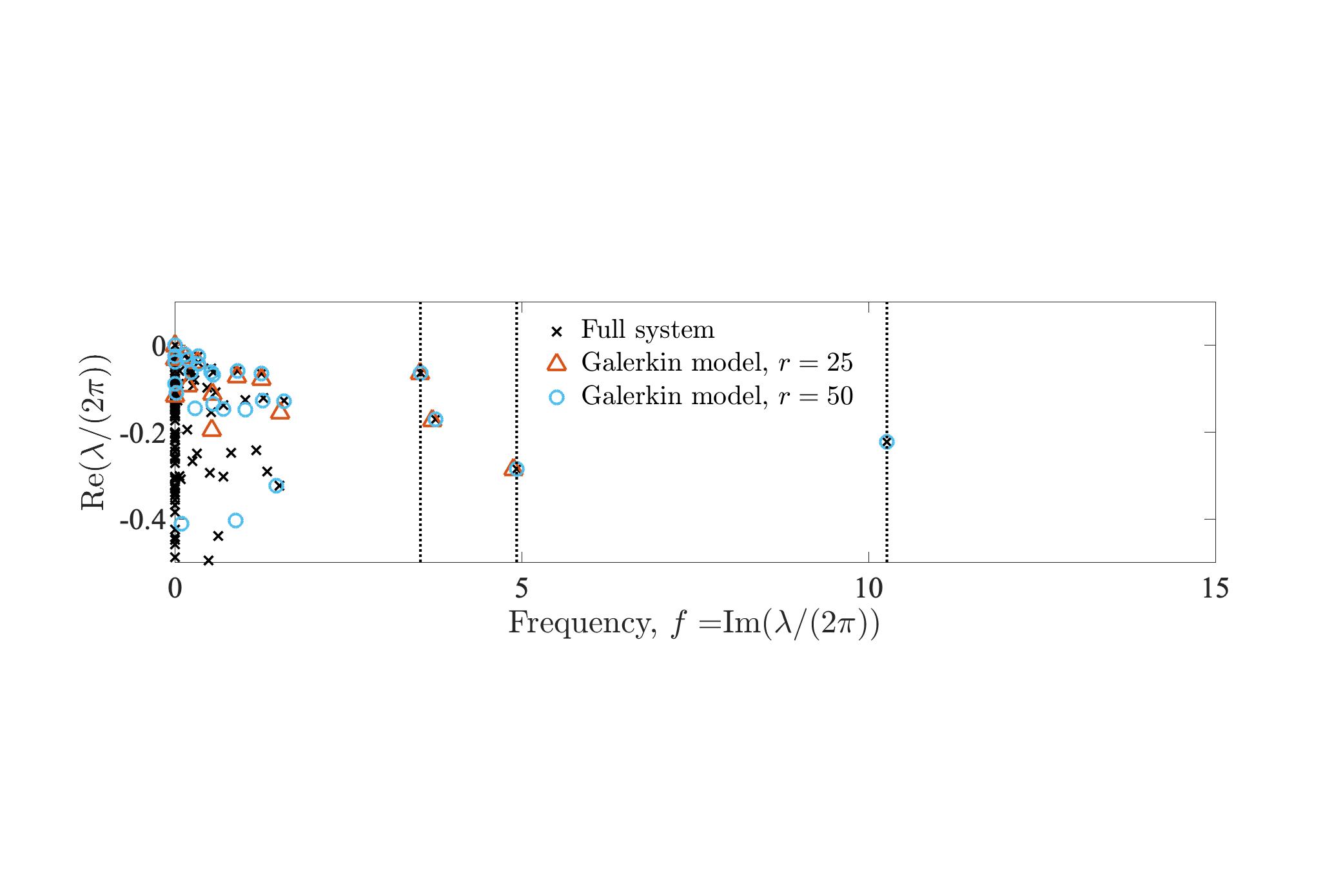}
  \caption{}
  \label{fig:spectra_tp}
\end{subfigure}
\caption{(a) Pseudospectra and (b) eigenspectra of the full random linear system ($\mA$) and Galerkin-POD ROMs with $r=25$ and $r=50$. Vertical dashed lines indicate frequencies $f_1$--$f_3$ that will be the focus of the subsequent analysis of the proposed GSE methodology.}
\label{fig:tp_results}
\end{figure}

The random system is generated with 200 states, and is evolved to collect $N_t = 120,000$ snapshots, with a timestep of $\Delta t = 0.02$, for an input $u(t)$ consisting of Gaussian white noise with unit variance applied using the same timestep.  As this is a linear system forced with white noise, its SPOD should agree with the properties of its pseudospectrum \cite{towne2018spectral} (i.e. with the results of resolvent analysis). For a linear system of the form given by Eq.~\ref{eq:ss}, the pseudospectrum corresponding to a specified (real) frequency $\omega$ is characterized by the operator norm (leading singular value, $\sigma_1$)
\begin{equation}
    \sigma_1 = \max_{\|\phi\| = 1}\| \mathcal{H}_\omega  \phi \| \defeq \| \mathcal{H}_\omega  \|,
\end{equation}
where the resolvent $\mathcal{H}_\omega$ is given by
\begin{equation}
    \mathcal{H}_\omega = \left(\omega \mI-\mA\right)^{-1}.
\end{equation}
The pseudospectrum of $\mA$ is shown in Fig.~\ref{fig:pseudo_tp}, along with the pseudospectra of POD-Galerkin ROMs identified from applying Eq.~\ref{eq:gplin} using POD modes identified from the data generated as described above, for ROMs using 25 and 50 POD modes (which specifies the size of $\mA_r$). Here the first 5,000 snapshots (corresponding to 100 time units) are excluded from the analysis to ensure that any transient effects are eliminated. Aside from numerous low-frequency peaks, we observe  three high-frequency peaks in the pseudospectrum of $\mA$, corresponding to frequencies $\{f_1,f_2,f_3\} = \{ 3.54,4.93,10.27\}$ shown on the plot.

It is observed that the ROM identified by projecting $\mA$ onto $r =50$ POD modes captures each of these peaks, while the $r=25$ model misses the peak corresponding to frequency $f_3$.  Fig.~\ref{fig:spectra_tp} shows the eigenvalues of $\mA$ and $\mA_r$ for each $r$ value. These eigenspectra demonstrate that these peaks in the pseudospectra arise from modal amplification due to the presence of eigenvalues at these frequencies $f_1$--$f_3$. Here and throughout, we express frequencies in oscillations per time unit (which is why the real and imaginary components of the eigenvalues $\lambda$ are divided by $2\pi$ to convert from radians per time unit). Note also that only positive frequencies are shown here, though spectra and pseudospectra are symmetric about the vertical axis as the eigenvalues of the real matrices $\mA$ and $\mA_r$ come in complex conjugate pairs.

The absence of a peak in the pseudospectrum at $f_3$ for the $r=25$ model is due to the lack of an eigenvalue of $\mA_r$ at this frequency. 
As we are interested in recovering these frequencies using the proposed GSE methodology, we henceforth focus on the $r=50$ ROM. 

\begin{figure}[!htb]
	\centering
	{\includegraphics[clip,trim=1cm 0cm 1cm 0cm,width = 0.8\linewidth]{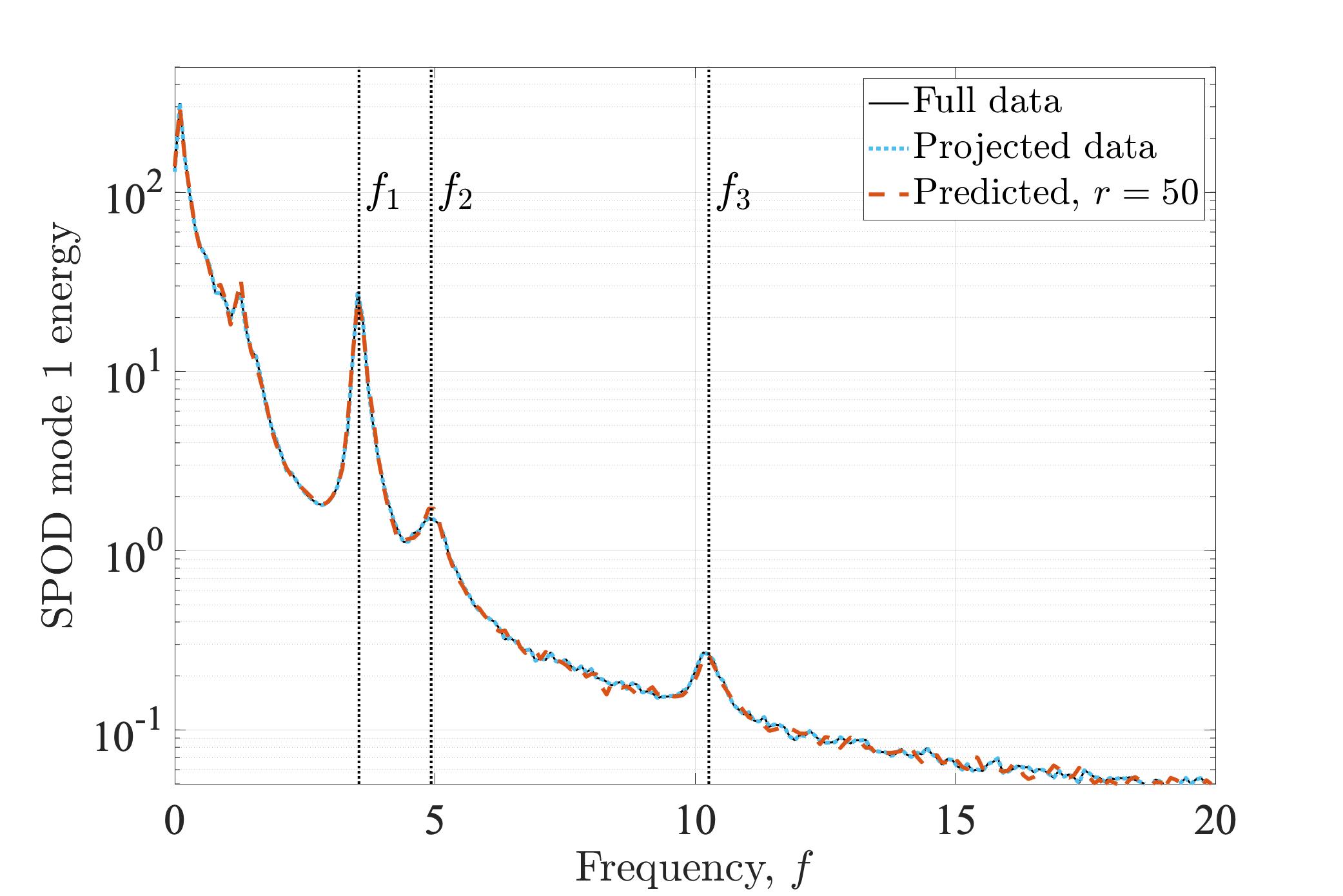}}
	\caption{Comparison between the SPOD spectra computed from the original data, this data directly projected onto the leading 50 POD modes, and using GSE with data generated from a ROM identified using these $r=50$ POD modes. In all cases SPOD is computed using $\Delta t = 0.02$ and a window size of 512 snapshots. The vertical lines indicate the pertinent frequencies identified from the pseudospectra in Fig.~\ref{fig:pseudo_tp}.
 }
	\label{fig:toy_SPOD_spectra1}
\end{figure}

We now apply SPOD to the data generated as described above, using DFT Hamming windows each containing 512 snapshots, with a 50\% overlap between adjacent windows. This results in a total of 448 DFT windows. In Fig.~\ref{fig:toy_SPOD_spectra1}, we show the leading SPOD energy levels for the original data, the data projected onto the first 50 POD modes, and data generated using the $r=50$ ROM. There is close agreement between all cases, and the three frequencies identified from the pseudospectra of the underlying operators $\mA$ and $\mA_r$ are all accurately
identified as peaks in the SPOD spectra. The agreement between the results for the original and projected data confirm that the original data can be closely approximated by 50 POD modes, suggesting that it should be possible for a ROM of this order to accurately capture the dynamics of the full system. Once the ROM is identified, following the GSE methodology described in Algorithm 1 it is simulated for the same number of timesteps as the original system, but with a different realization of white noise. We emphasize that this ROM does not directly utilize the temporal information of the original data, only the associated POD modes and knowledge of the original equations.

\begin{figure}[!htb]
	\centering
	{\includegraphics[clip,trim=1cm 0cm 1cm 0cm,width = 0.8\linewidth]{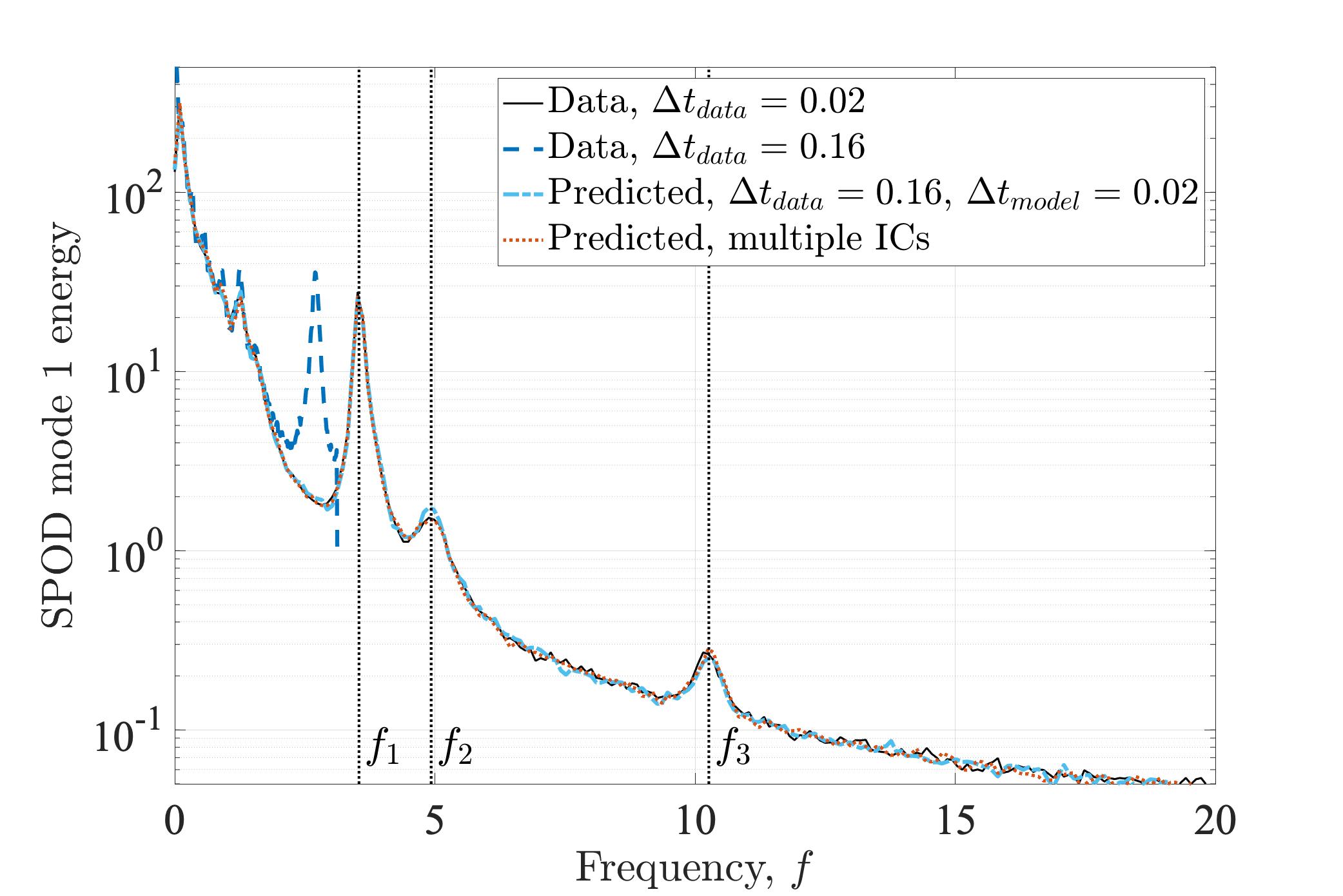}}
	\caption{Comparison between the SPOD spectra identified from finely ($\Delta t_{data} = 0.02$) and coarsely ($\Delta t_{data} = 0.16$) sampled data, and predicted/recovered from the coarsely sampled data using GSE. The GSE data is generated using either a single trajectory, or many different trajectories (multiple initial conditions). }
	\label{fig:toy_SPOD_spectra2}
\end{figure}

To emphasize this point more directly, in Fig.~\ref{fig:toy_SPOD_spectra2} we consider a case where data is sampled at a coarser timestep 
 of $\Delta t = 0.16$. As shown, performing SPOD directly on this dataset cannot recover any of the frequencies $f_1$--$f_3$, with the associated Shannon-Nyquist sampling frequency of $1/(2\Delta t) =  3.125$ smaller than $f_1 = 3.54$. 
Moreover, due to aliasing this SPOD incorrectly identifies a peak at $f_1/2$. This SPOD of the data collected at this coarser timestep uses the same total number of snapshots and DFT window size. 
While this coarsely-sampled dataset cannot directly identify the correct frequency content, it can still be utilized to identify a POD subspace to apply the GSE method. We show that application of this technique generated predicted data that can accurately recover the SPOD energy content, including correct identification of the location and size of each of the peaks corresponding to $f_1$--$f_3$. Once the ROM is identified, we apply two variants of GSE. In the first case, we again use the ROM to generate a single time-series of data, with the same timestep $\Delta t = 0.02$ and total number of snapshots $N_t=120,000$ as originally considered (again using a different realization of white noise). In the second case, we run a total of 448 simulations of the ROM, each for only 1024 snapshots of data (again with $\Delta t = 0.02$). Within these short simulations, the first 512 snapshots are discarded to minimize initial transient effects, while the next 512 snapshots are used for a single DFT window. This allows us to apply SPOD using the same number and size of windows as the other cases. While not required here, this variant which uses many short simulations of the ROM could be particularly useful in cases where the ROM has nonphysical growth or decay in energy over long time intervals. 

Of course, in this section the example considered was small enough such that simulating and storing data for the full system is not computationally burdensome. However, it has been demonstrated that the proposed GSE method can indeed accurately recover spectral information for the system, even when the initial data is not fully time-resolved. 

We conclude this section by noting that in the case of linear systems in particular, there exist alternative methods to obtain ROMs that can be more accurate than POD-Galerkin projection, such as balanced truncation \cite{moore1981ieeetac}, which accounts for the controllability and observability of the linear system.

\section{Galerkin spectral estimation of flow over two collinear plates}\label{results_twoplates}

In this section, we apply GSE to incompressible, two-dimensional, viscous flow over two flat plates that are collinear and arranged perpendicular to the oncoming flow. Unlike the simple example considered in Sect.~\ref{results_toy}, here the governing Navier--Stokes equations are nonlinear and high-dimensional. 

\begin{figure}[!htb]
	\centering
	{\includegraphics[clip,trim=0cm 0cm 0cm 0cm,width = 0.75\linewidth]{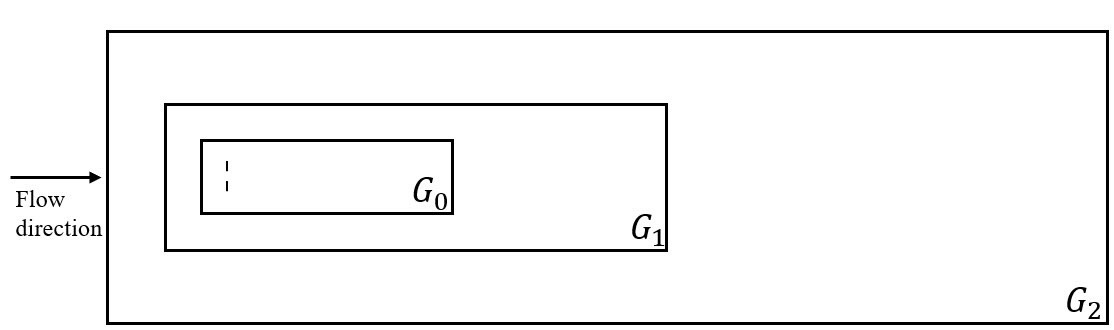}}
	\caption{Spatial domain illustrating the separate grids and placement of the two flat plates simulated. Note that results are shown here for the grid described by $G_1$.}
	\label{fig:twoplates_diagram}
\end{figure}

Direct numerical simulations are performed using an immersed boundary projection method (IBPM) \cite{taira2007immersed,colonius2008fast}, where no-slip boundary conditions associated with the solid bodies are enforced through the use of Lagrangian multipliers. This formulation is solved using a fractional step (projection) method \cite{chorin1968numerical,temam1969approximation}, where here a nullspace method allows for the divergence-free condition to be enforced without needing to solve for the pressure separately \cite{colonius2008fast}. 
The flow state is solved on three nested uniform grids, as shown in Fig.~\ref{fig:twoplates_diagram}. 
The use of uniform grids allows for computational speedup with a fast sine transform \cite{colonius2008fast}, while the use of multiple grids with different resolutions reduces the total computational cost, by not requiring the fine resolution needed near the body to expend across the entire computational domain.

The finest grid ($G_0$) has a grid spacing $\Delta x = \Delta y = 0.02c$, where $c$ is the chord length of a single plate. Grid $G_0$ has a total extent of $27c\times 7 c$, with $1200
\times 350 = 420,000$ total grid points. The two collinear plates have a gap of $c$ (corresponding to a dimensionless gap ratio $g^* =1$) between them, and are located $2c$ from the front of grid $G_0$. Each of the two larger grids $G_1$ and $G_2$ is twice the total size of the previous in both the horizontal (streamwise) and vertical directions, with half the spatial resolution. The larger grids are offset such that the majority of the domain is downstream of the plates. The resolution of the grid for the Reynolds numbers considered here has been validated in a range of past studies \cite{brunton2013reduced,brunton2014state,dawson2016lift,dawson2017reduced}. For the analysis, we will use velocity data from the $G_1$ grid, where $\Delta x = \Delta y = 0.04 c$. 

We consider flow at three Reynolds numbers, Re = 40, 80, and 100. 
While this particular flow does not appear to have been studied in past works, the qualitative observations of the dynamical features of each flow are broadly consistent with prior results observed for cylinder arrays \cite{williamson1985evolution,zhou2016wake, deng2020low}. 

\begin{figure}[!htb]
  \centering
\begin{subfigure}[!htb]{.6\textwidth}
  \centering
  \includegraphics[trim=2cm 10cm 4cm 12cm,clip,width=1\linewidth]{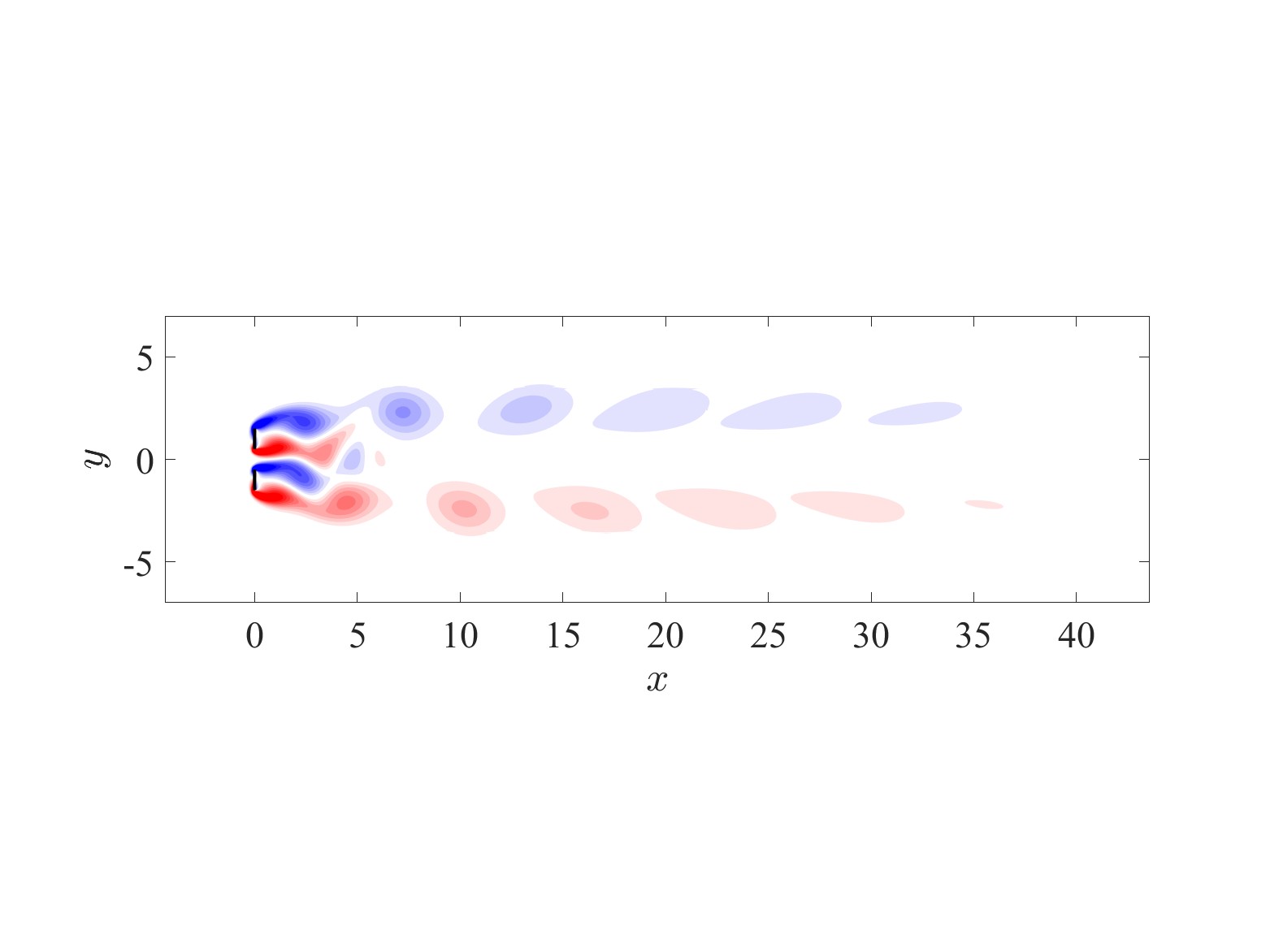}
  \caption{}
  \label{fig:tp_vort_Re40}
\end{subfigure}
\begin{subfigure}[!htb]{.35\textwidth}
  \centering
  \includegraphics[trim=0cm 0cm 2cm 0cm,clip,width=1\linewidth]{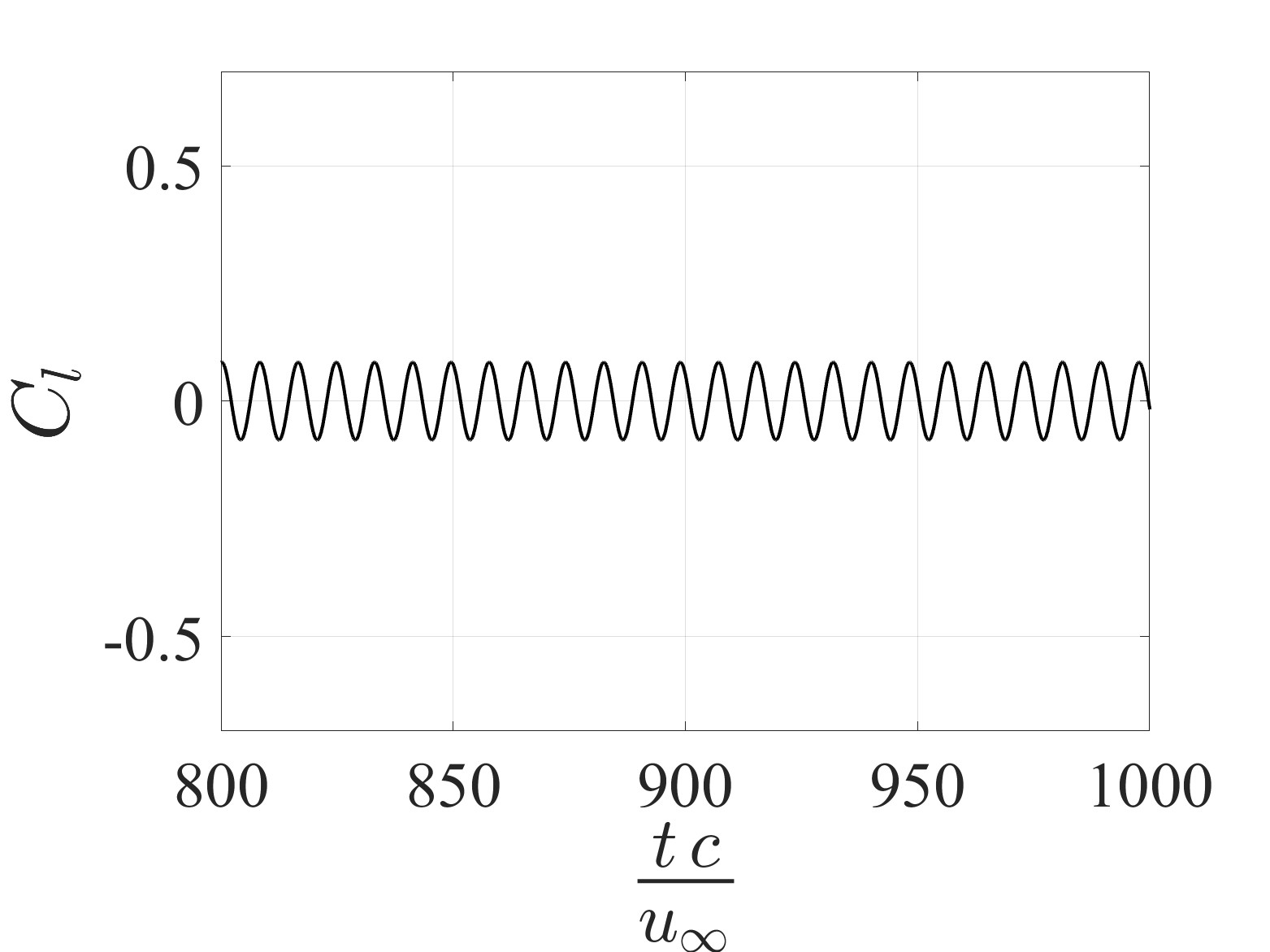}
  \caption{}
  \label{fig:cl_Re40}
\end{subfigure}
\begin{subfigure}[!htb]{.6\textwidth}
  \centering
  \includegraphics[trim=2cm 10cm 4cm 12cm,clip,width=1\linewidth]{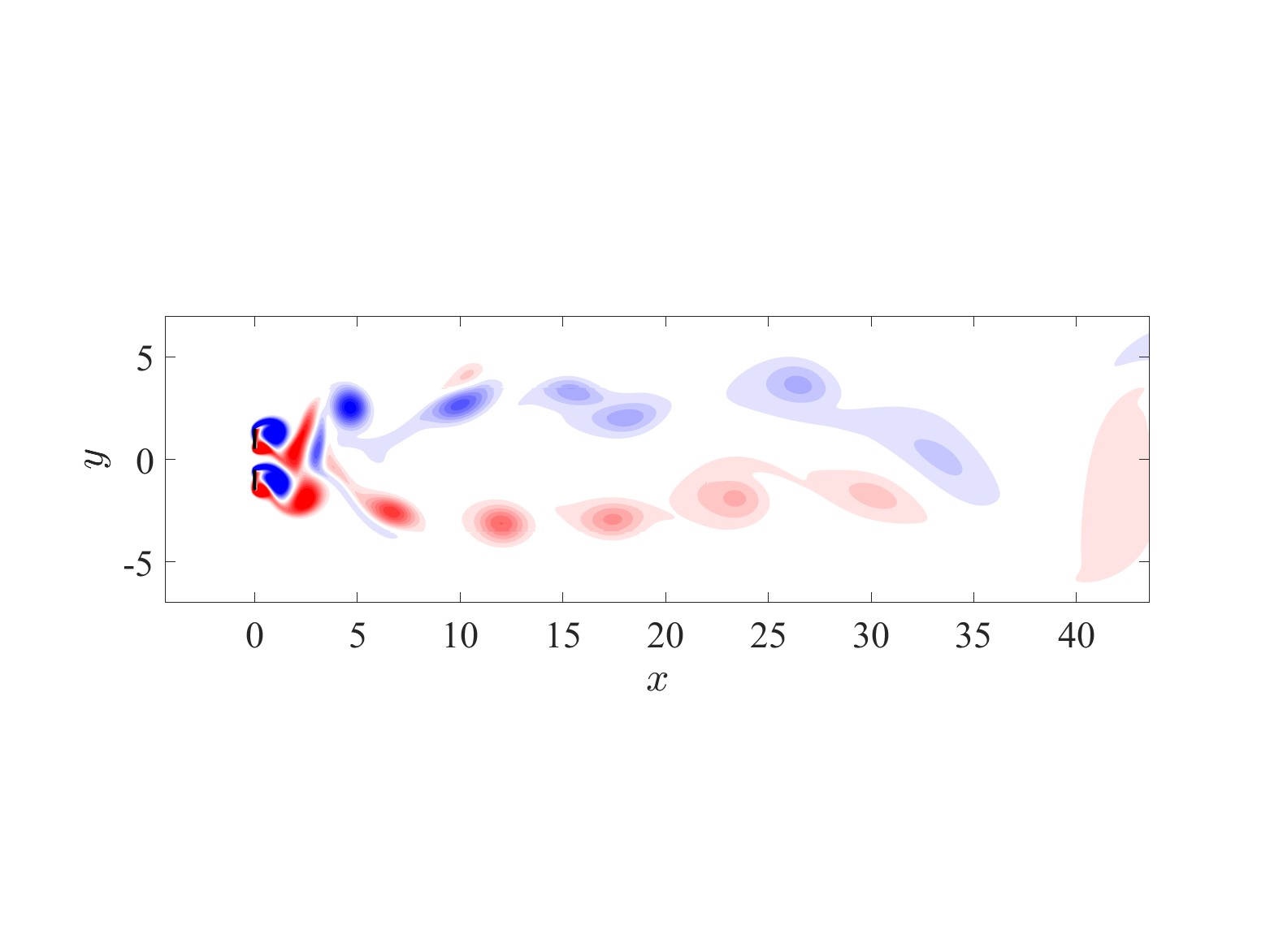}
  \caption{}
  \label{fig:tp_vort_Re80}
\end{subfigure}
\begin{subfigure}[!htb]{.35\textwidth}
\centering
\includegraphics[trim=0cm 0cm 2cm 0cm,clip,width=1\linewidth]{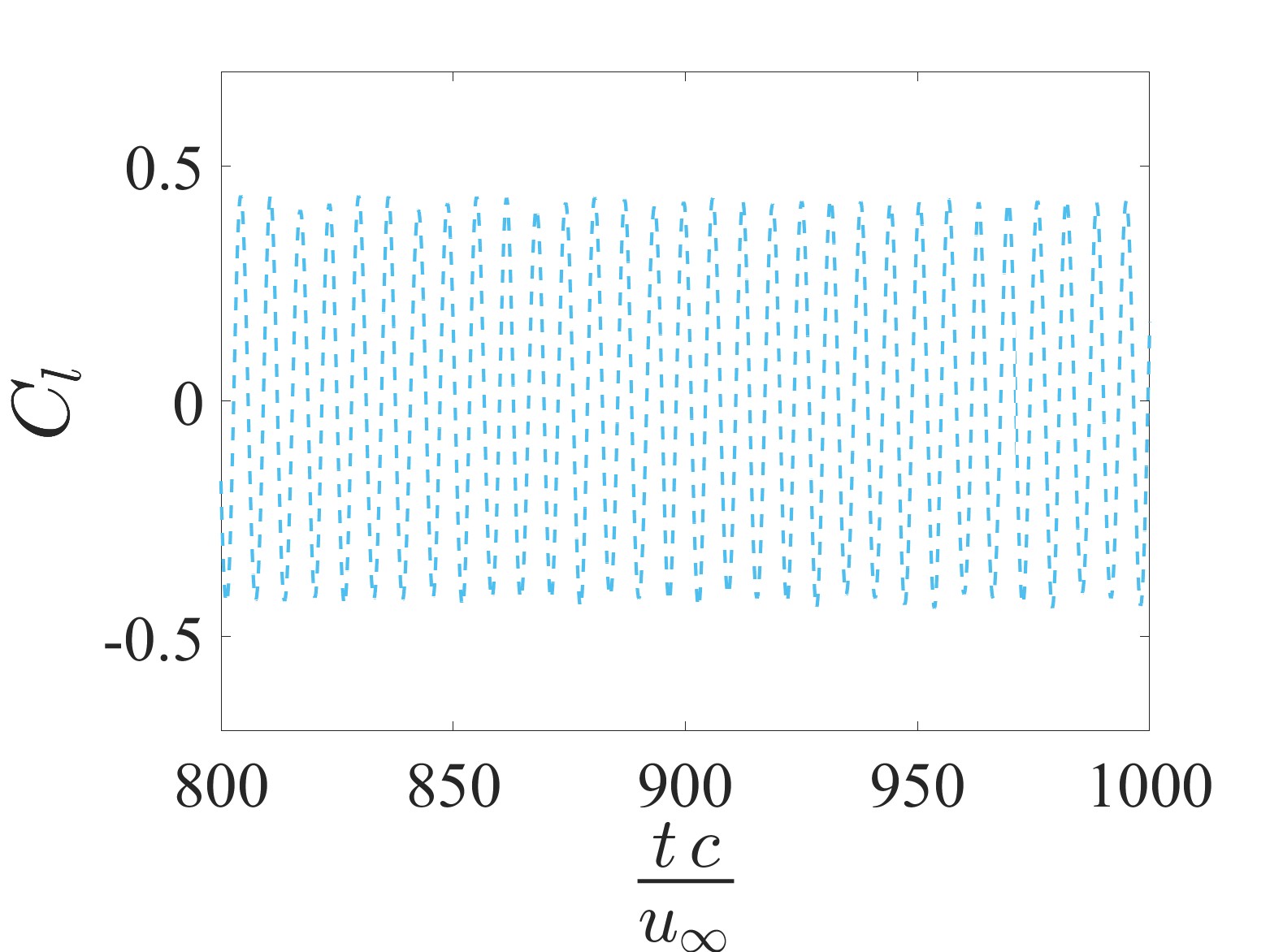}
  \caption{}
  \label{fig:cl_Re80}
\end{subfigure}
\begin{subfigure}[!htb]{.6\textwidth}
  \centering
  \includegraphics[trim=2cm 10cm 4cm 12cm,clip,width=1\linewidth]{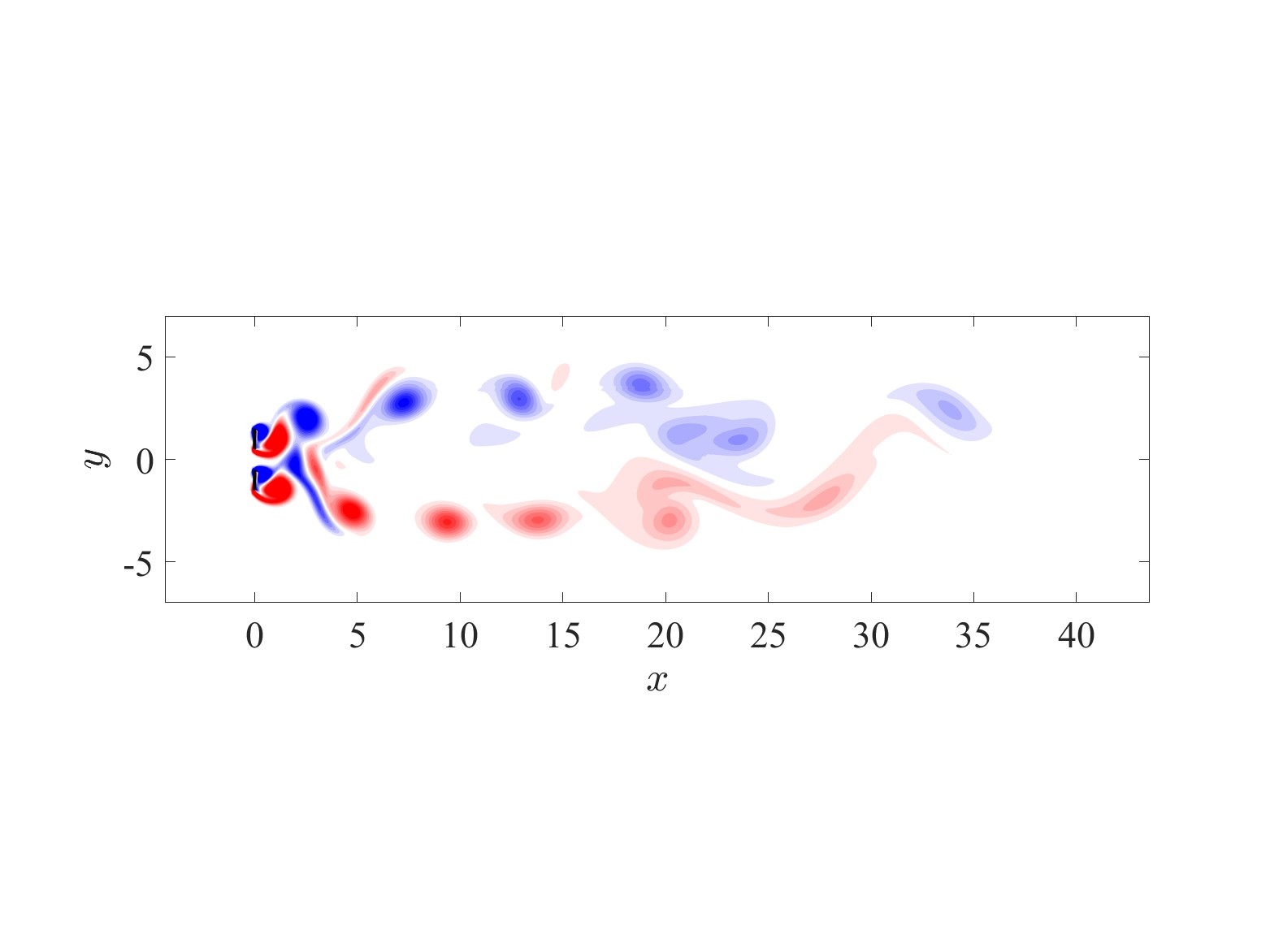}
  \caption{}
  \label{fig:tp_vort_Re100}
\end{subfigure}
\begin{subfigure}[!htb]{.35\textwidth}
\centering
\includegraphics[trim=0cm 0cm 2cm 0cm,clip,width=1\linewidth]{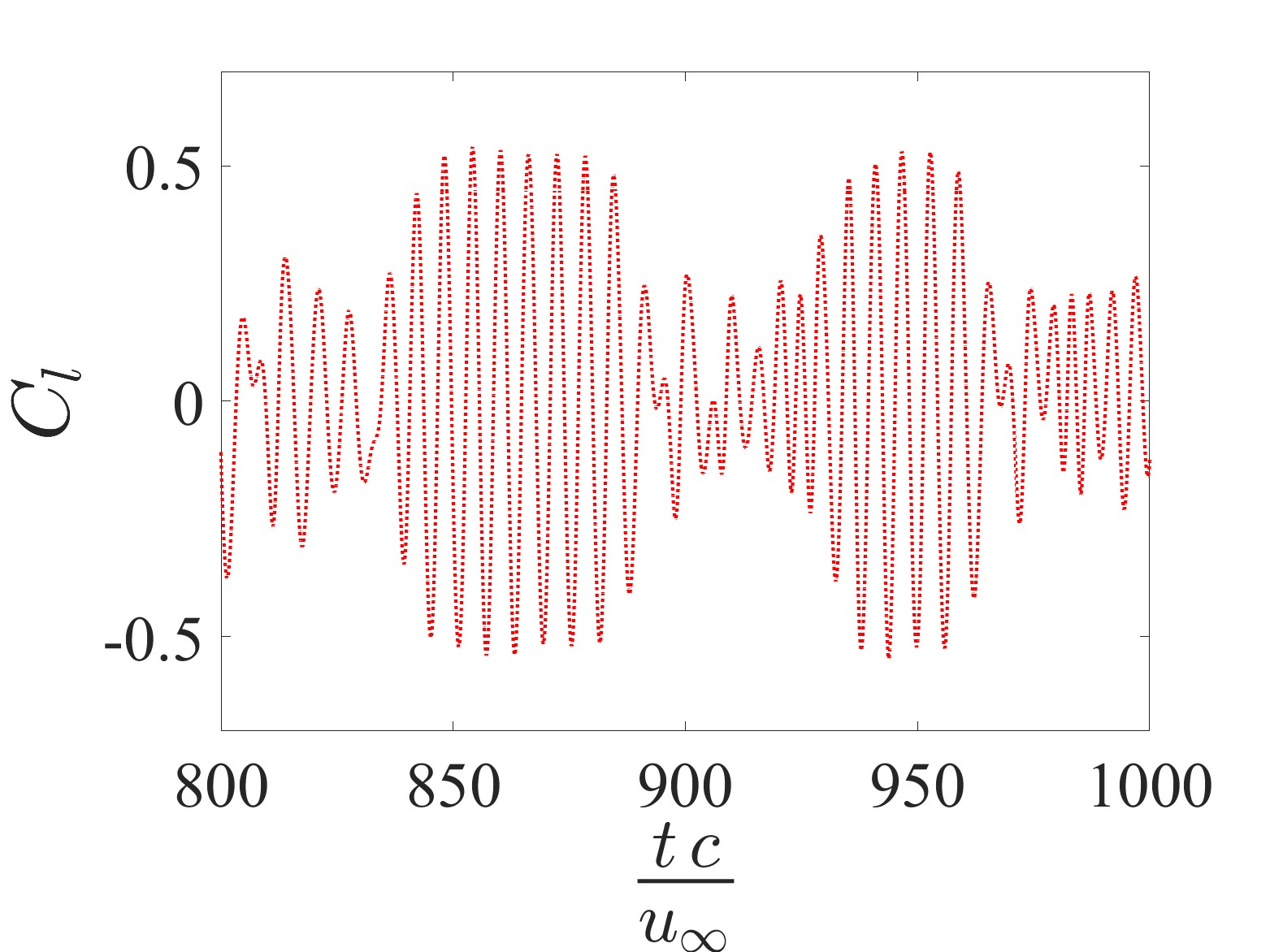}
  \caption{}
  \label{fig:cl_Re100}
\end{subfigure}
\caption{Snapshots of vorticity computed for flow over two flat plates separated at a gap of $d = c$ for (a) Re = 40, (c) Re = 80, and (e) Re = 100. Lift coefficient shown for (b) Re = 40, (d) Re = 80, and (f) Re = 100. Contours are shown within a range of [-2.5, 2.5].}
\label{fig:twoplates_vort}
\end{figure}

Instantaneous vorticity snapshots are shown in Fig.~\ref{fig:twoplates_vort}. For the Re = 40 case, the wake (Fig.~\ref{fig:tp_vort_Re40}) is regular and periodic. Vorticity is shed from the top and bottom surfaces of each plate in phase, so the formation of positive vorticity from the bottom (and negative vorticity from the top) of each plate occurs approximately simultaneously. The vorticity shed from the inner plate edges appear to largely cancel out as we move downstream, leaving the vorticity shed from the outer edges, resulting in a vortex street of alternating positive and negative vortices. This is somewhat reminiscent of the von K\'arm\'an vortex street that forms for flow past a single bluff body, though with further vertical separation between the vortices. Consistent with the snapshot of the wake,  the lift coefficient $C_l = \frac{F_y}{1/2\rho U_{\infty}^2 c}$ (summed across both bodies) is periodic. 

The wakes for the Re = 80 and 100 cases shown in Figs.~\ref{fig:tp_vort_Re80} and~\ref{fig:tp_vort_Re100} respectively at first glance appear to be qualitatively similar to each other based on the snapshots shown. They both exhibit in-phase vortex shedding from the plates, with stronger and more distinct vortices forming nearer to the plates than in the Re = 40 case. Further downstream, unlike in the Re = 40 case, the vortices forming from the outer edges of the plates exhibit meandering, pairing, and merging. Such vortex pairing is similar to that observed in Ref.~\cite{williamson1985evolution}, though in some cases that was due to the merging of same-sign vortices shed from both (typically further apart) bluff bodies simultaneously, rather than the case here of merging between consecutively-shed same-sign vortices from a single body.

While the wake snapshots look similar, the forces for the Re = 80 and 100 cases shown in Figs.~\ref{fig:cl_Re80} and~\ref{fig:cl_Re100} respectively show that the two systems are dynamically quite distinct. The lift coefficient for the Re=80 case is approximately periodic (with a larger amplitude of oscillation than the Re=40 case), while the Re=100 case exhibits a lift force that is more irregular and aperiodic. The reason for this difference is because the near-wake behavior Re=80 case consistently features periodic in-phase shedding, whereas the Re=100 case features intermittent behavior where the shedding from each plate is not always in-phase or at the same frequency. 

\begin{figure}[!htb]
  \centering
  \includegraphics[clip, trim=2cm 0cm 3cm 2cm,width=0.6\linewidth]{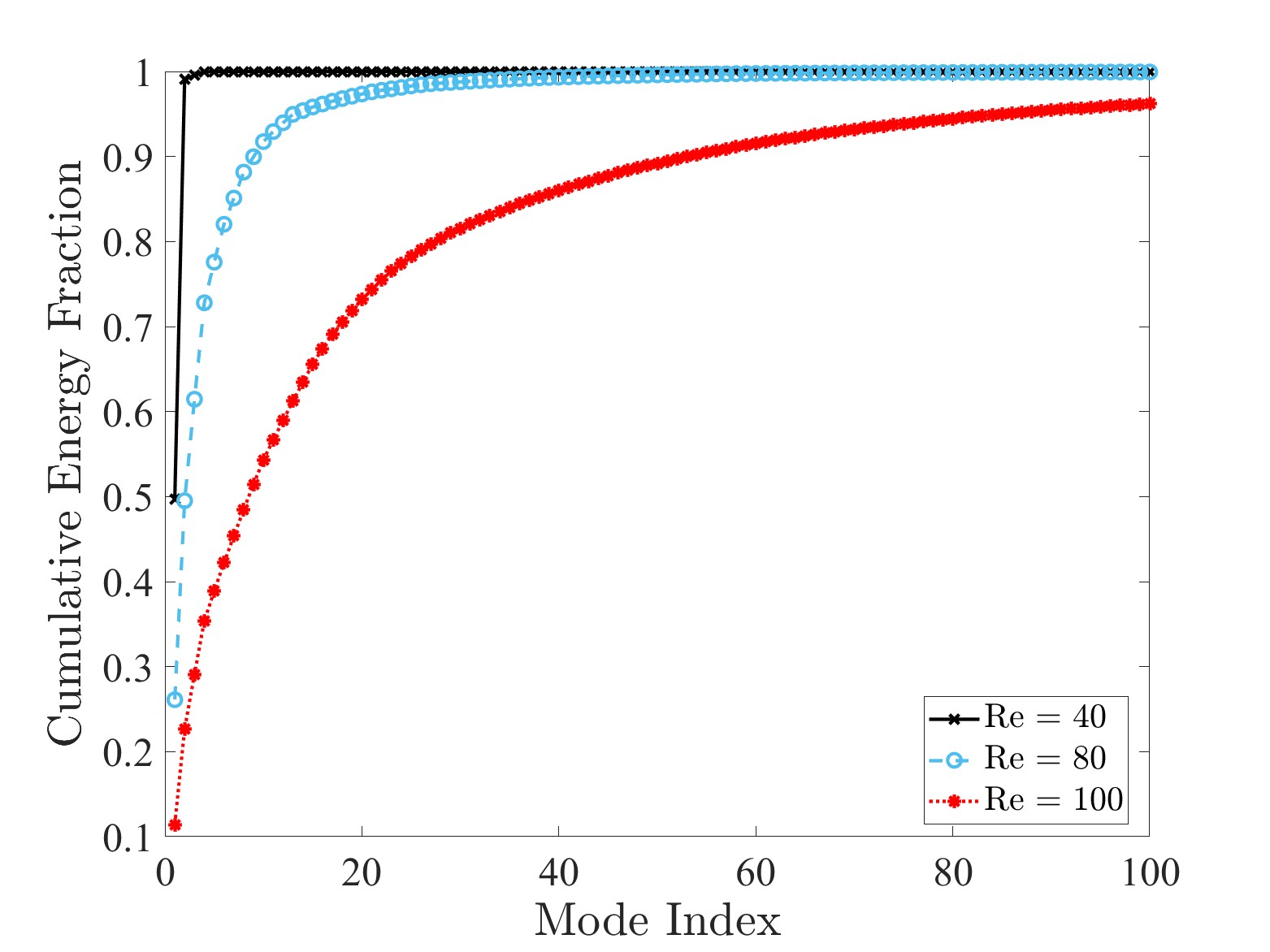}
  \label{fig:pod_cum_energy_compare}
\caption{The singular values computed from performing POD on the velocity snapshots for the Re = 40, 80, and 100 case shown as the cumulative energy fraction.}
\label{fig:POD_singular_values_all_Res}
\end{figure}

The dynamical complexity of the wake at these three Reynolds numbers can also be quantified by considering the (space-only) POD of velocity field data for snapshots collected for each case. Fig.~\ref{fig:POD_singular_values_all_Res}  plots the 
cumulative energy fraction of the POD mode energy content for each case. This POD computation utilizes 512 snapshots for the Re = 40 and 80 cases, and 1024 snapshots for the Re = 100 case. As well as these snapshots, we also utilize the symmetry of the geometry to generate additional data flipped in the $y=0$ plane. In particular, for each snapshot with horizontal (streamwise, $u$) and vertical ($v$) velocity components $(u^j,v^j)$ collected (where $j$ is the snapshot index) we also append to the data $(u^j_{sym},v^j_{sym})$, where
\begin{align}
u^j_{sym}(x,y) &= u^j(x,-y),\\
v^j_{sym}((x,y) &= -v^j(x,-y).
\end{align}
Note that this would not be appropriate if the flow exhibited a persistent symmetry-breaking bias in the wake (i.e. a pitchfork bifurcation as described for the fluidic pinball in \cite{deng2020low}). We do not observe such behavior for the cases considered here, though it was observed for this configuration at Re = 60 \cite{almashjary2021reduced}. 
As expected based on prior discussion of the flow physics, in Fig.~\ref{fig:POD_singular_values_all_Res} as the Reynolds number increases a larger number of modes must be retained to capture a certain fraction of the kinetic energy. For the subsequent results shown and the POD-Galerkin models used to obtain them, the number of POD modes retained for models at different Reynolds number was chosen to capture at least 96\% of the kinetic energy of the flowfield. In what follows, we use $r = 6$, 30, and 80 modes for the Re = 40, 80, and 100 cases respectively.

The goal of this analysis is to determine the extent to which GSE can predict the dynamics of these wakes, in terms of the energy content and leading SPOD mode shapes associated with each frequency, from velocity field snapshots that are not resolvent in time. For example, while the snapshots of the flow at Figs.~\ref{fig:cl_Re80} and~\ref{fig:cl_Re100} look qualitatively similar, we would hope that GSE could predict that the Re=80 case features strongly periodic behavior, whereas the Re=100 case is more intermittent and complex. 

As was the case for the example considered in Sect.~\ref{results_toy}, we first consider GSE as applied to time-resolved data (Sect.~\ref{results_twoplates_uniform}), before considering data that is subsampled in time (Sect.~\ref{results_twoplates_sub}), such that the direct SPOD cannot accurately capture the true dynamics of the system. 

\subsection{Galerkin spectral estimation of time-resolved data}\label{results_twoplates_uniform}

\begin{figure}[!htb]
\begin{subfigure}{0.5\textwidth}
  \centering
  \includegraphics[clip, trim=2cm 0cm 3cm 2cm,width=1.0\linewidth]{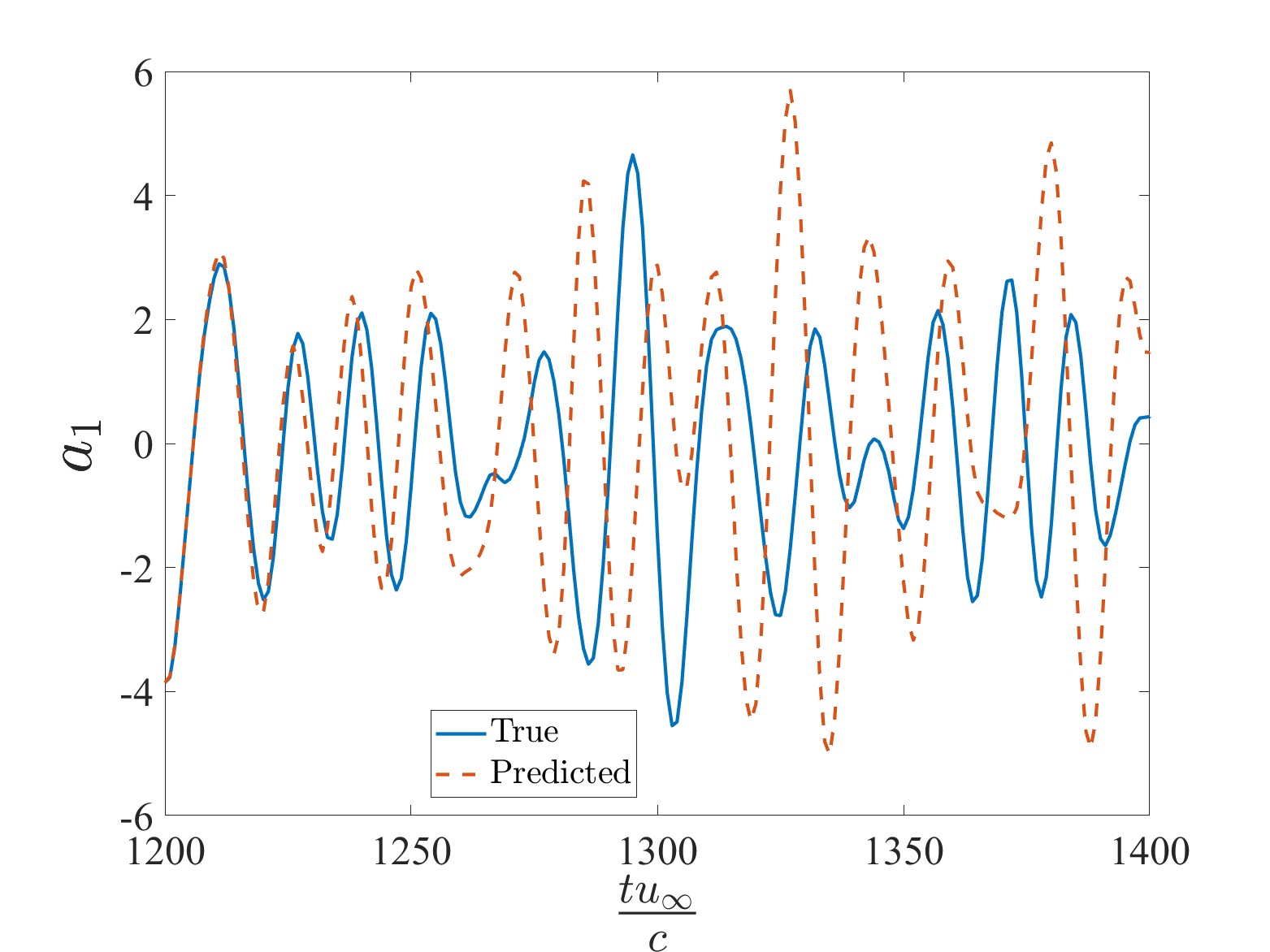}
  \caption{}
  \label{fig:pod_dns_twoplates_mode1}
\end{subfigure}
\begin{subfigure}{0.5\textwidth}
  \centering
  \includegraphics[clip, trim=2cm 0cm 3cm 2cm,width=1.0\linewidth]{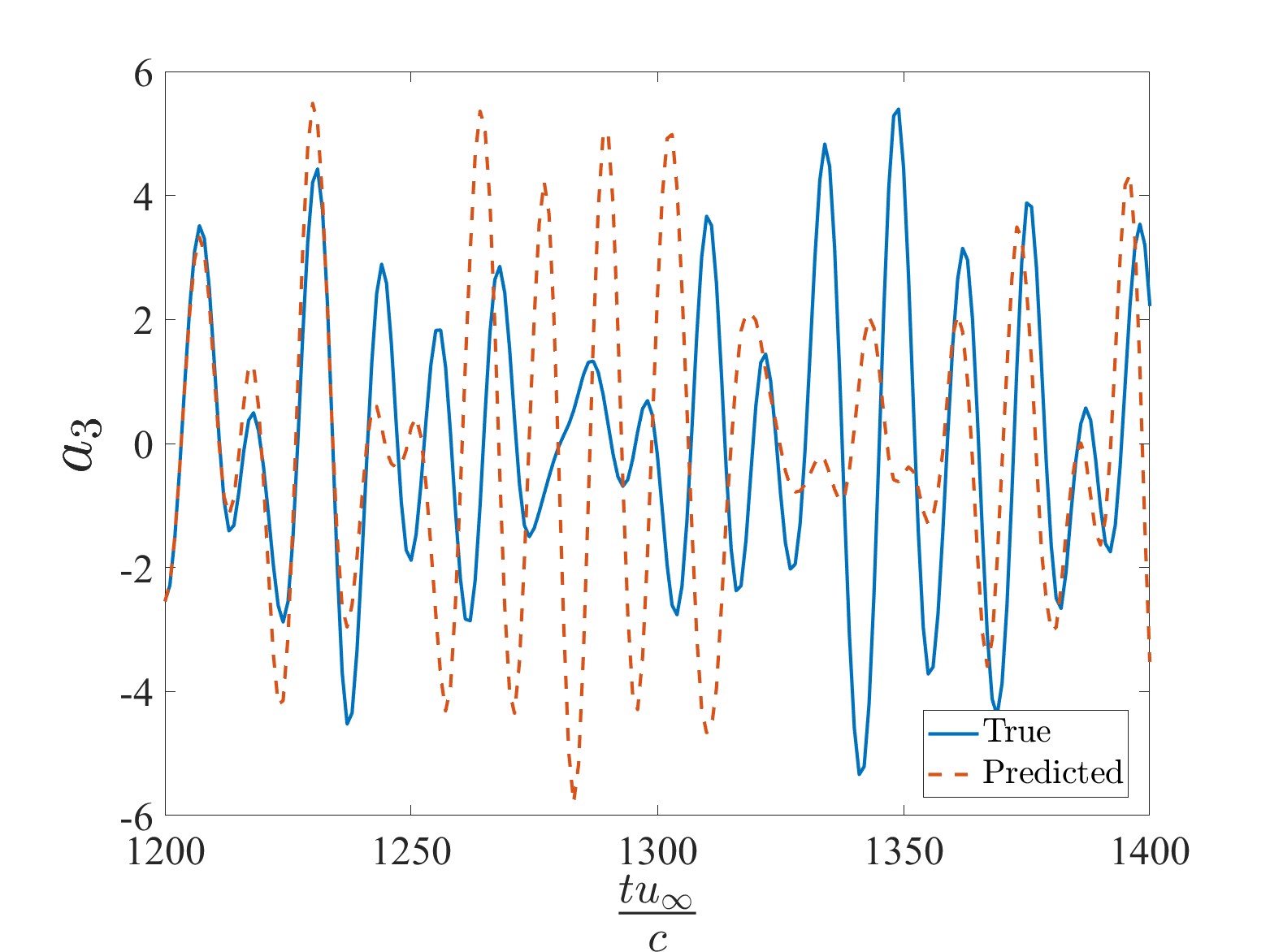}
  \caption{}
  \label{fig:pod_dns_twoplates_mode3}
\end{subfigure}
\begin{subfigure}{0.5\textwidth}
  \centering
  \includegraphics[clip, trim=4cm 10cm 4cm 12cm,width=1.0\linewidth]{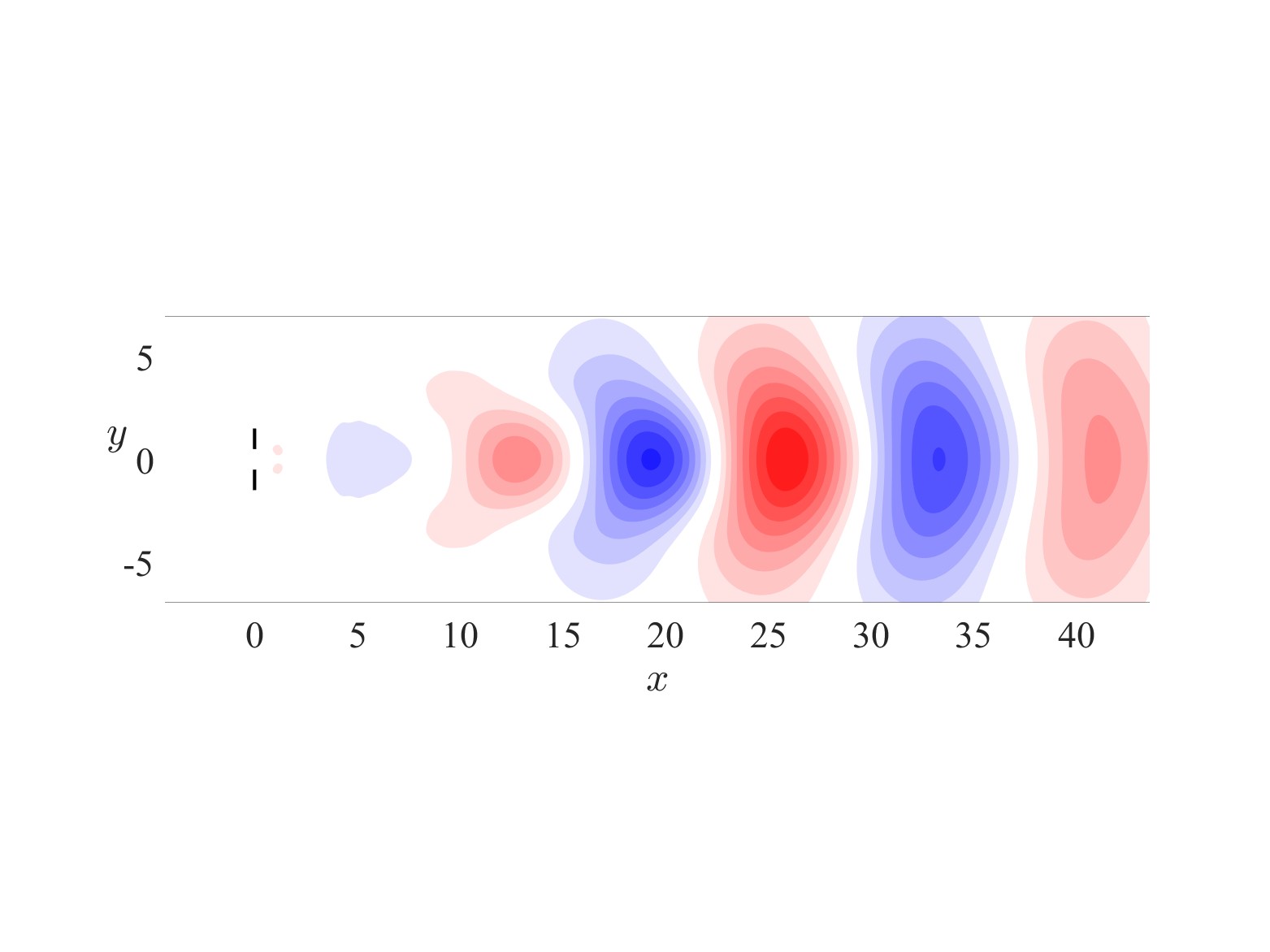}
  \caption{}
  \label{fig:POD_dns_twoplates_mode1}
\end{subfigure}
\begin{subfigure}{0.5\textwidth}
  \centering
  \includegraphics[clip, trim=4cm 10cm 4cm 10cm,width=1.0\linewidth]{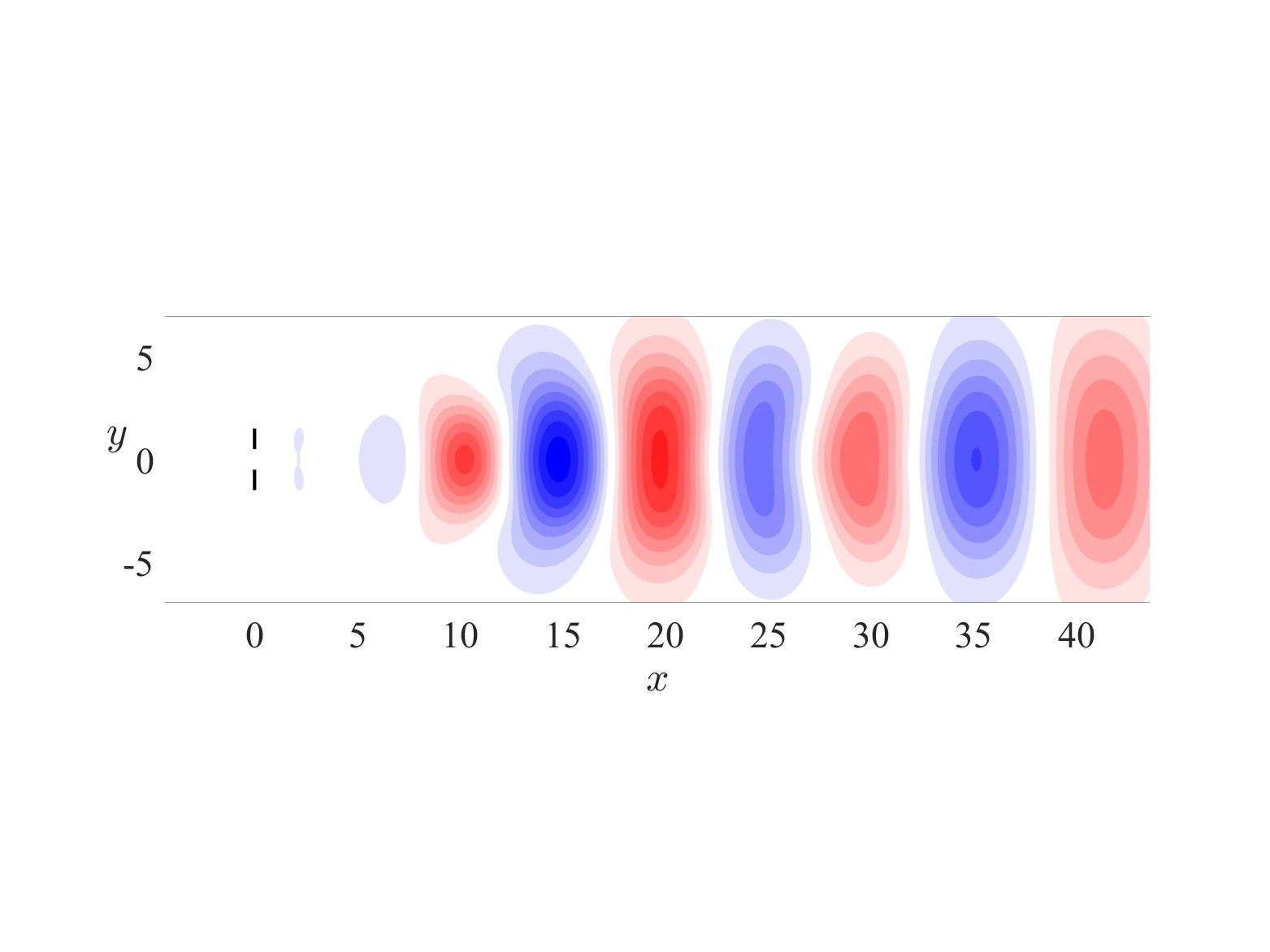}
  \caption{}
  \label{fig:POD_dns_twoplates_mode3}
\end{subfigure}
\caption{True and predicted (a) first and (b) third POD mode coefficients for the Re = 100 case, where the predictions are made from a nonlinear Galerkin projection model of order $r = 80$. (c) and (d) show the streamwise velocity component of POD modes 1 and 3, respectively.}
\label{fig:PODCoeffs_all_twoplates_Re100}
\end{figure}

This section applies GSE and compares the estimated spectra to that computed directly from the data. Data is collected with a timestep $\Delta t = 1 u_\infty/c$, which is sufficient to resolve the pertinent dynamics for all cases. 
A portion of the true and POD-Galerkin model-predicted POD mode coefficients for Re = 100 case are shown in Fig.~\ref{fig:PODCoeffs_all_twoplates_Re100}, along with the corresponding POD modes. We plot the POD-Galerkin model predictions for a single trajectory. 
It is observed in general that the Galerkin projection model diverge after approximately 50 convective time units. 
In what follows, we will perform GSE using both a single trajectory and multiple different trajectories of data, with each initial condition coming from the original data. This multiple-trajectory case should ensure that the predicted data used for SPOD remains physically realistic.

\begin{figure}
  \centering
\begin{subfigure}{.6\textwidth}
  \centering
  \includegraphics[trim=0cm 0cm 0cm 0cm,clip,width=1\linewidth]{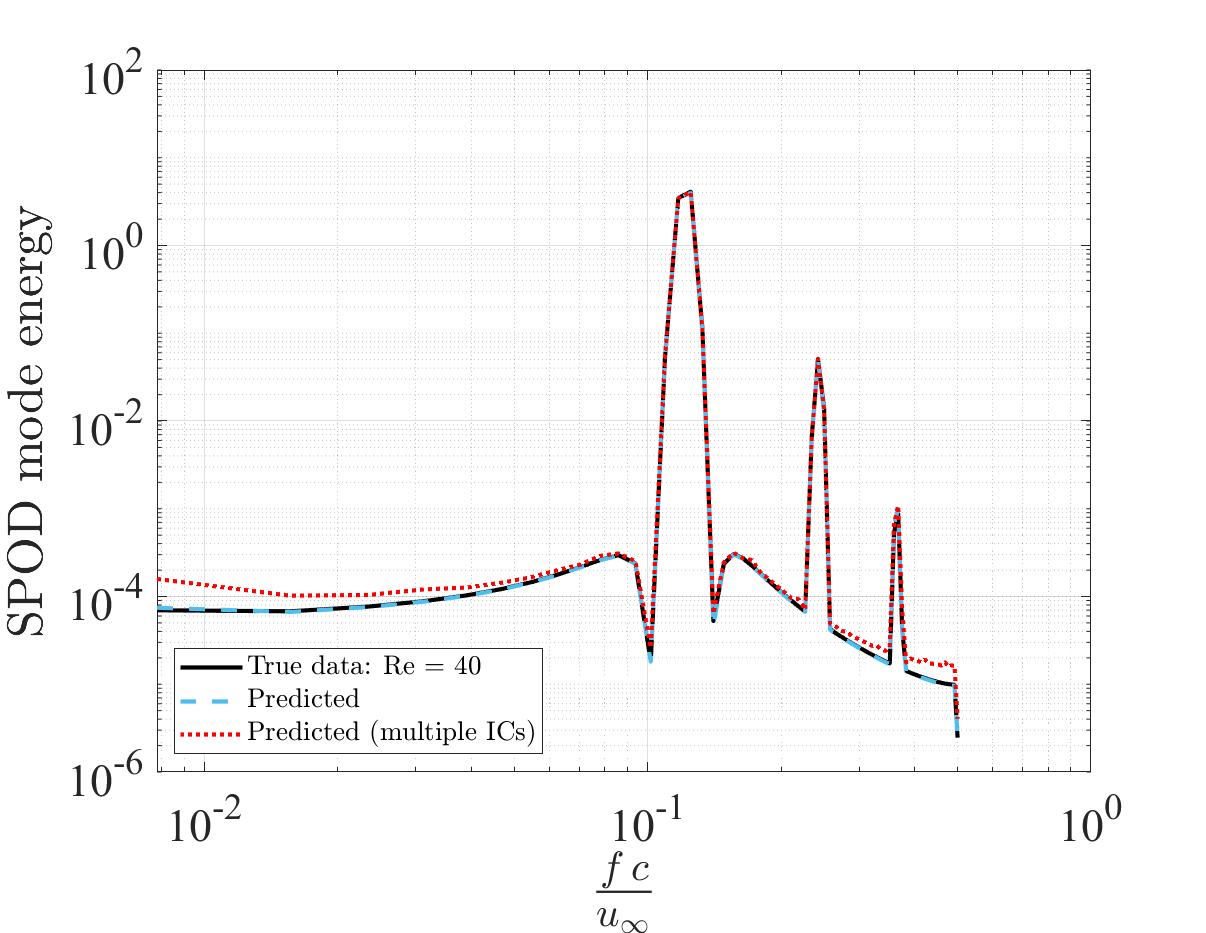}
  \caption{}
  \label{fig:twoplates_spectra_Re40}
\end{subfigure}
\begin{subfigure}{.6\textwidth}
  \centering
  \includegraphics[trim=0cm 0cm 0cm 0cm,clip,width=1\linewidth]{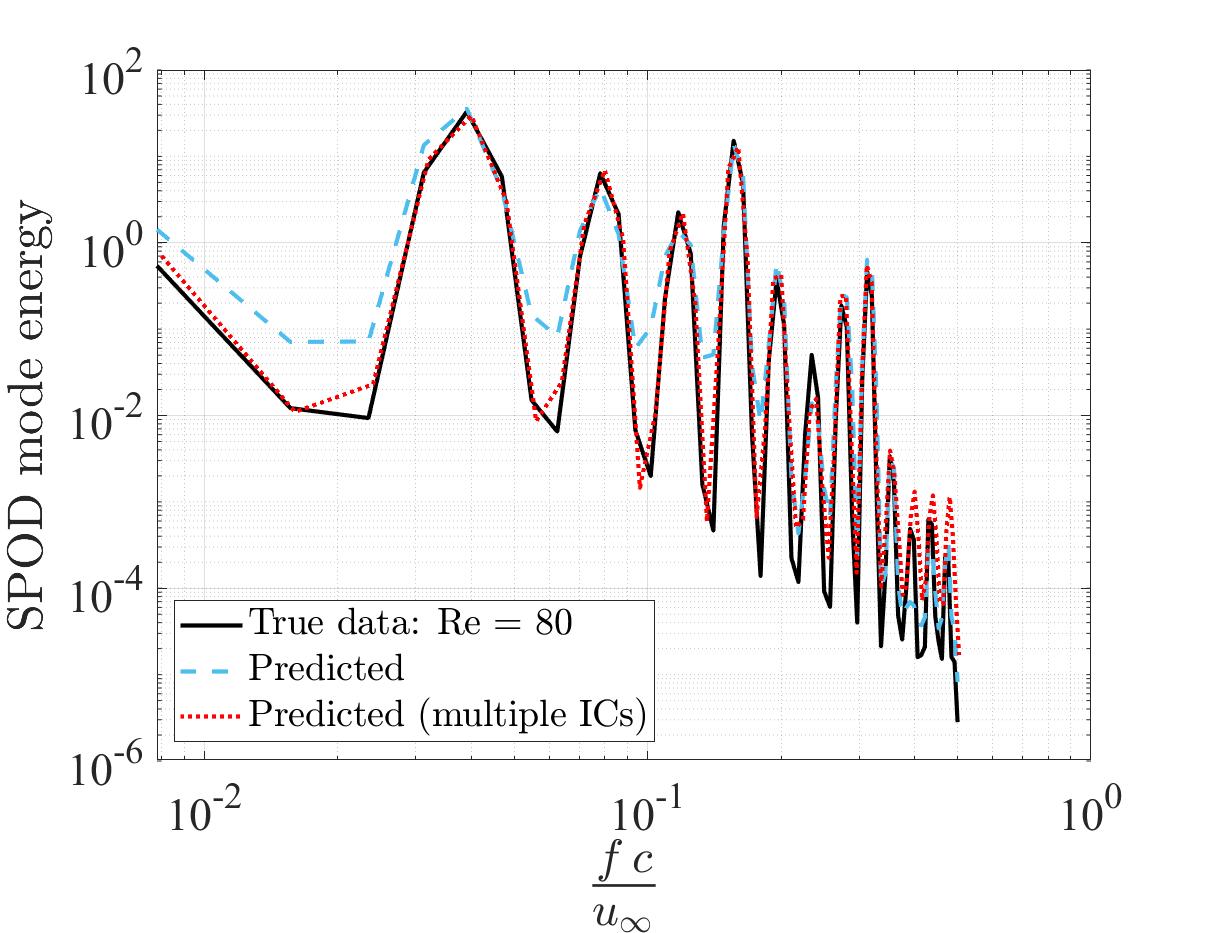}
  \caption{}
  \label{fig:twoplates_spectra_Re80}
\end{subfigure}
\begin{subfigure}{.6\textwidth}
  \centering
  \includegraphics[trim=0cm 0cm 0cm 0cm,clip,width=1\linewidth]{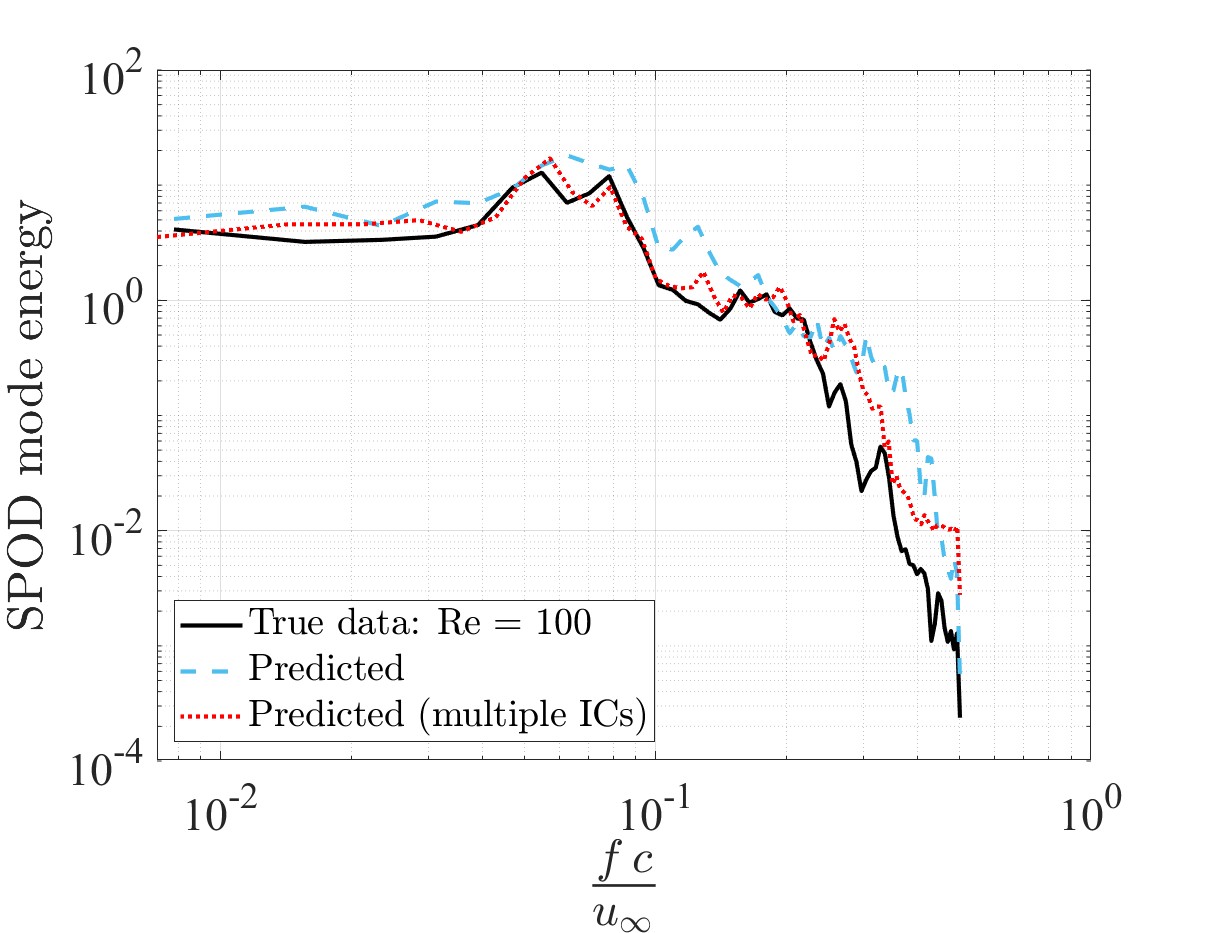}
  \caption{}
  \label{fig:twoplates_spectra_Re100}
\end{subfigure}
\caption{
Comparison between SPOD spectra computed directly from data and using GSE for a single and multiple reconstructed trajectories (with different initial conditions, for flow over two flat plates at (a) Re = 40, (b) Re = 80, and (c) Re = 100.}
\label{fig:spectra_DNS_twoplates_multiICs}
\end{figure}

Figure~\ref{fig:spectra_DNS_twoplates_multiICs} shows the SPOD spectrum comparisons between the original data  and GSE estimates. 
 When performing GSE, we generate 4096 total snapshots for the Re = 40 and 80 cases, and 8196 snapshots for the Re = 100 case, and perform SPOD using a window size of 128 snapshots (with $\Delta t = 1$ as per the original data). 

It is observed that GSE is able to capture the 
behavior of the true SPOD spectra in all cases. At Re=40, the SPOD spectrum contains a single isolated peak at $fc/U_{\infty} \approx 0.12$, and smaller peaks at harmonics of this frequency. The Galerkin model captures this spectrum almost exactly. The Re = 80 case features a larger number of peaks, owing to the more complex wake dynamics present. At Re = 100, the spectrum is more broadband, with an energy roll-off at dimensionless frequencies greater than approximately 0.1. For the Re = 80 and 100 cases, the GSE method using multiple trajectories is slightly more accurate at capturing the true SPOD spectrum than the case using a single trajectory of predicted data. 
A more comprehensive analysis of the SPOD spectra and associated modes (shown in Tables \ref{tab:SPOD_modes_Re40}--\ref{tab:SPOD_modes_Re100}) and flow physics corresponding to frequency peaks will be discussed following the consideration of temporally under-resolved data in Sect.~\ref{results_twoplates_sub}.

\subsection{Galerkin spectral estimation using non-time-resolved data}\label{results_twoplates_sub}

In this section, we apply GSE to recover spectral content from data that is initially under-resolved in time, analogous to the cases considered for the forced linear system in Sect.~\ref{results_toy}.

\begin{figure}
  \centering
\begin{subfigure}{.6\textwidth}
 \centering
    \includegraphics[trim=0cm 0cm 0cm 0cm,clip,width=1\linewidth]{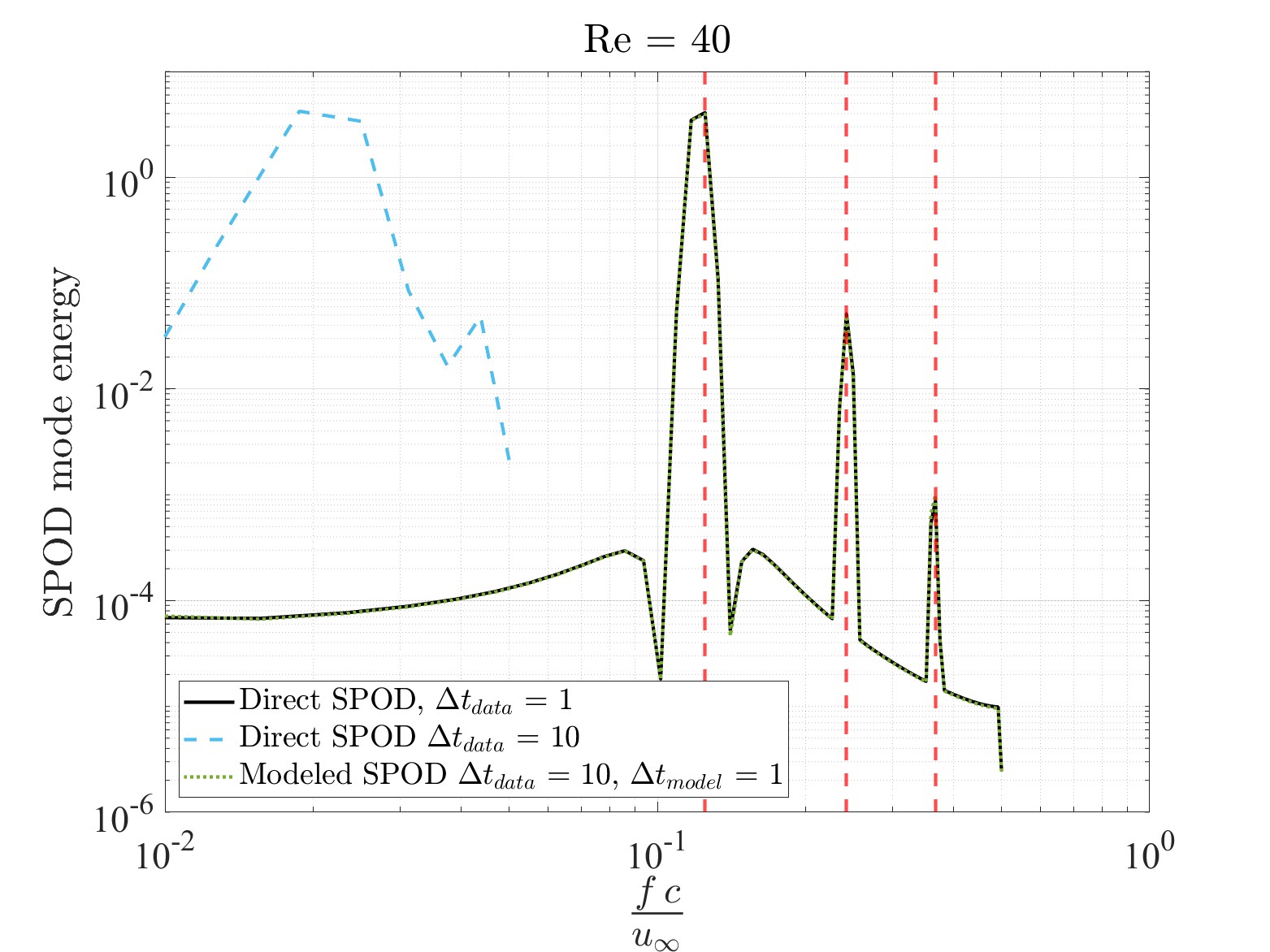}
  \caption{}
  \label{fig:twoplates_spectra_SS_dt10_Re40}
\end{subfigure}
\begin{subfigure}{.6\textwidth}
  \centering
    \includegraphics[trim=0cm 0cm 0cm 0cm,clip,width=1\linewidth]{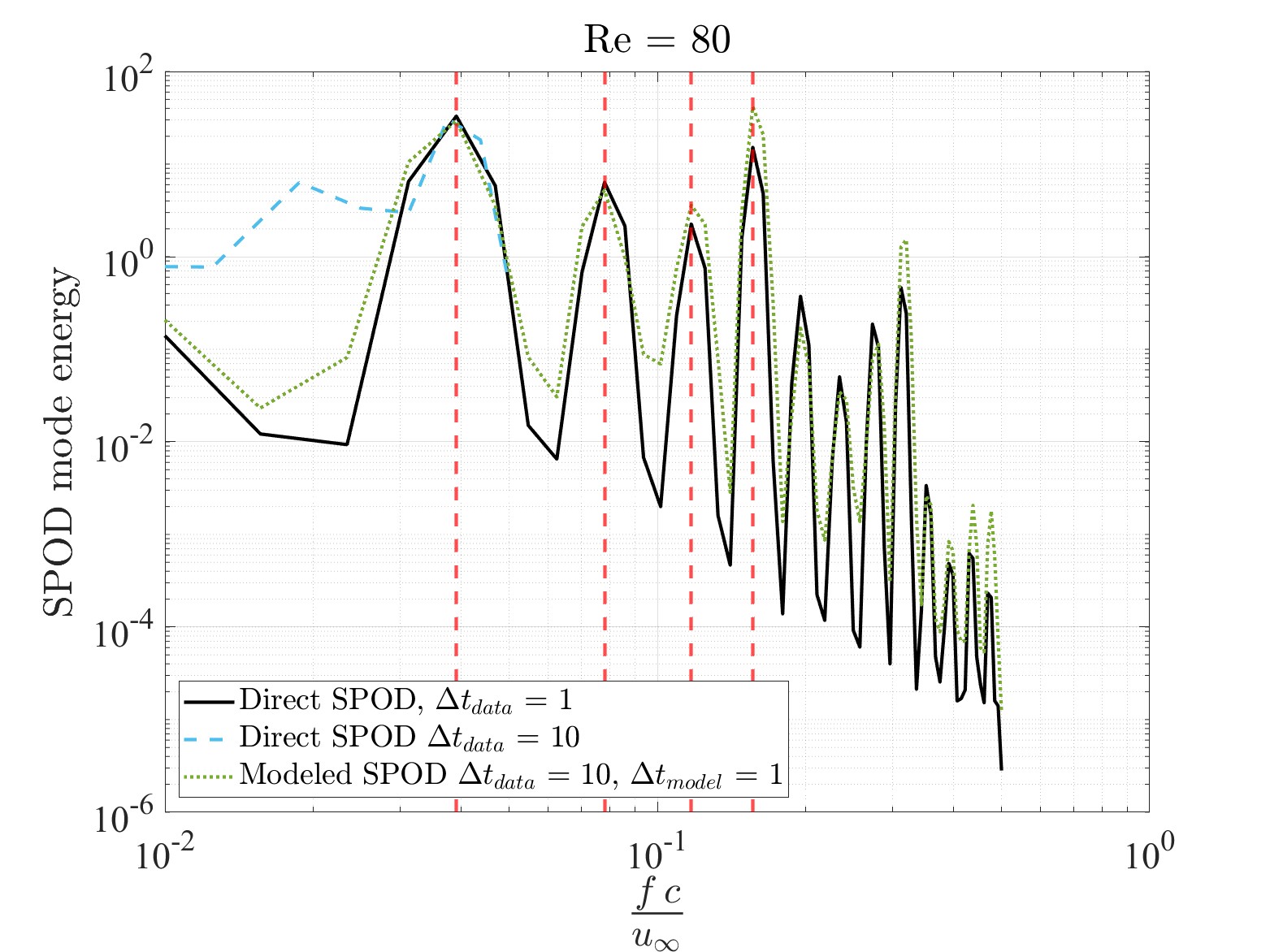}
  \caption{}
  \label{fig:twoplates_spectra_SS_dt10_Re80}
\end{subfigure}
\begin{subfigure}{.6\textwidth}
  \centering
    \includegraphics[trim=0cm 0cm 0cm 0cm,clip,width=1\linewidth]{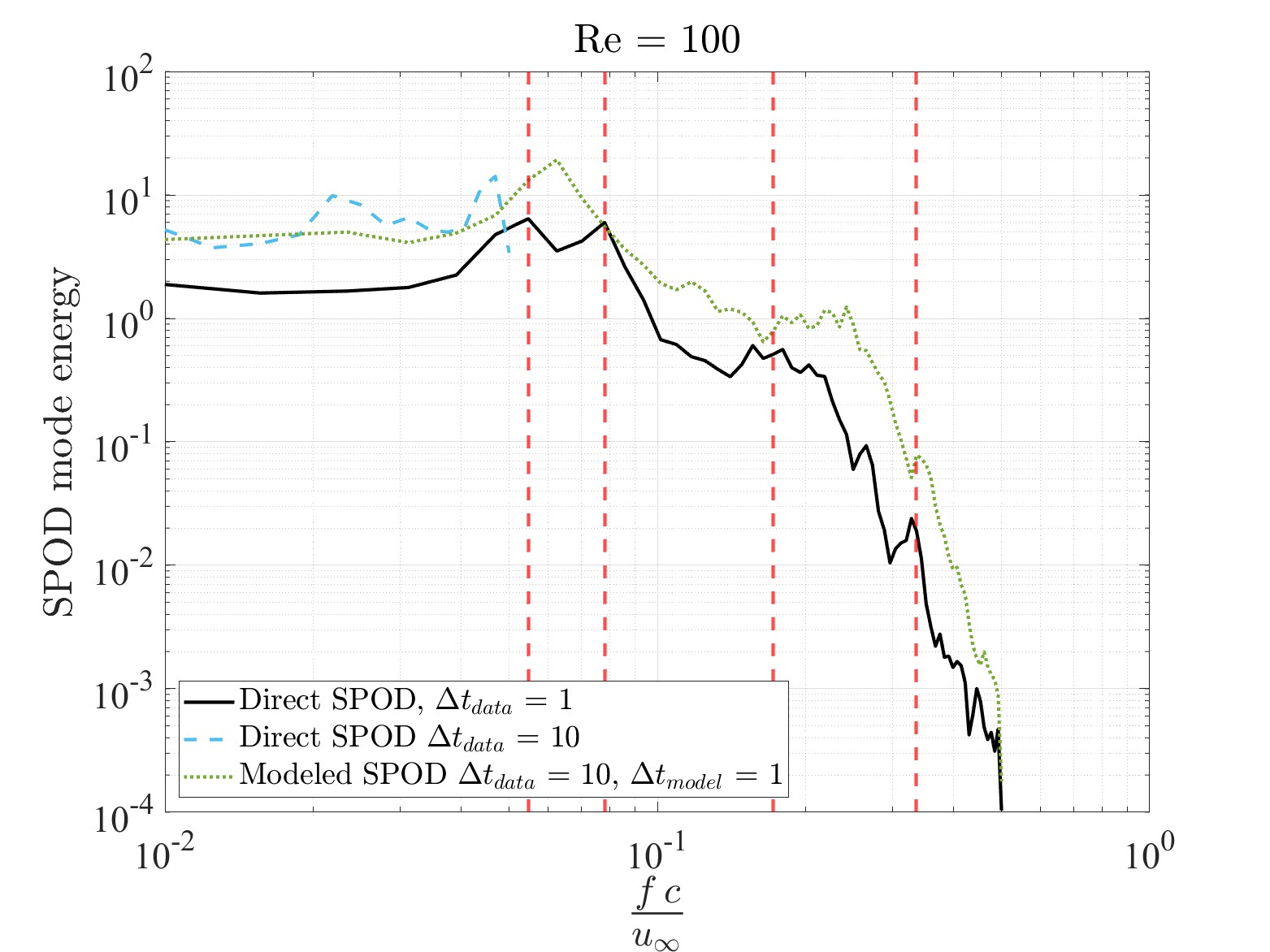}
  \caption{}
  \label{fig:twoplates_spectra_SS_dt10_Re100}
\end{subfigure}
\caption{Comparison between SPOD spectra computed directly from data sampled at $\Delta t = 1$ and $\Delta t =10$,   and GSE reconstruction of the latter, at (a) Re = 40, (b) Re = 80, and (c) Re = 100. The red dashed lines mark frequencies of SPOD modes plotted in Tables~\ref{tab:SPOD_modes_Re40}--\ref{tab:SPOD_modes_Re100}.}
\label{fig:spectra_DNS_twoplates_subsamp_compare}
\end{figure}

Figure~\ref{fig:spectra_DNS_twoplates_subsamp_compare} 
compares SPOD performed directly on data that is under-resolved in time, with GSE reconstruction of the SPOD spectra. 
For all Reynolds numbers, we observe that the original under-resolved data is unable to accurately capture any of the energy peaks in the SPOD spectra, except for the low-frequency peak at Re = 80. In contrast, GSE is able to reconstruct the full spectra, with a similar accuracy to the cases shown in Fig.~\ref{fig:spectra_DNS_twoplates_multiICs} where time-resolved data was used from the outset. 
 Note that for the direct SPOD of the underresolved data, the smaller total amount of data necessitated the use of a much smaller window size (16 snapshots). However,  the lack of time resolution means that the high frequency dynamics would still not be directly available even with more data and larger DFT windows.

\begin{sidewaystable}
\sidewaystablefn%
\begin{center}
\begin{minipage}{\textheight}
\caption{True and GSE-reconstructed SPOD modes for the three leading peak frequencies for Re = 40, for initial data with two different temporal sampling rates. Transverse ($v$) velocity contours are shown within a range of [-0.01, 0.01].}\label{tab:SPOD_modes_Re40}
\begin{tabular*}{\textheight}{@{\extracolsep{\fill}}lccc@{\extracolsep{\fill}}}
\toprule%
Re = 40 & True SPOD	& GSE, $\Delta t_{data} = 1$ & GSE, $\Delta t_{data} = 10$ \\
\midrule
$\frac{fc}{u_\infty} = 0.125$ & \includegraphics[clip, trim=2cm 10cm 2cm 11.35cm,width=0.25\linewidth]{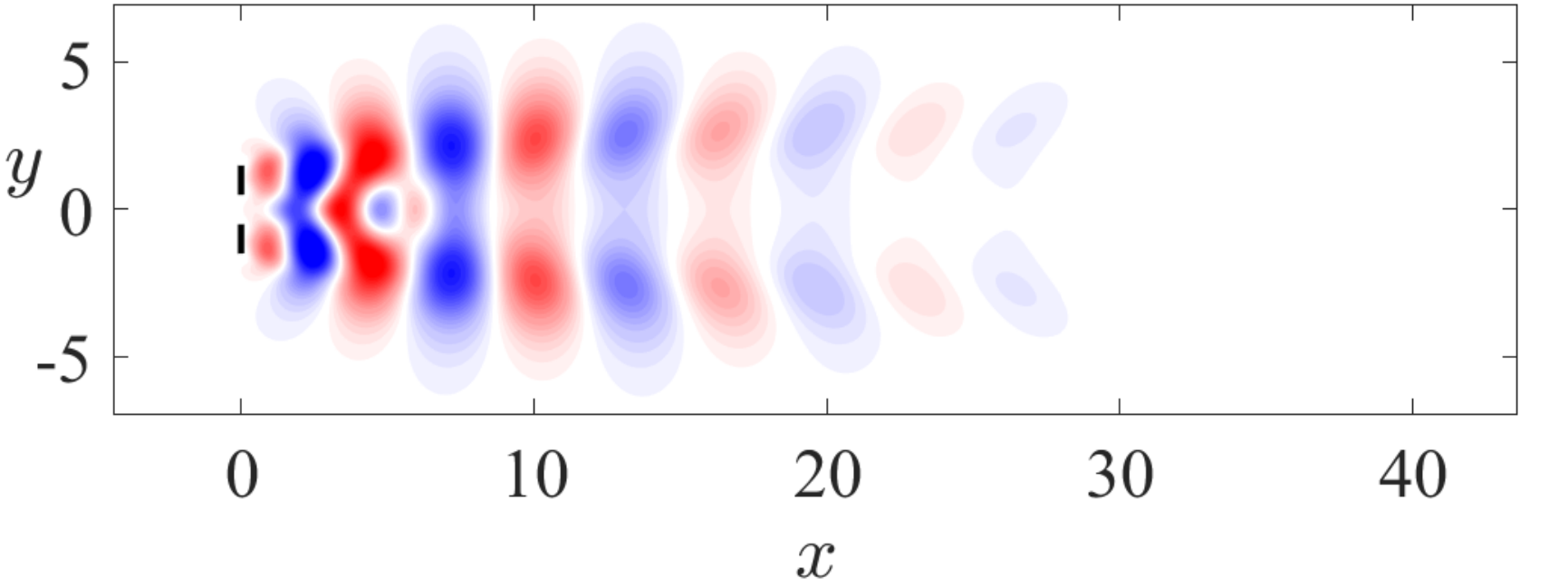} & \includegraphics[clip, trim=2cm 10cm 2cm 11.35cm,width=0.25\linewidth]{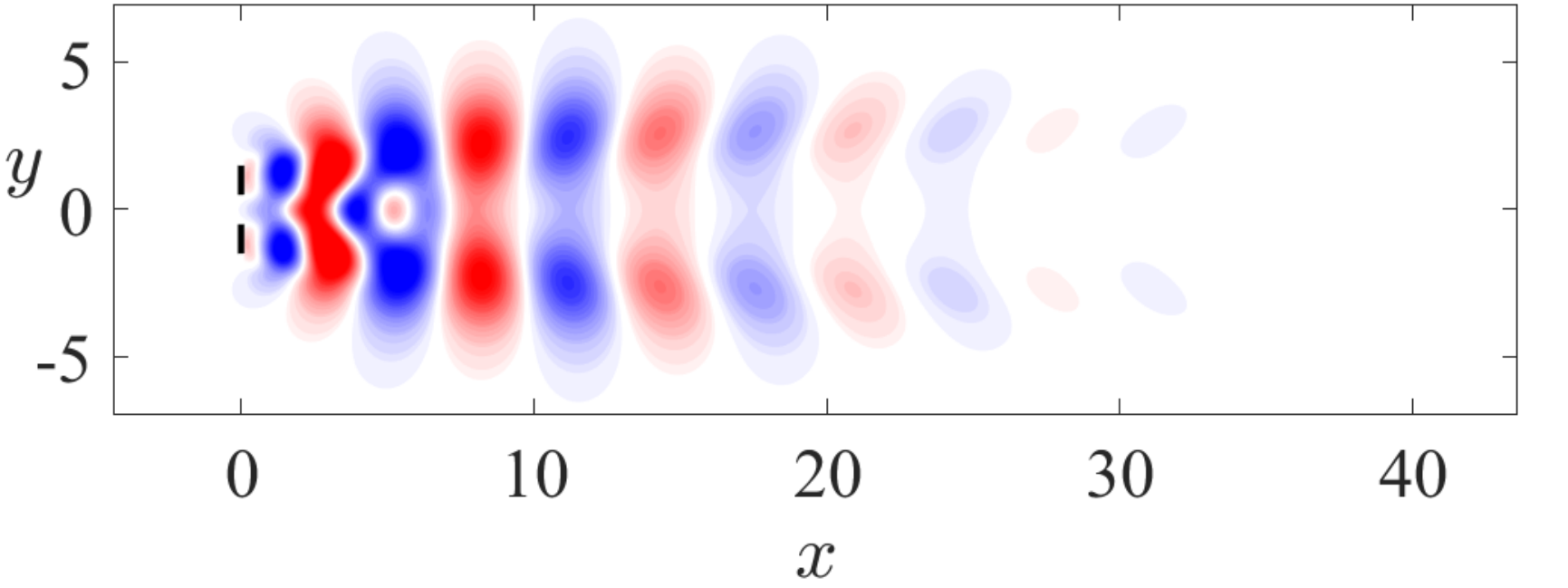} & \includegraphics[clip, trim=2cm 10cm 2cm 11.35cm,width=0.25\linewidth]{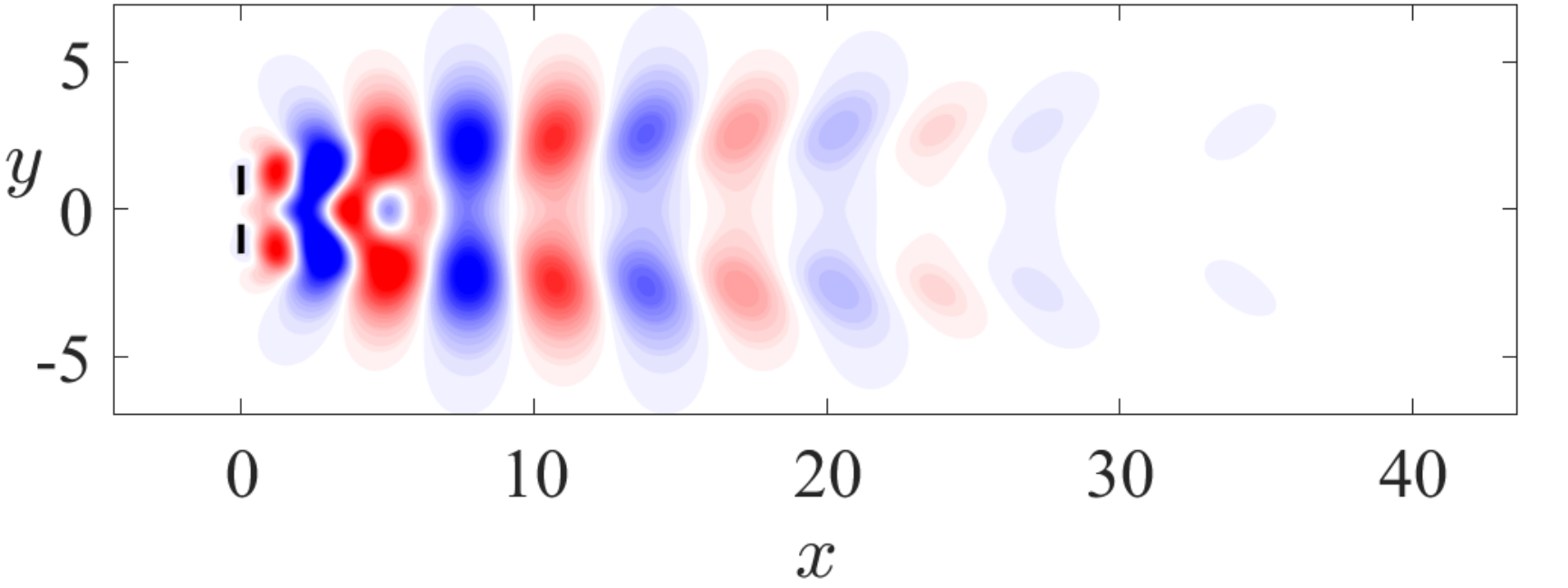} \\ 
$\frac{fc}{u_\infty} = 0.2422$ & \includegraphics[clip, trim=2cm 10cm 2cm 11.35cm,width=0.25\linewidth]{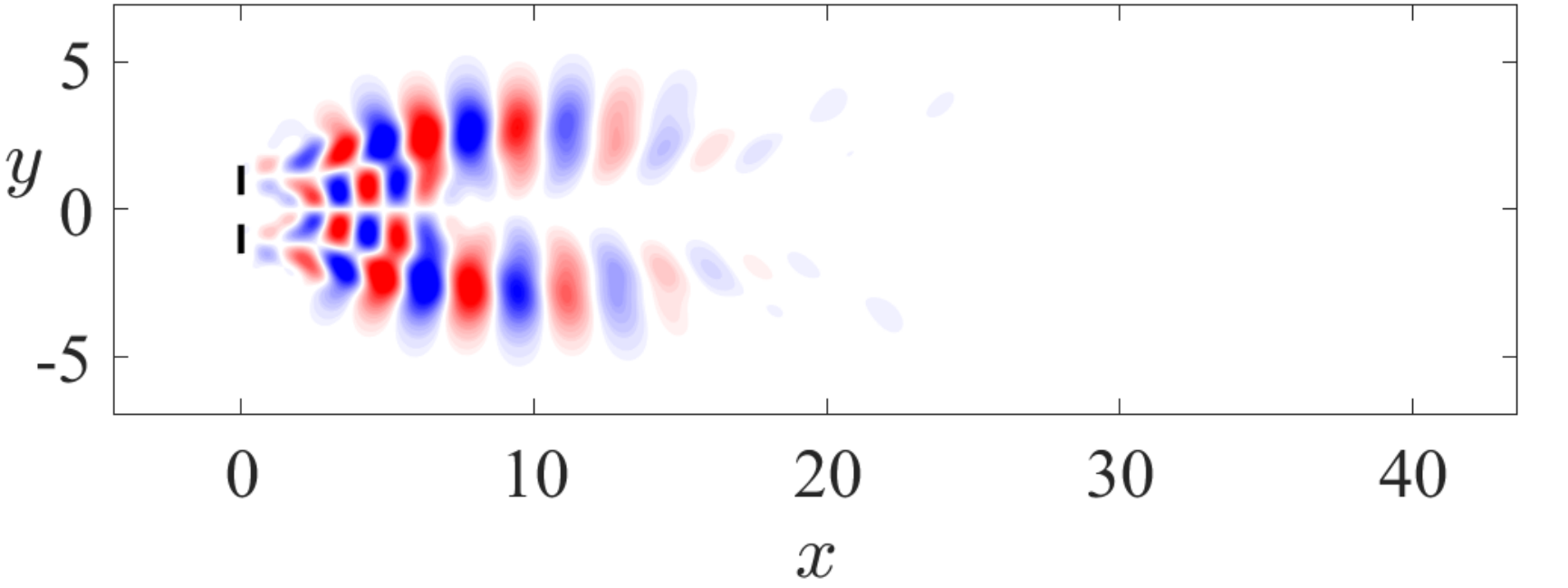} & \includegraphics[clip, trim=2cm 10cm 2cm 11.35cm,width=0.25\linewidth]{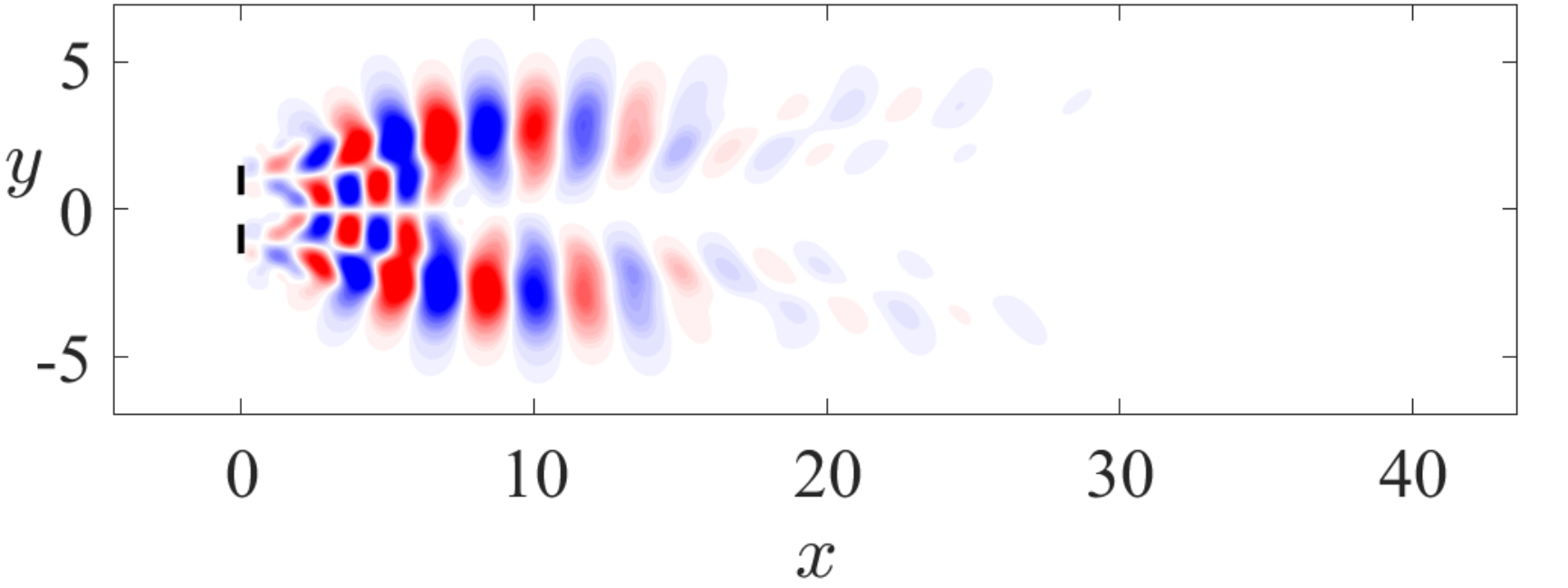} & \includegraphics[clip, trim=2cm 10cm 2cm 11.35cm,width=0.25\linewidth]{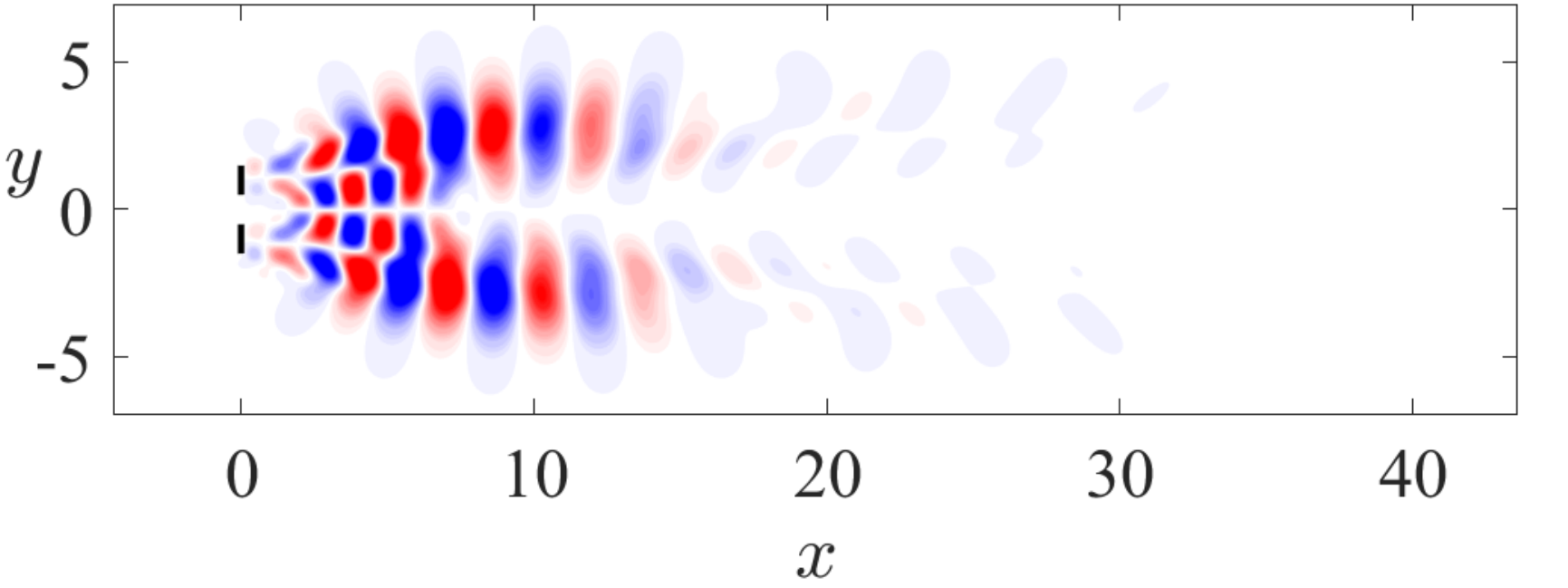} \\
$\frac{fc}{u_\infty} = 0.3672$ & \includegraphics[clip, trim=2cm 10cm 2cm 11.35cm,width=0.25\linewidth]{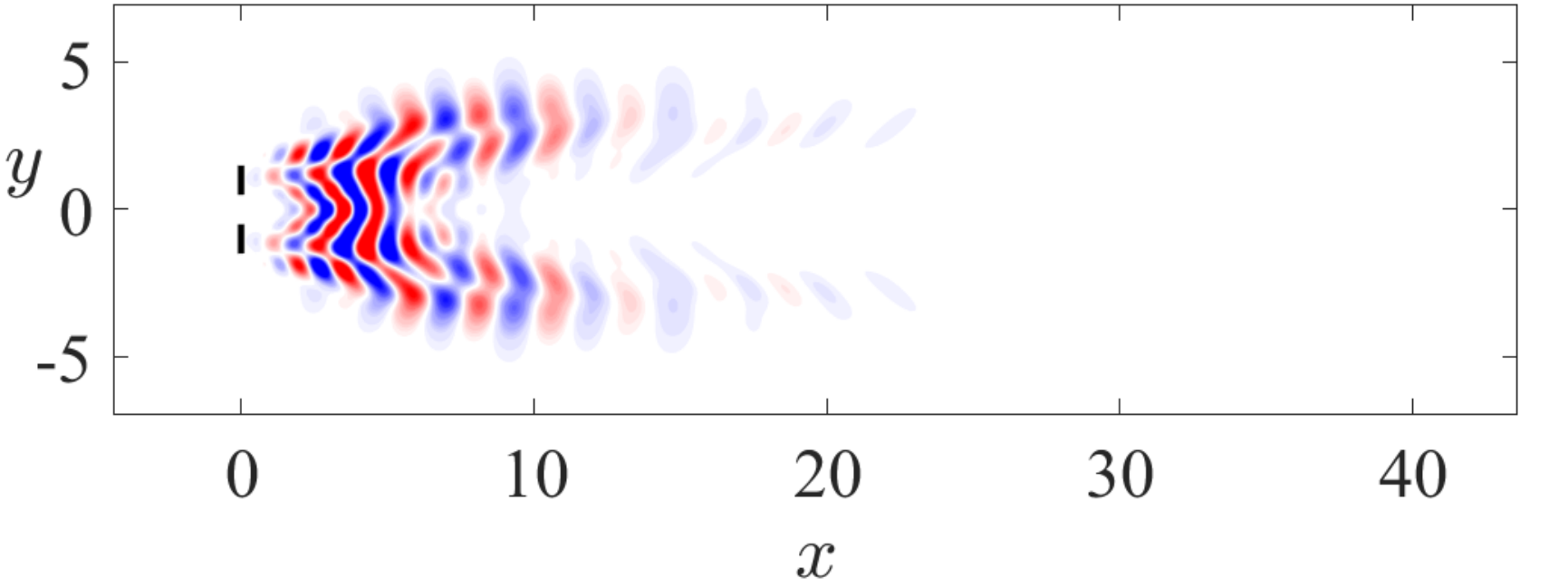} & \includegraphics[clip, trim=2cm 10cm 2cm 11.35cm,width=0.25\linewidth]{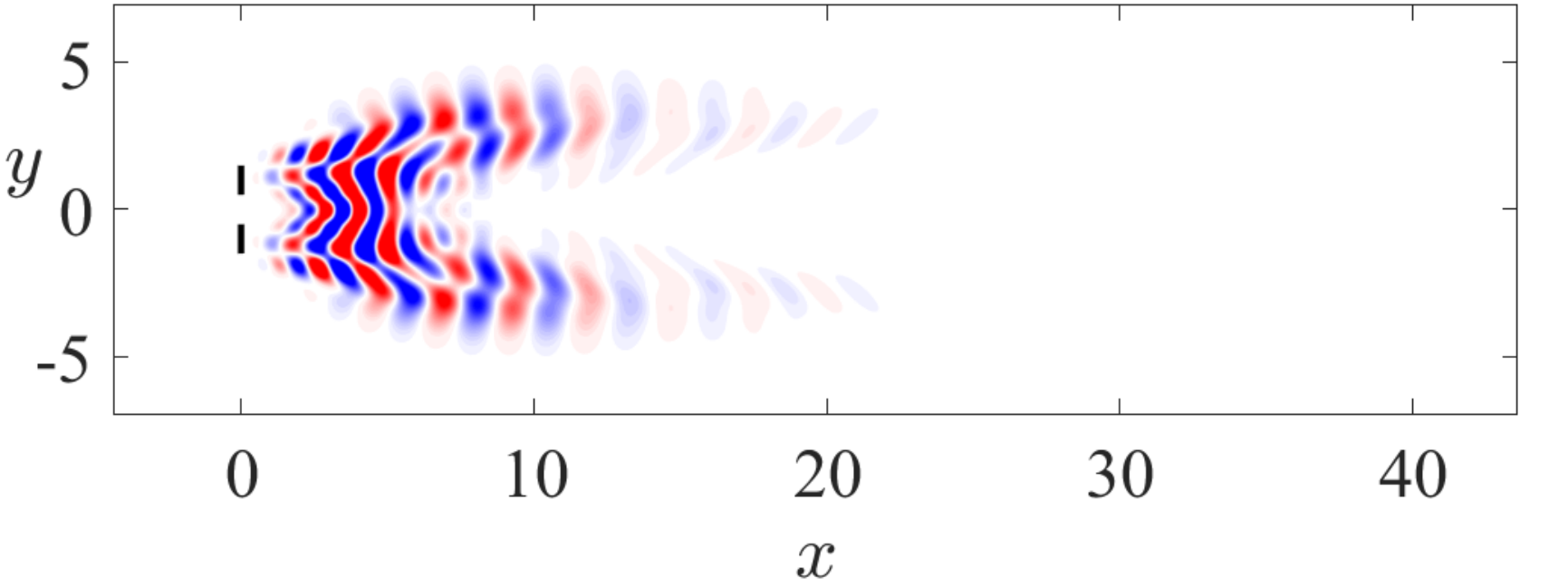} & \includegraphics[clip, trim=2cm 10cm 2cm 11.35cm,width=0.25\linewidth]{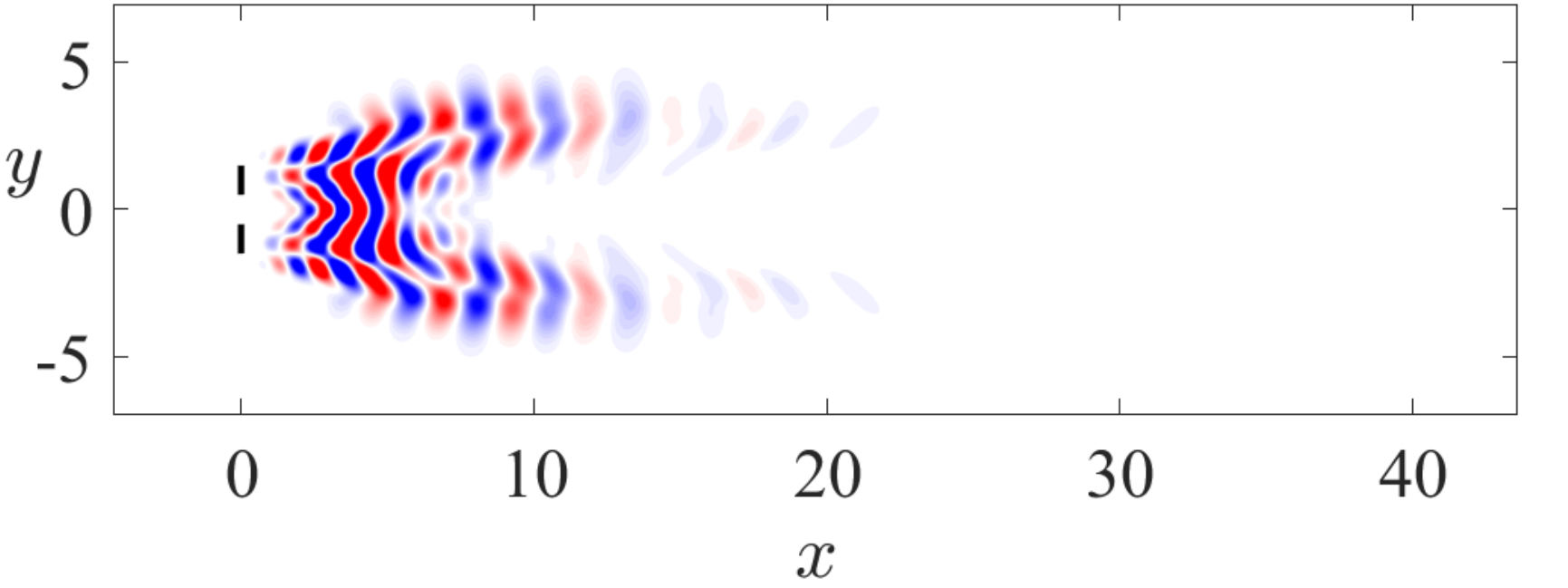} \\ 
\botrule
\end{tabular*}
\end{minipage}
\end{center}
\end{sidewaystable}

We now analyze a selection of true and reconstructed SPOD modes, at frequencies indicated by vertical dashed lines in Fig.~\ref{fig:spectra_DNS_twoplates_subsamp_compare}. 
The true and GSE-reconstructed SPOD modes for the Re = 40 case are shown in Table \ref{tab:SPOD_modes_Re40}, indicating that GSE can also accurately reconstruct the leading SPOD modes at the frequencies corresponding to the peaks in the SPOD spectrum. We observe a symmetric mode at the dominant peak concentrated in the near wake, consistent with a mode representing in-phase shedding from each plate. Note that all of the GSE-computed SPOD modes shown here have been phase-aligned with the true SPOD modes (computed from the flowfield data).

For the Re = 80 case, Table \ref{tab:SPOD_modes_Re80}  shows that GSE is again able to accurately reconstruct the leading SPOD modes at the selected frequencies. The symmetric mode at the highest frequency $fc/U_{\infty} = 0.156$ (corresponding to the second highest peak in the SPOD energy spectrum) looks visually similar to the fundamental mode for the Re = 40 case, suggesting that it is association with in-phase vortex shedding. 
The mode corresponding to the peak at $fc/U_{\infty} = 0.039$ and its first harmonic are largest further downstream of the bodies, and are reminiscent of modes behind a single bluff body. These frequencies are one quarter and one half of the highest frequency, suggesting that they are associated with period doubling and quadrupling behavior associated with the pairing, merging, and meandering of the shed vortices as they more further downstream. 

While the Re = 100 case does not have as distinct peaks in its energy spectrum, GSE is still able to quite accurately reconstruct the true SPOD modes for most cases, as shown in Table \ref{tab:SPOD_modes_Re100}. The exception is the mode at the highest frequency of 0.3359, where GSE predicts a mode that looks similar to the $fc/U_{\infty} = 0.1719$ mode. It is speculated that this is because this frequency is associated with a lower energy than the others, so the true mode may not be included in the span of the initial POD modes used for GSE. This highlights one limitation of the GSE method: any SPOD mode reconstructed must be spanned by the initial POD subspace used to identify the ROM. The mode at a frequency of 0.1719 again resembles the in-phase shedding mode observed for the other cases, though it is more strongly localized near the bodies for the Re = 100 case. This indicates that the shed vortices lose their spatial coherence more quickly for this higher Reynolds number. Similar to the Re = 80 case, the low frequency modes consist of larger structures reminiscent of those that represent the convection of vortical structures behind a single bluff body.


\begin{sidewaystable}
\sidewaystablefn%
\begin{center}
\begin{minipage}{\textheight}
\caption{
True and GSE-reconstructed SPOD modes at  four selected peak frequencies for Re = 80, for initial data with two different temporal sampling rates. Transverse ($v$) velocity contours are shown within a range of [-0.004, 0.004] }\label{tab:SPOD_modes_Re80}
\begin{tabular*}{\textheight}{@{\extracolsep{\fill}}lccc@{\extracolsep{\fill}}}
\toprule%
Re = 80 & True SPOD	& GSE, $\Delta t_{data} = 1$ & GSE, $\Delta t_{data} = 10$ \\
\midrule
$\frac{fc}{u_\infty} = 0.0391$ & \includegraphics[clip, trim=2cm 10cm 2cm 11.35cm,width=0.25\linewidth]{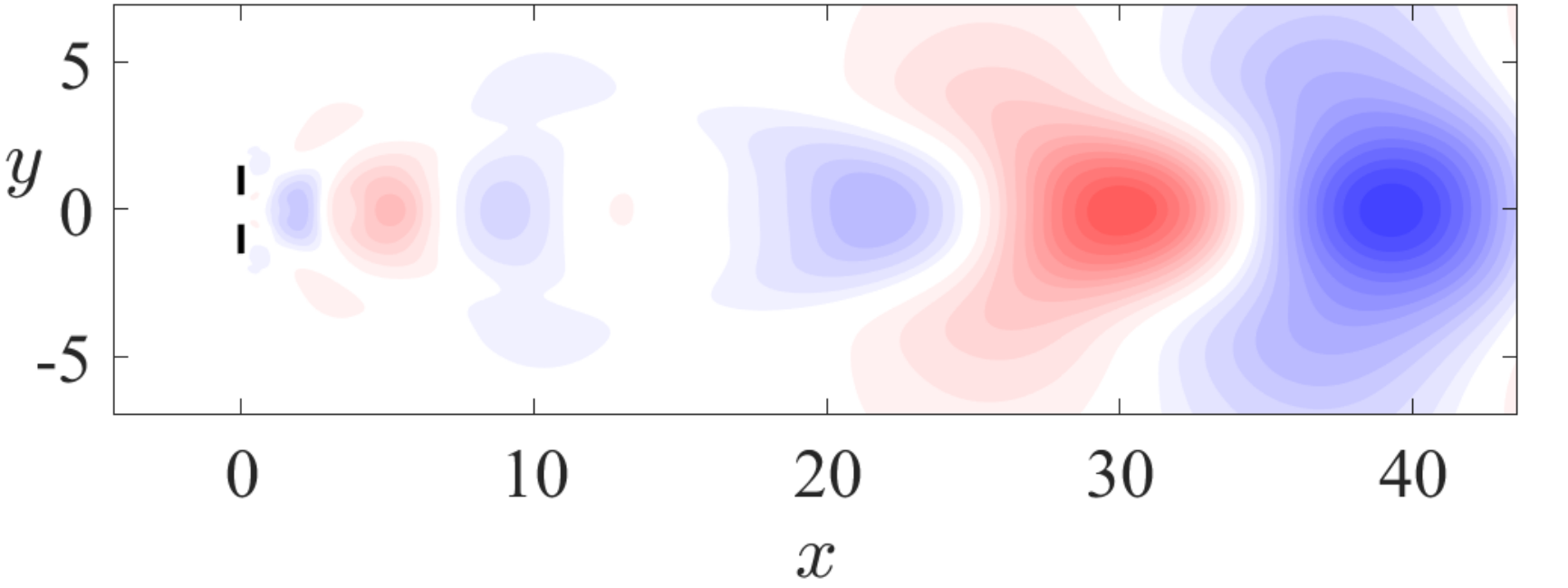} & \includegraphics[clip, trim=2cm 10cm 2cm 11.35cm,width=0.25\linewidth]{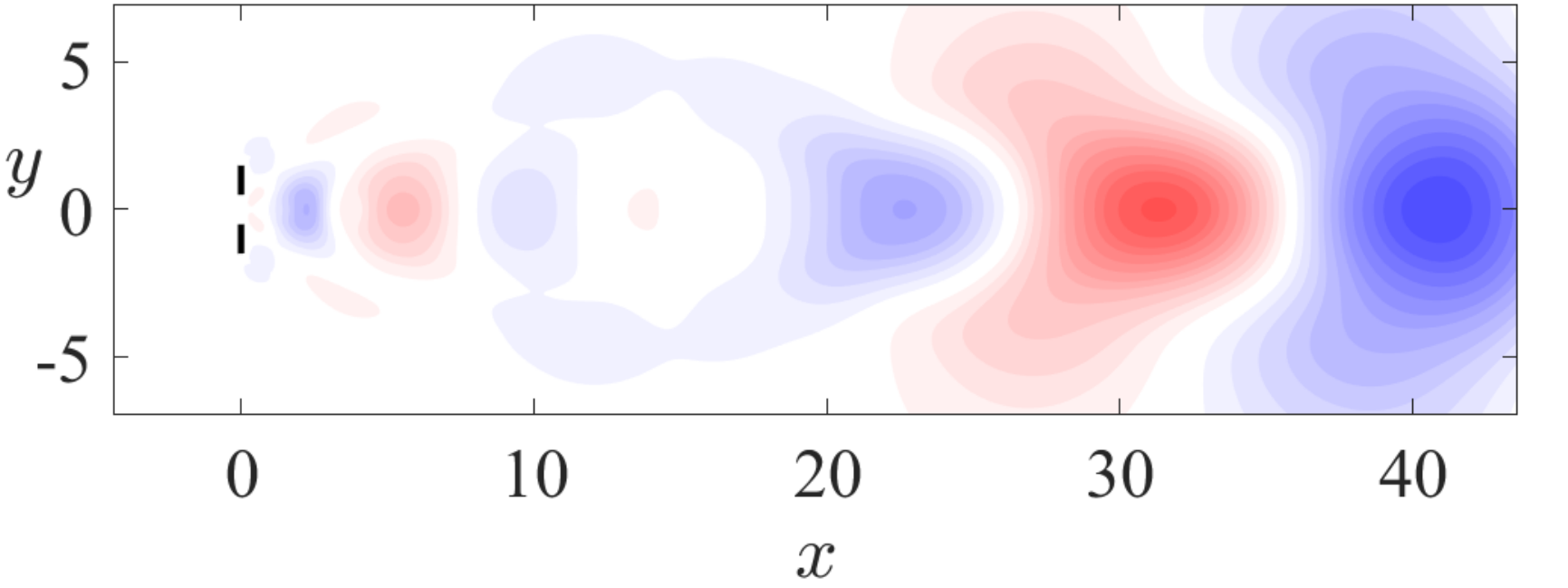} & \includegraphics[clip, trim=2cm 10cm 2cm 11.35cm,width=0.25\linewidth]{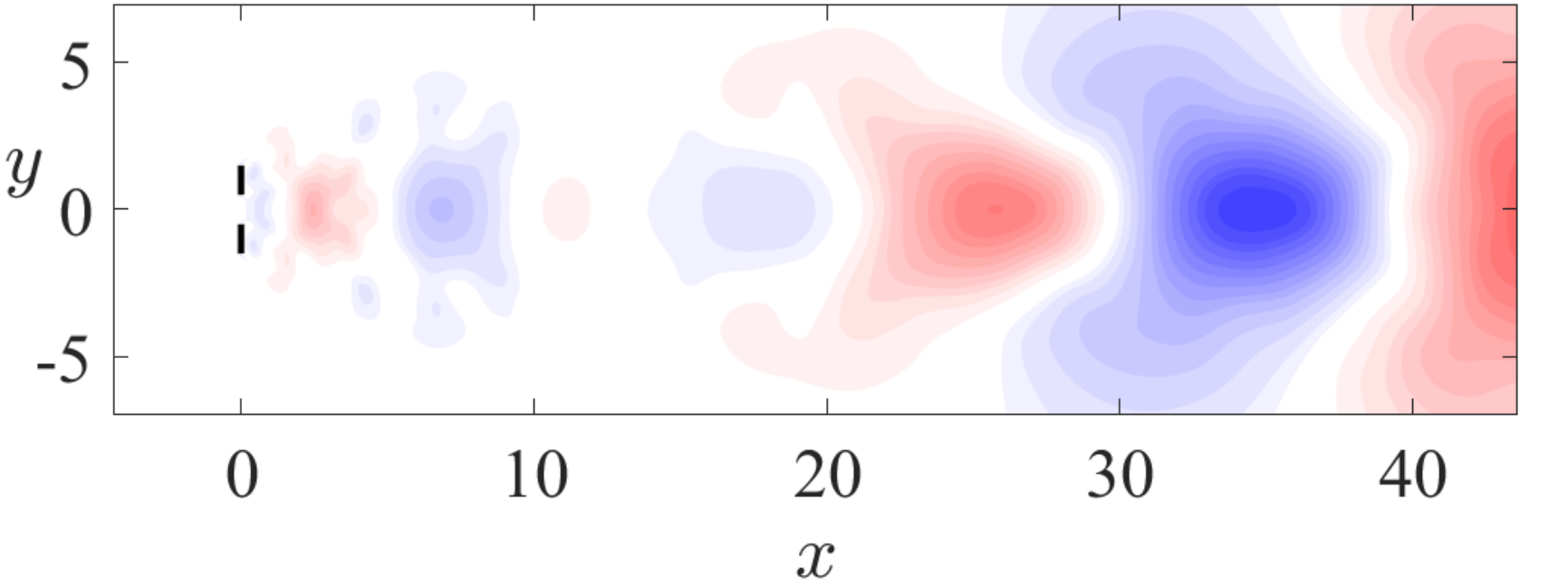} \\ 
$\frac{fc}{u_\infty} = 0.0781$ & \includegraphics[clip, trim=2cm 10cm 2cm 11.35cm,width=0.25\linewidth]{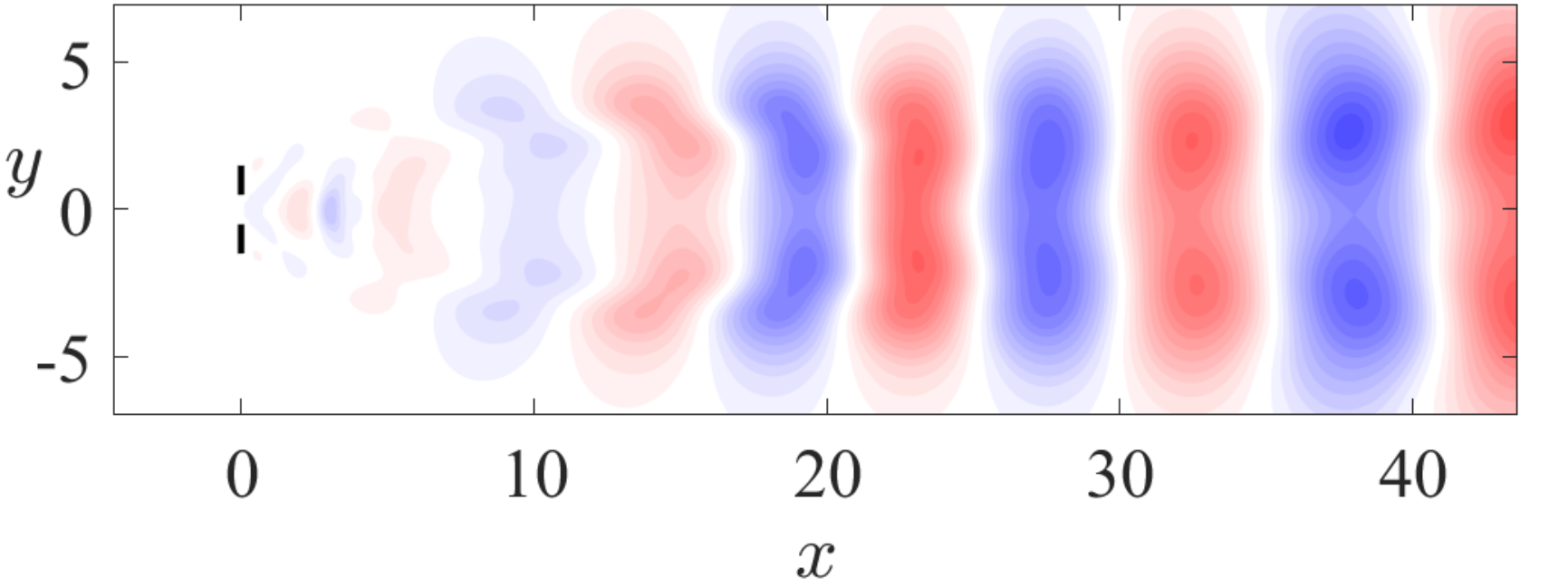} & \includegraphics[clip, trim=2cm 10cm 2cm 11.35cm,width=0.25\linewidth]{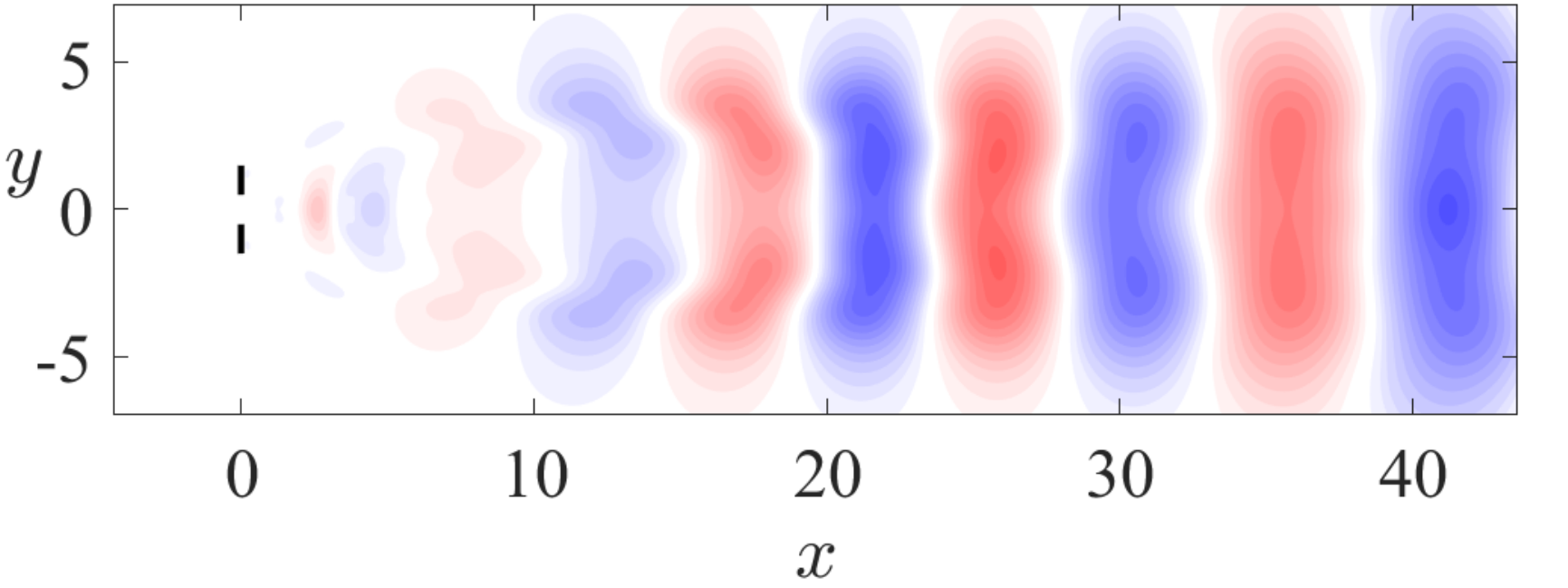} & \includegraphics[clip, trim=2cm 10cm 2cm 11.35cm,width=0.25\linewidth]{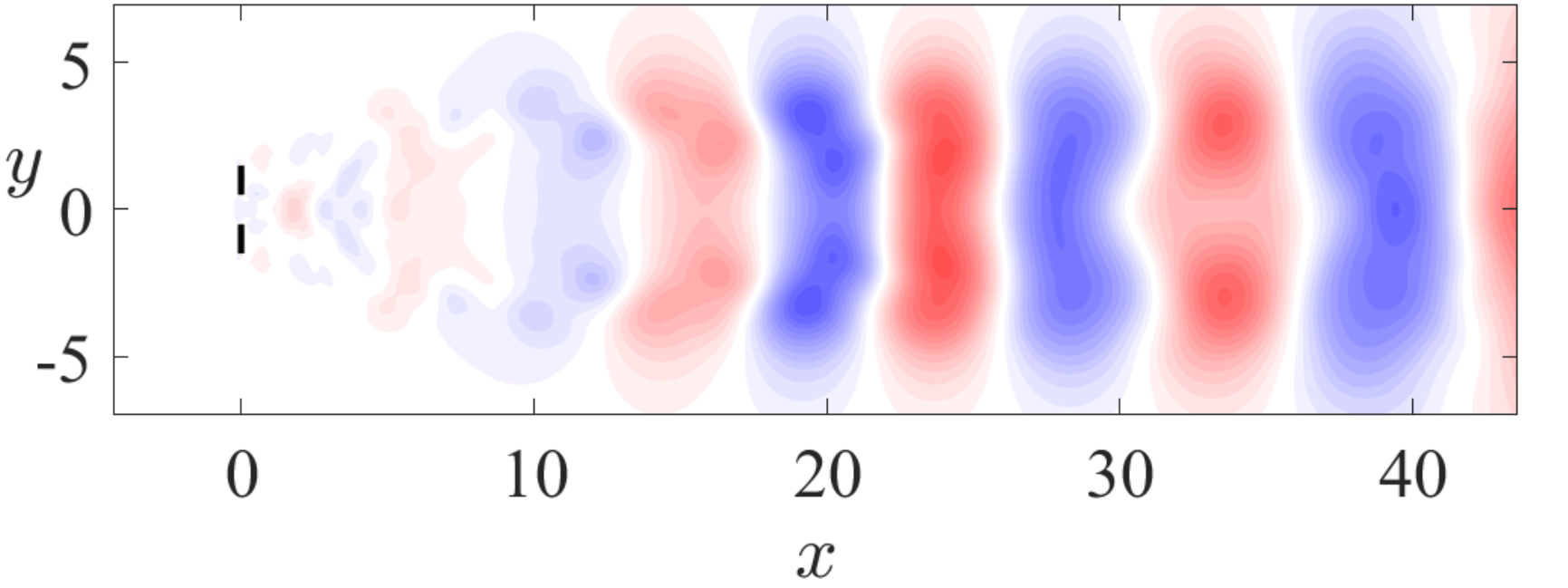} \\ 
$\frac{fc}{u_\infty} = 0.1172$ & \includegraphics[clip, trim=2cm 10cm 2cm 11.35cm,width=0.25\linewidth]{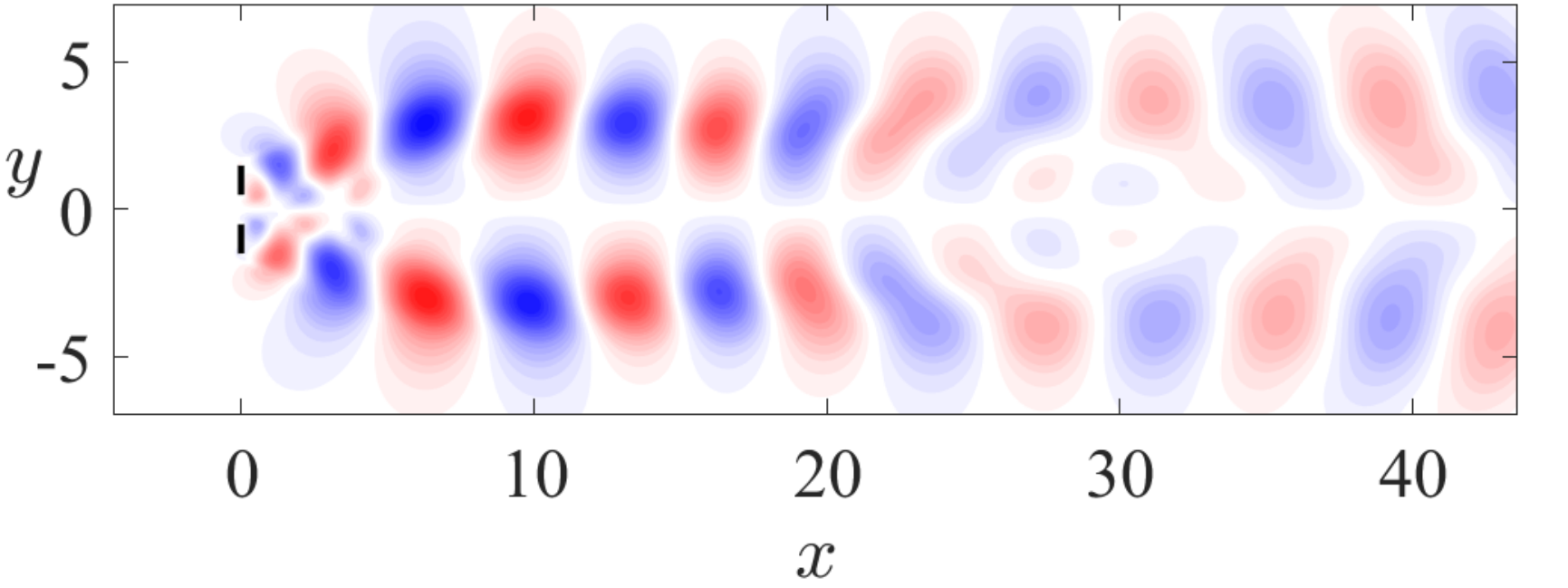} & \includegraphics[clip, trim=2cm 10cm 2cm 11.35cm,width=0.25\linewidth]{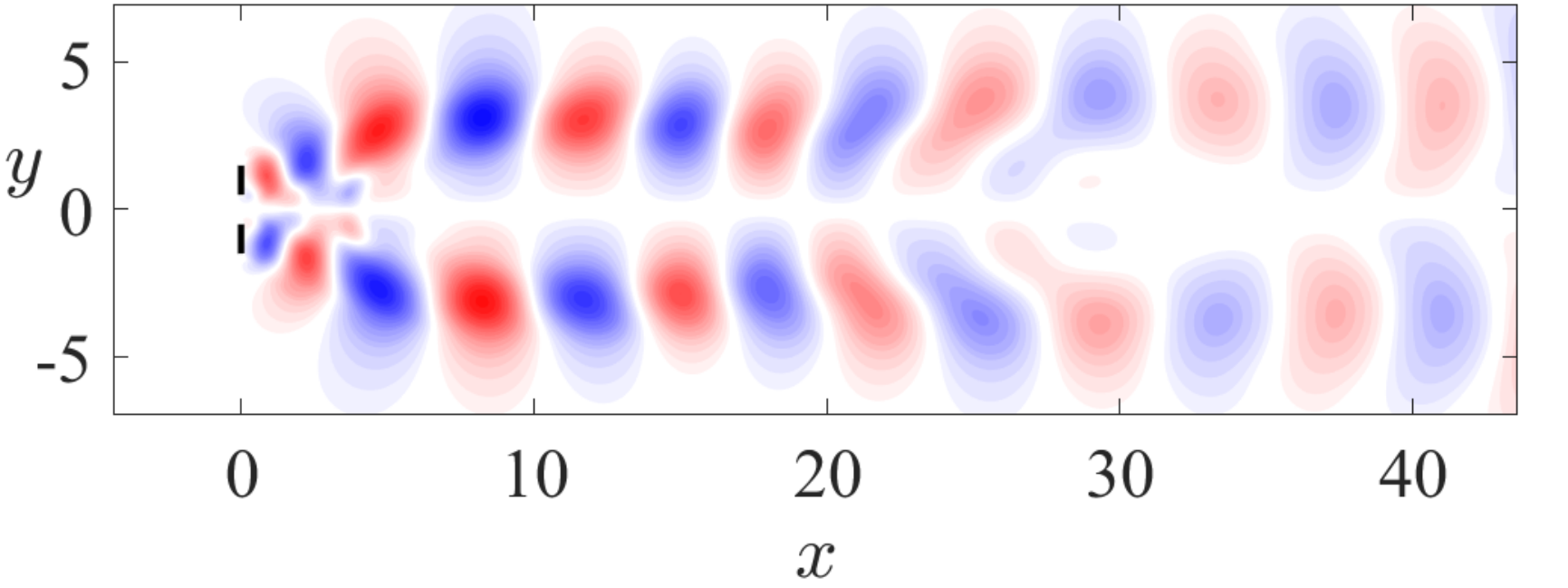} & \includegraphics[clip, trim=2cm 10cm 2cm 11.35cm,width=0.25\linewidth]{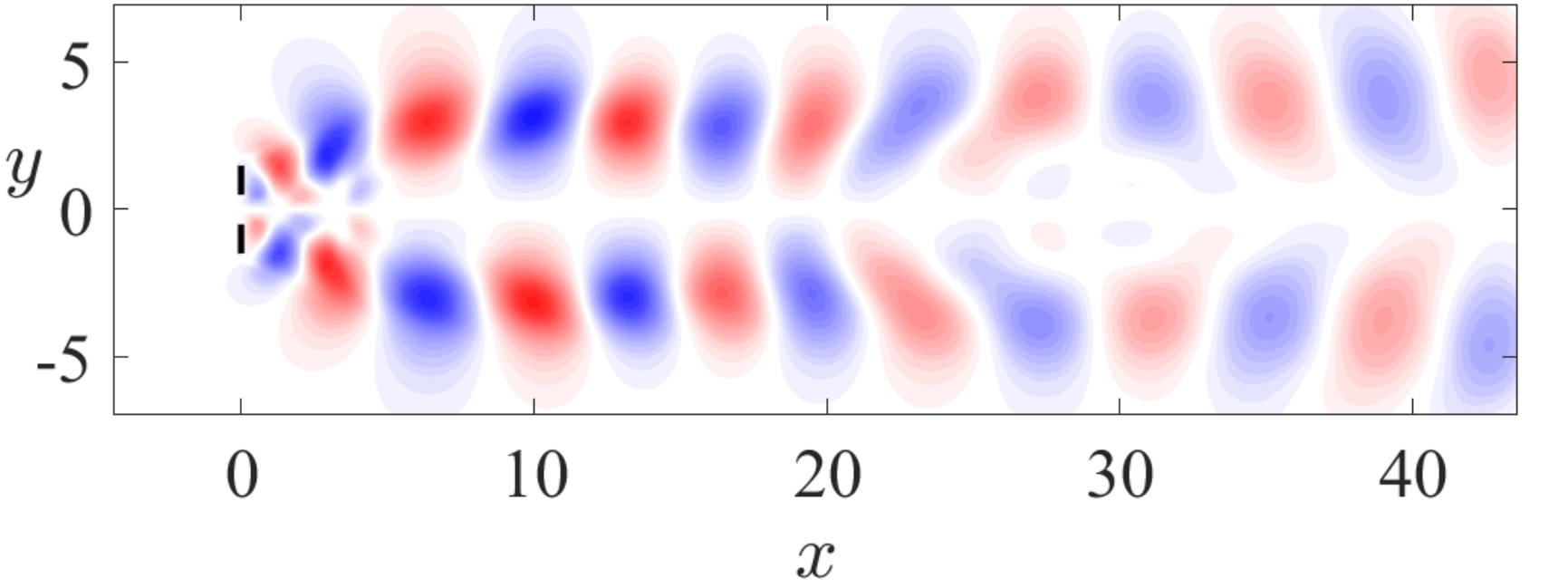} \\ 
$\frac{fc}{u_\infty} = 0.1562$ & \includegraphics[clip, trim=2cm 10cm 2cm 11.35cm,width=0.25\linewidth]{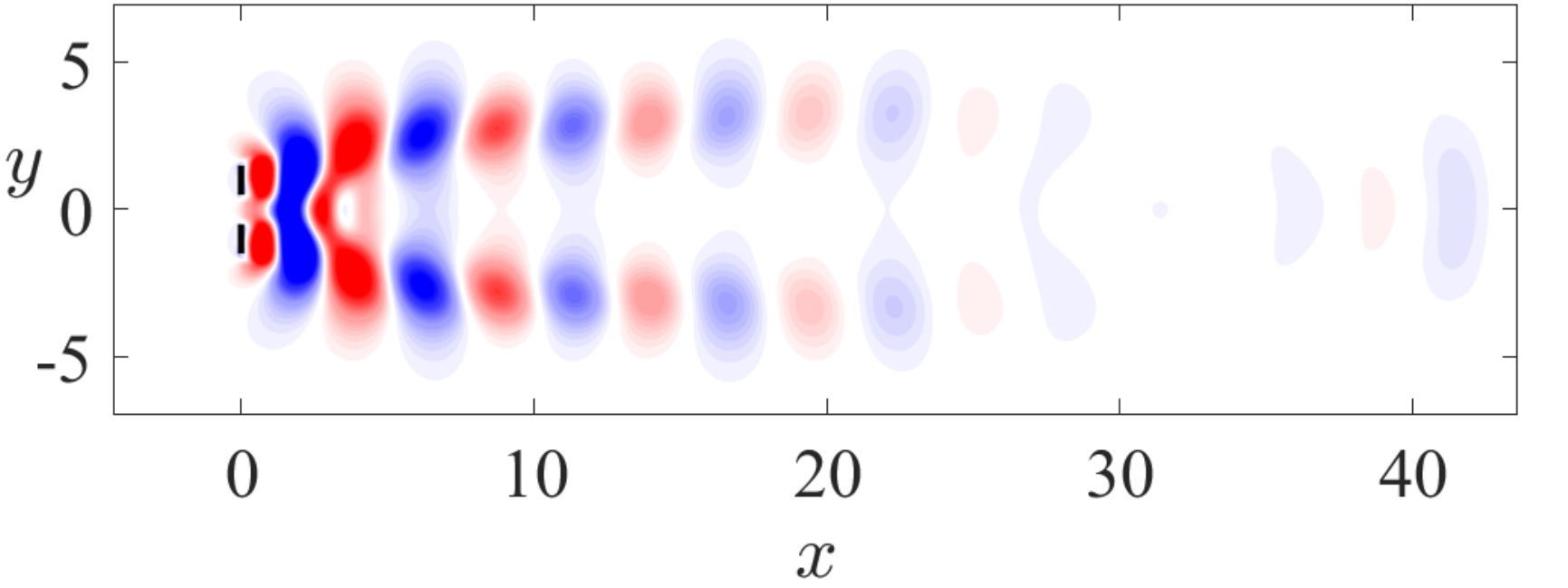} & \includegraphics[clip, trim=2cm 10cm 2cm 11.35cm,width=0.25\linewidth]{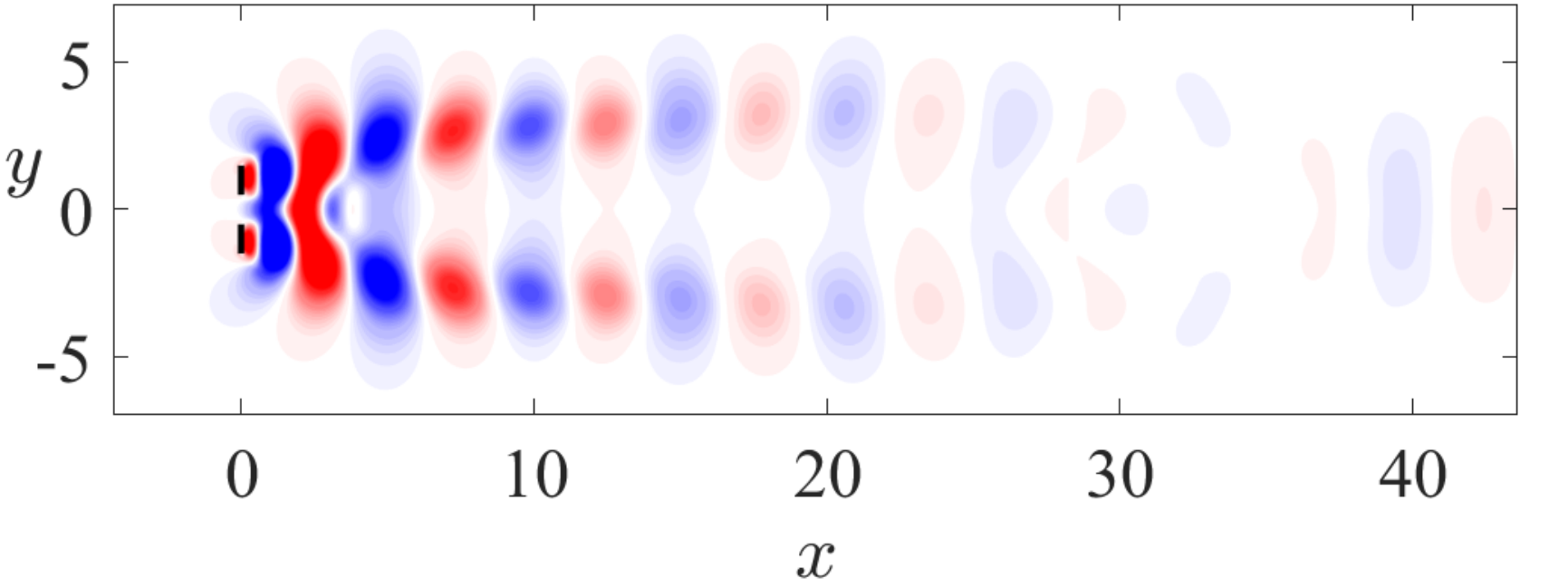} & \includegraphics[clip, trim=2cm 10cm 2cm 11.35cm,width=0.25\linewidth]{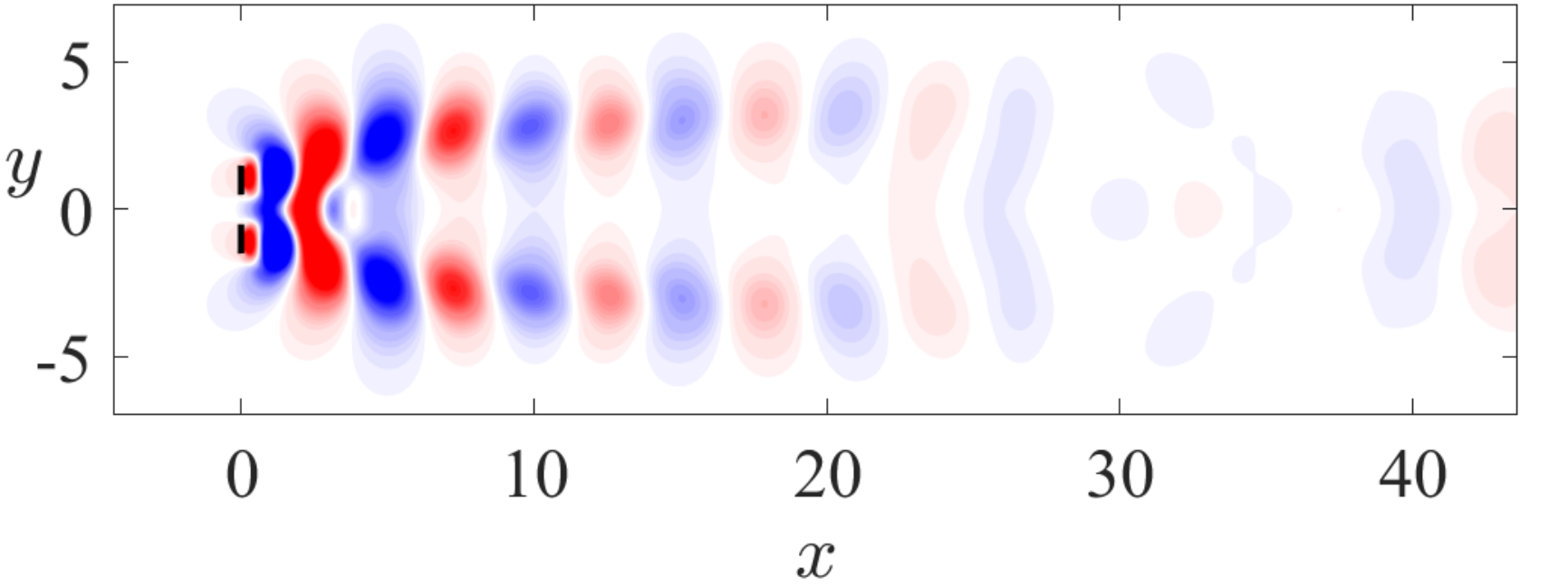} \\
\botrule
\end{tabular*}
\end{minipage}
\end{center}
\end{sidewaystable}

\begin{sidewaystable}
\sidewaystablefn%
\begin{center}
\begin{minipage}{\textheight}
\caption{
True and GSE-reconstructed SPOD modes at  four selected frequencies for Re = 100, for initial data with two different temporal sampling rates. Transverse ($v$) velocity contours are shown within a range of [-0.006, 0.006].
}\label{tab:SPOD_modes_Re100}
\begin{tabular*}{\textheight}{@{\extracolsep{\fill}}lccc@{\extracolsep{\fill}}}
\toprule%
Re = 100 & True SPOD	& GSE, $\Delta t_{data} = 1$ & GSE, $\Delta t_{data} = 10$ \\
\midrule
$\frac{fc}{u_\infty} = 0.0547$ & \includegraphics[clip, trim=2cm 10cm 2cm 11.35cm,width=0.25\linewidth]{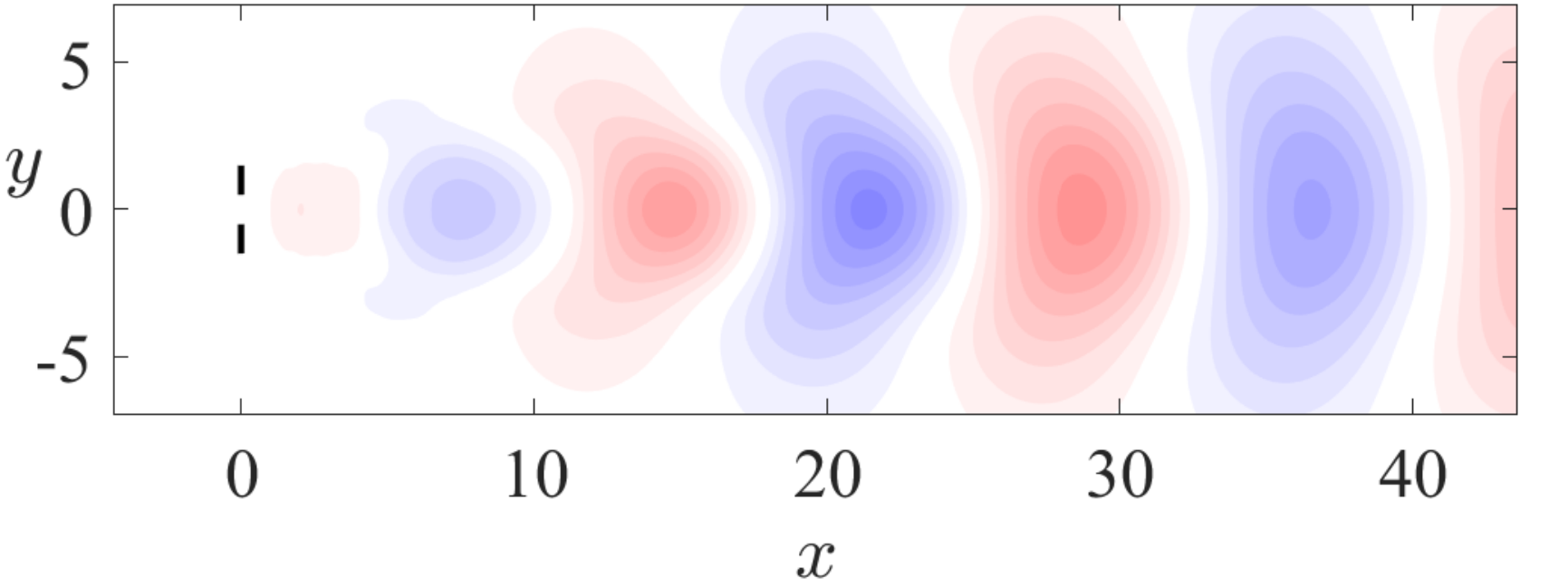} & \includegraphics[clip, trim=2cm 10cm 2cm 11.35cm,width=0.25\linewidth]{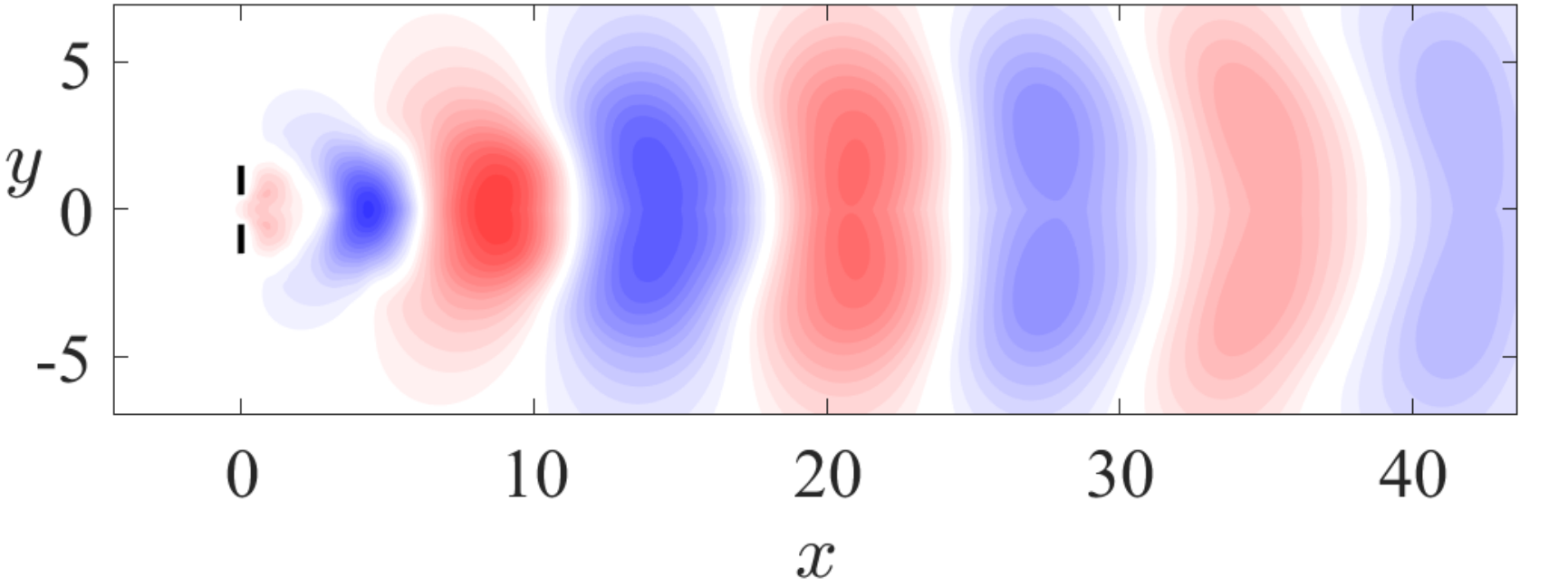} & \includegraphics[clip, trim=2cm 10cm 2cm 11.35cm,width=0.25\linewidth]{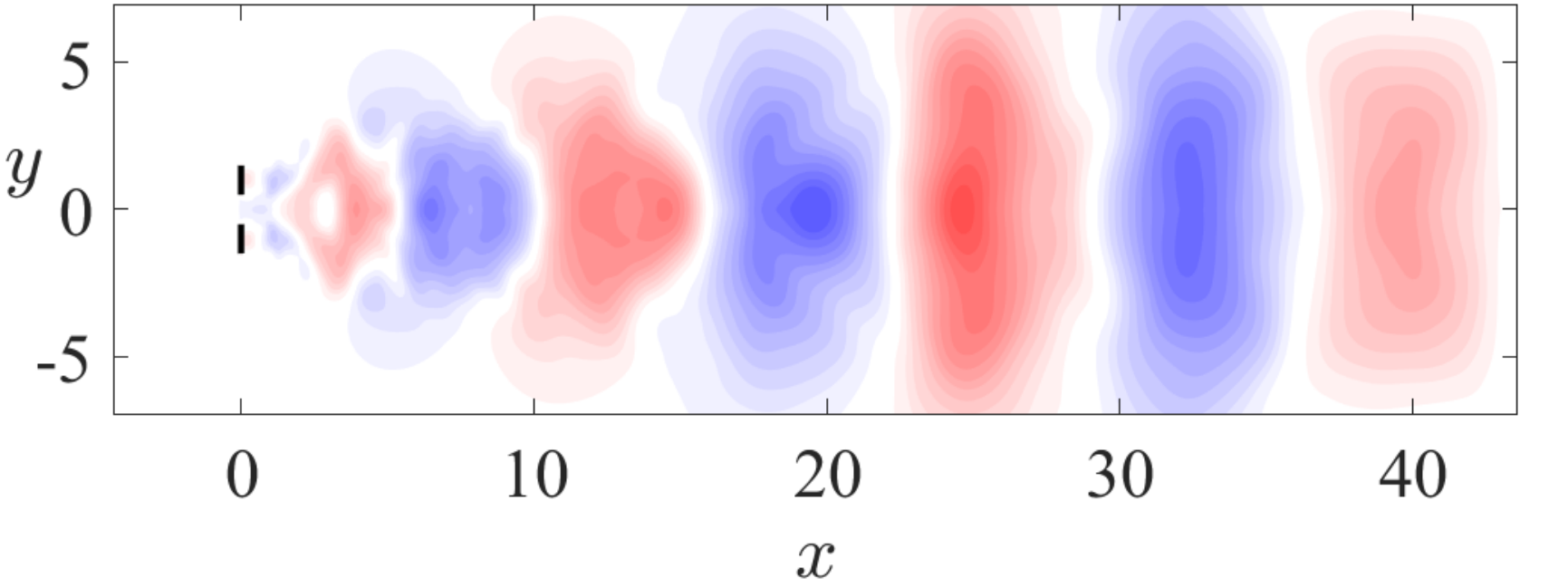} \\
$\frac{fc}{u_\infty} = 0.0781$ & \includegraphics[clip, trim=2cm 10cm 2cm 11.35cm,width=0.25\linewidth]{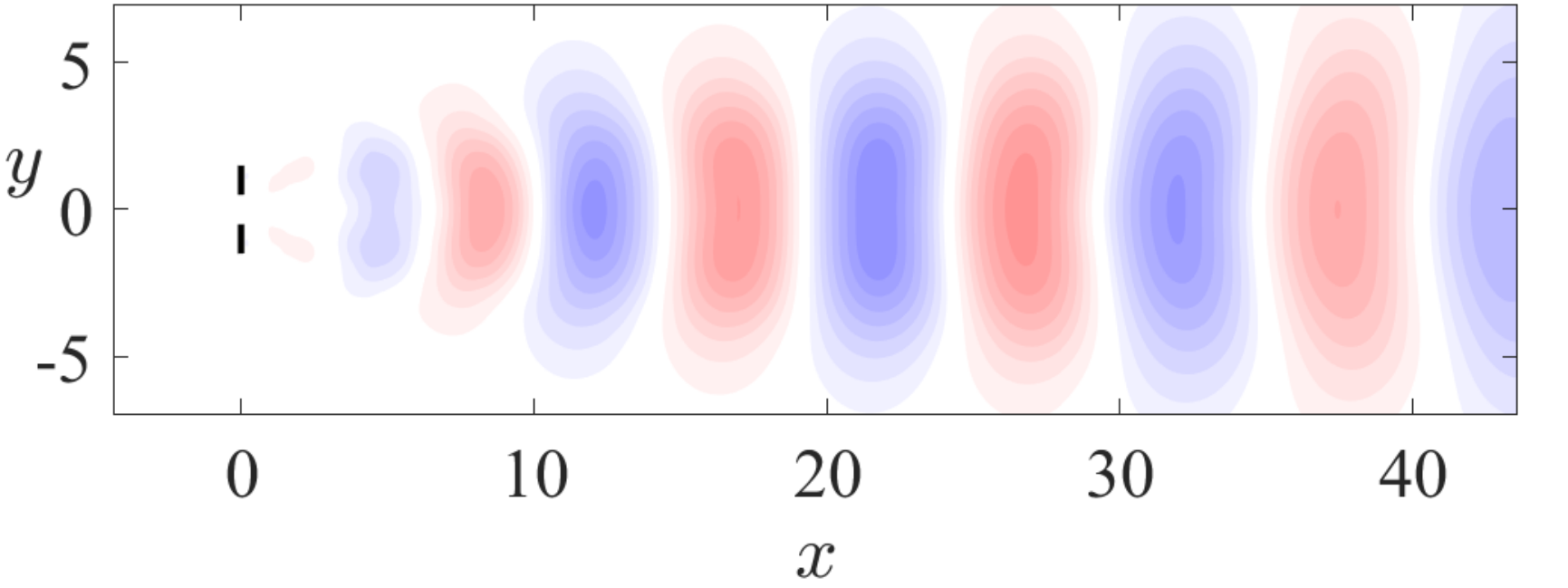} & \includegraphics[clip, trim=2cm 10cm 2cm 11.35cm,width=0.25\linewidth]{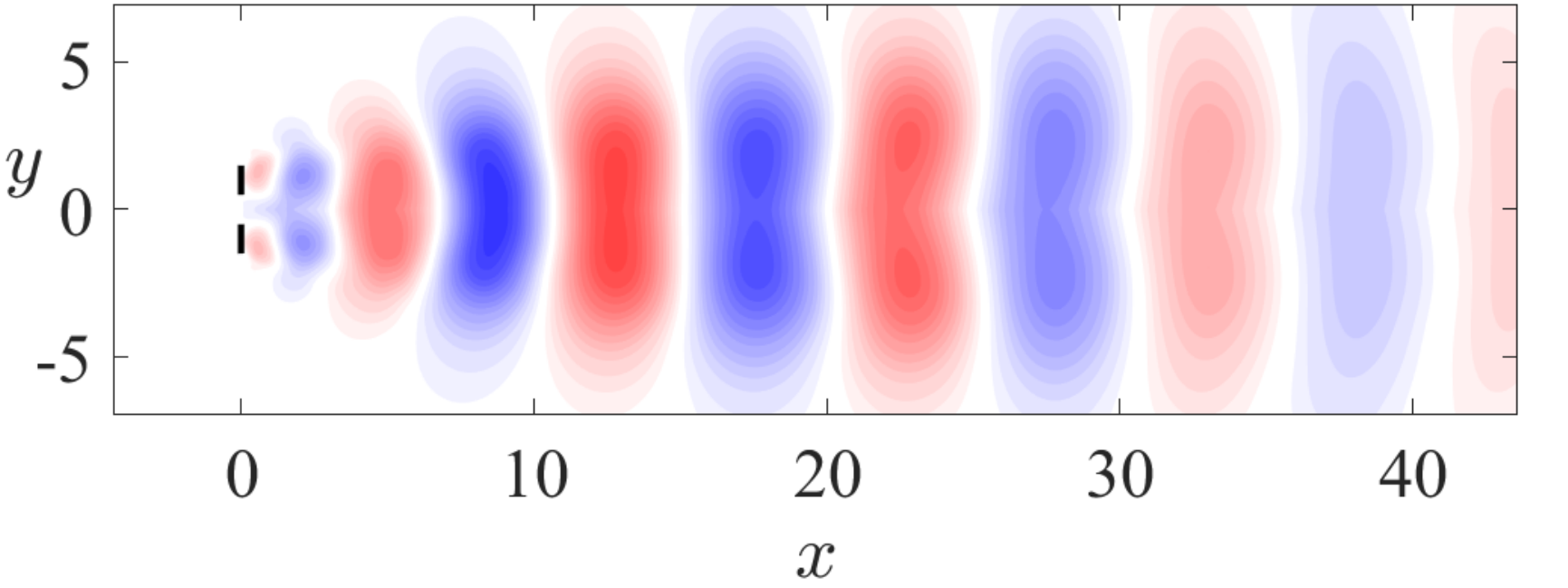} & \includegraphics[clip, trim=2cm 10cm 2cm 11.35cm,width=0.25\linewidth]{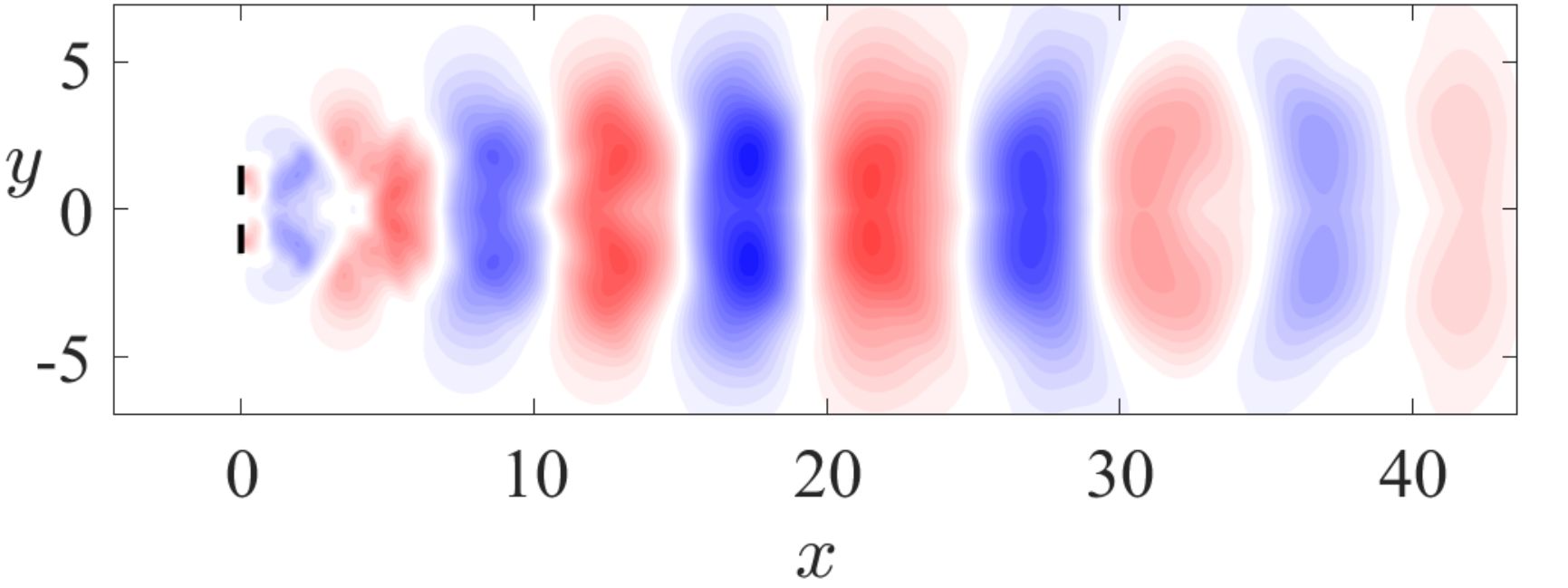} \\
$\frac{fc}{u_\infty} = 0.1719$ & \includegraphics[clip, trim=2cm 10cm 2cm 11.35cm,width=0.25\linewidth]{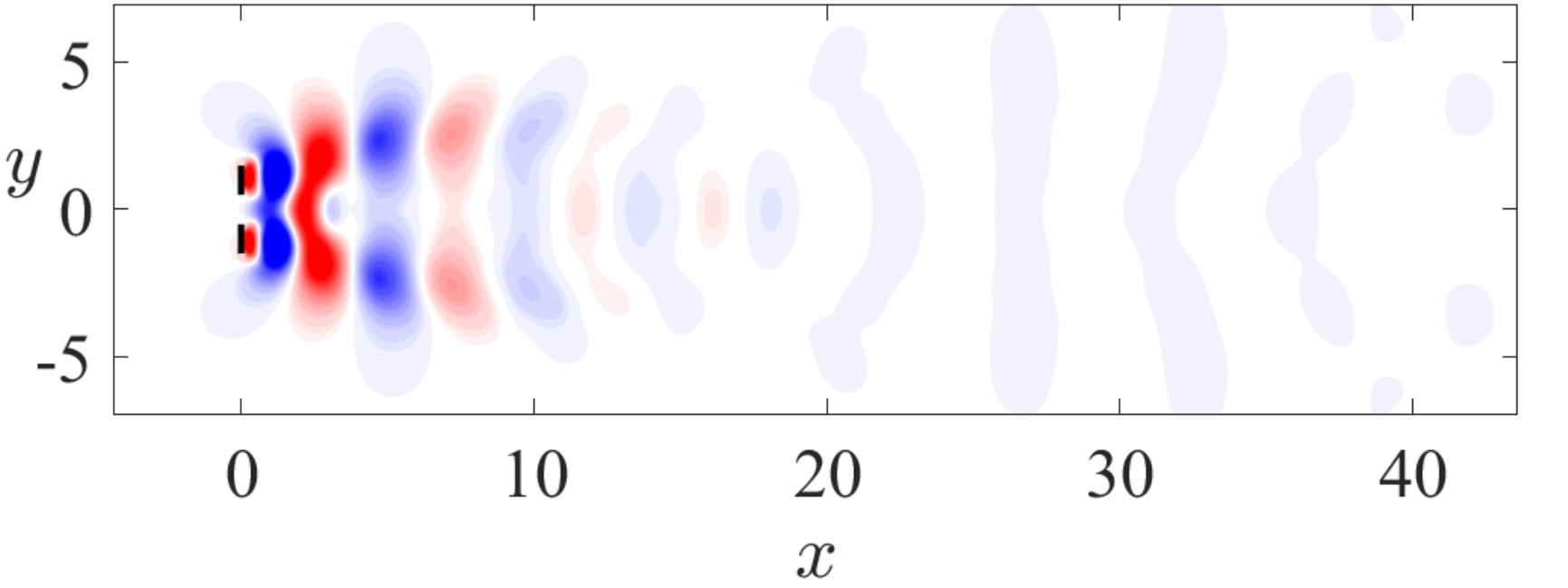} & \includegraphics[clip, trim=2cm 10cm 2cm 11.35cm,width=0.25\linewidth]{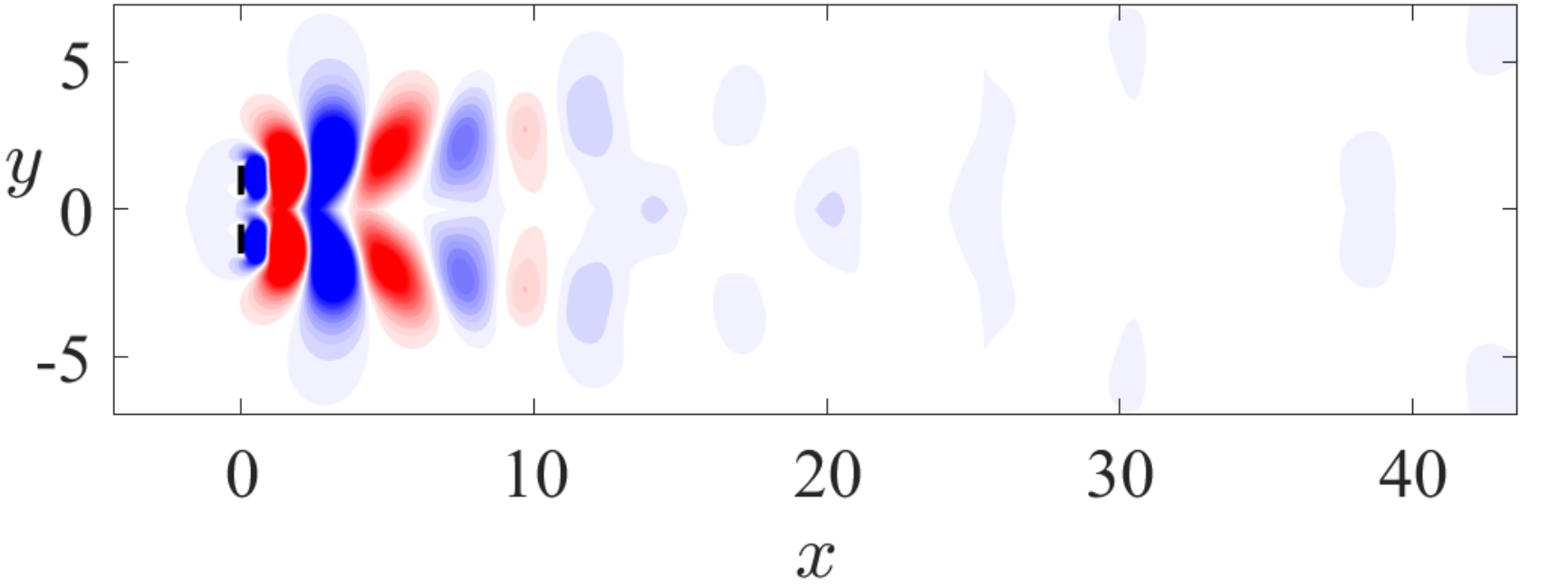} & \includegraphics[clip, trim=2cm 10cm 2cm 11.35cm,width=0.25\linewidth]{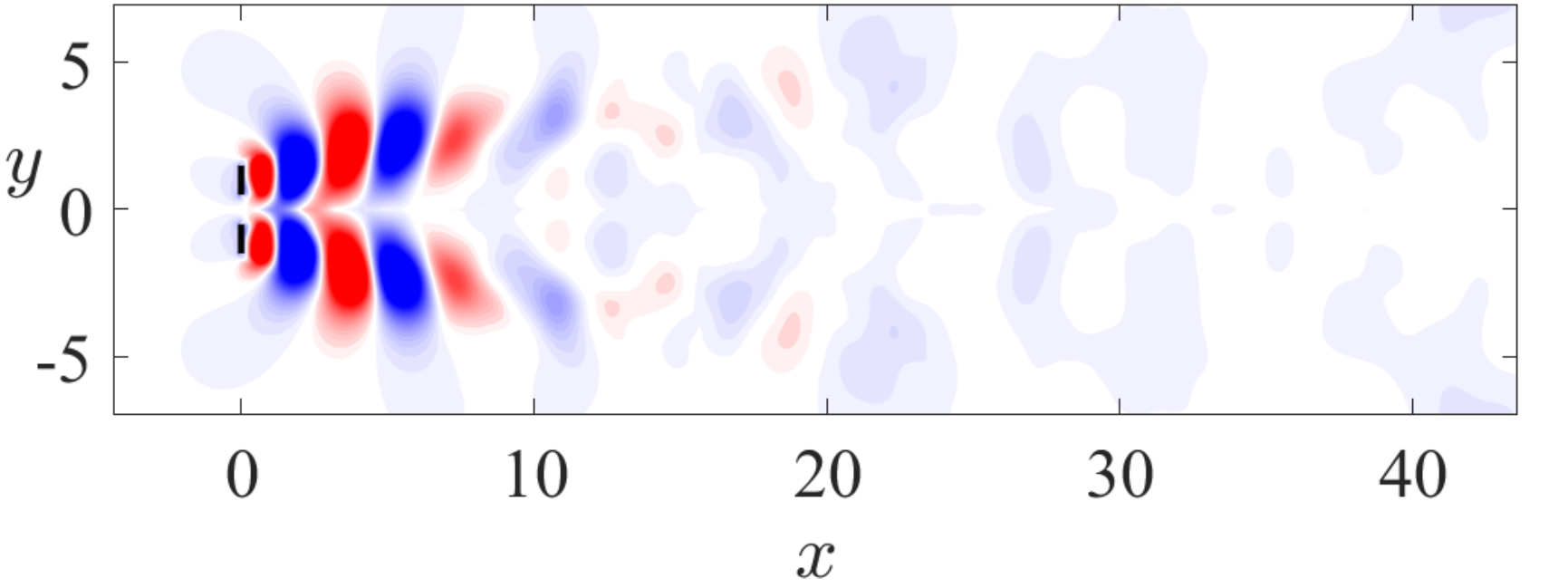} \\
$\frac{fc}{u_\infty} = 0.3359$ & \includegraphics[clip, trim=2cm 10cm 2cm 11.35cm,width=0.25\linewidth]{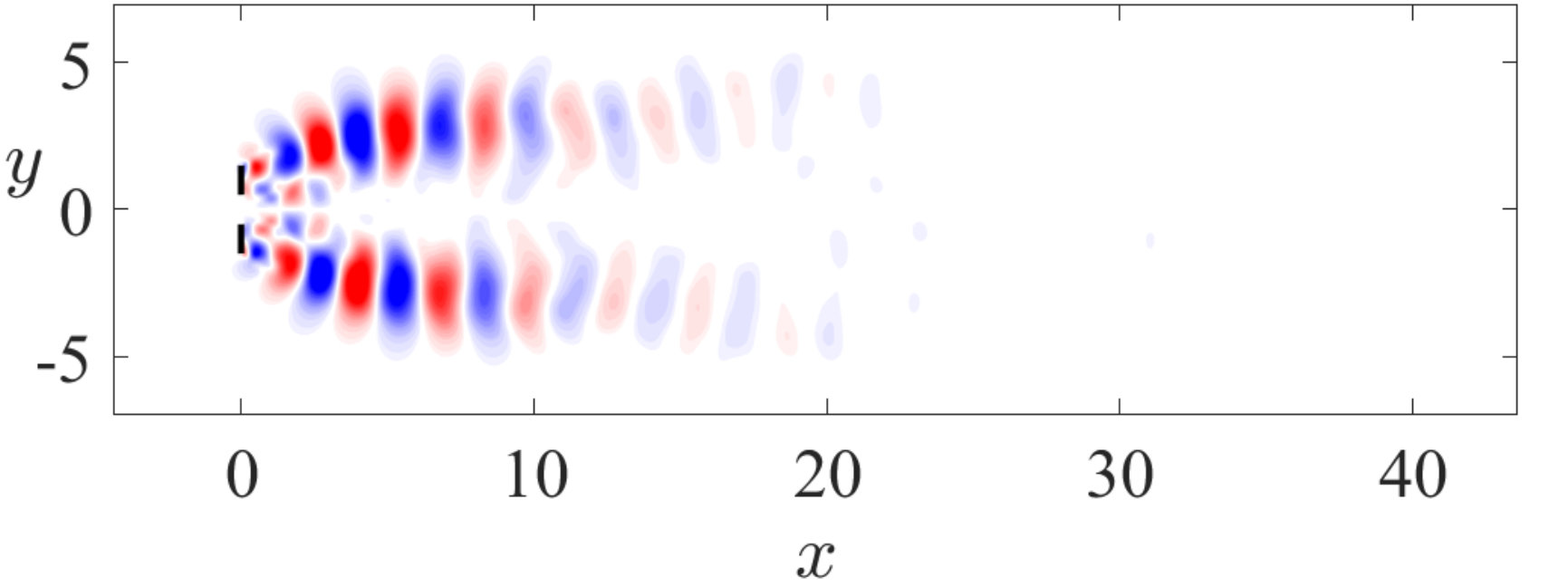} & \includegraphics[clip, trim=2cm 10cm 2cm 11.35cm,width=0.25\linewidth]{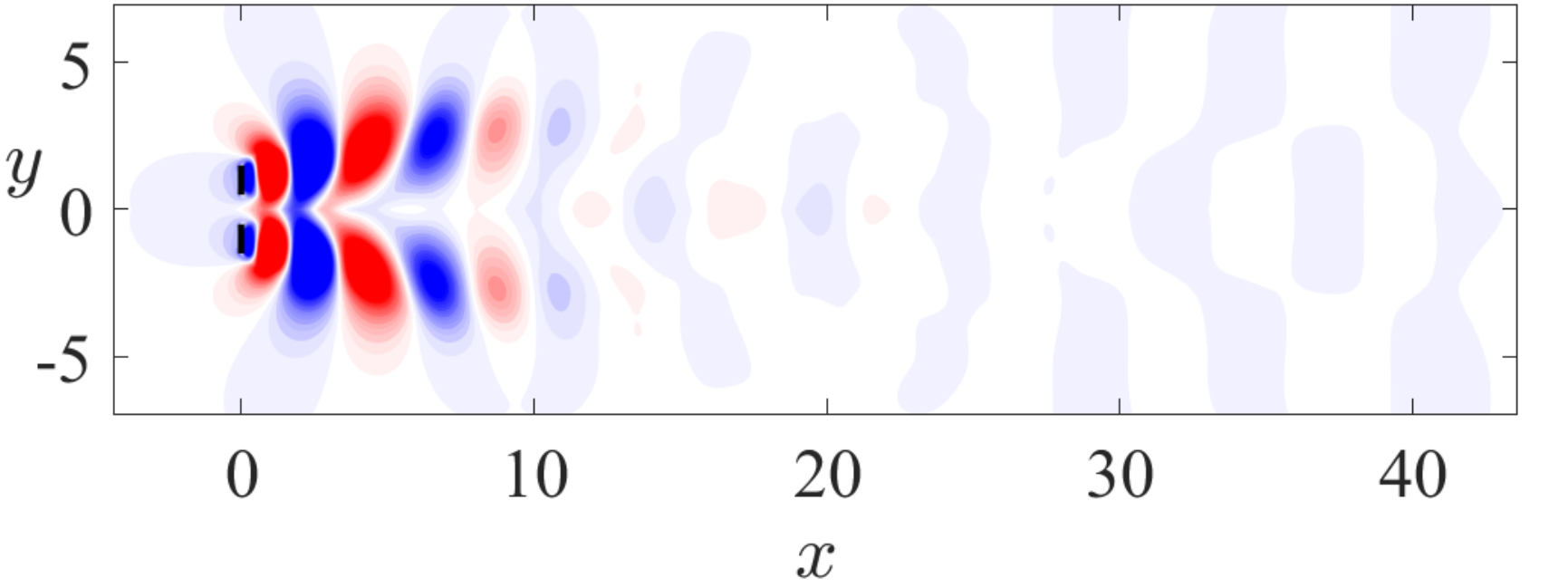} & \includegraphics[clip, trim=2cm 10cm 2cm 11.35cm,width=0.25\linewidth]{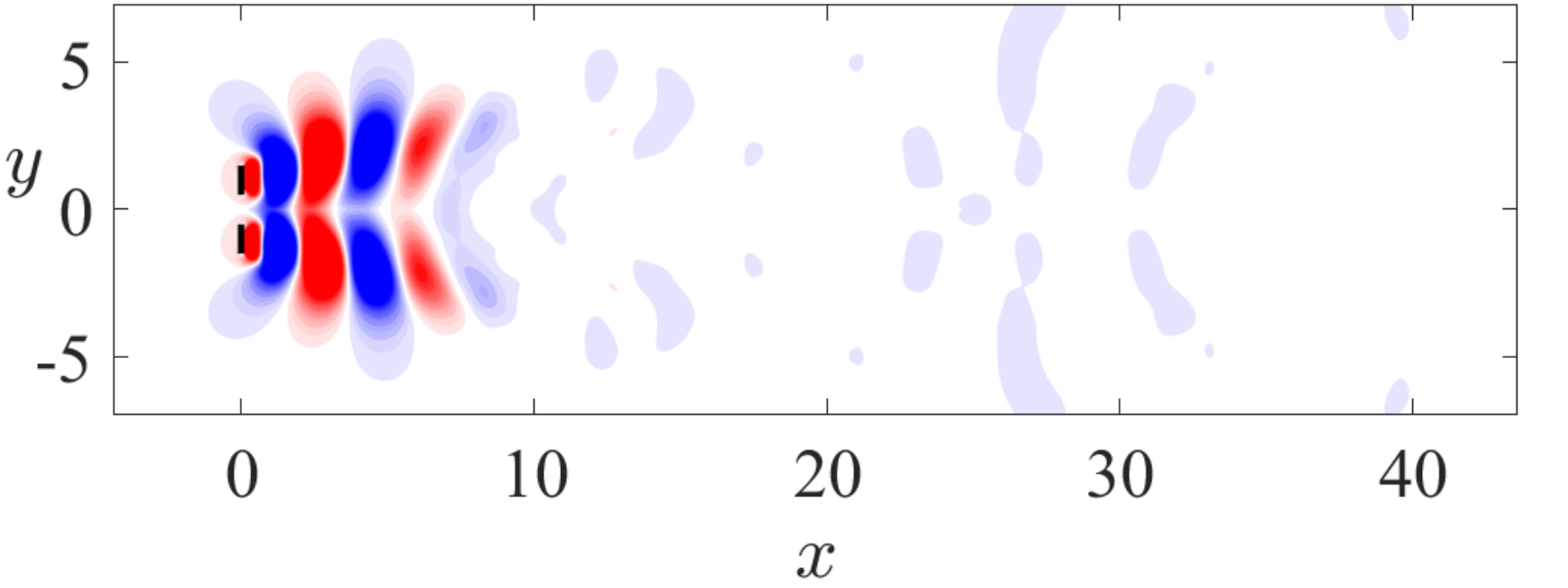} \\ 
\botrule
\end{tabular*}
\end{minipage}
\end{center}
\end{sidewaystable}

Table~\ref{tab:spod_error} shows the error  in the GSE estimation of the first SPOD mode for the leading peak in the SPOD spectrum for each case, for three different initial sampling rates. It can be observed that the error increases with Reynolds number, but is relatively independent of the sampling timestep used for the initial data for GSE. This is unsurprising, since GSE does not utilize this temporal information, and only requires that the space-only POD modes used to obtain the POD-Galerkin model are sufficiently converged. While not shown here, we have also verified that accurate reconstruction of SPOD using GSE can also be achieved from non-uniform, randomly sampled snapshots of data.

\begin{table}[!htb]
\begin{center}
\begin{minipage}{174pt}
\caption{Percent error in GSE-computed SPOD modes}\label{tab:spod_error}
\begin{tabular}{@{}lccc@{}}
\toprule
\text{Re} & $\Delta t_{data} = 1$  & $\Delta t_{data} = 5$ & $\Delta t_{data} = 10$ \\
\midrule
40 & 3.5$\times 10^{-4}$ & 3.7$\times 10^{-4}$ & 4$\times 10^{-4}$ \\
80 & 2.1 & 3.5 & 3 \\
100 & 13 & 5.7 & 9.1 \\
\botrule
\end{tabular}
\end{minipage}
\end{center}
\end{table}

\section{Discussion and Conclusions}
\label{sec:conclusion}

We have demonstrated that the proposed GSE methodology can be utilized for reconstructing both energy spectra and leading SPOD modes for vortex-dominated wake flows exhibiting a range of dynamical complexity. 
 In essence, this method combines two methodologies that are both derived from basic principals of the proper orthogonal decomposition. On one hand, space-only POD identifies a tailored subspace that can be used to form a reduced-order model for a given system. Spectral POD identifies modes tied to a specific temporal frequency, which we are using POD-Galerkin ROMs to reconstruct. 
 From another point of view, the proposed GSE method amounts to a way of utilizing traditional POD-Galerkin ROMs to estimate additional statistical properties of the underlying system. In cases where the true SPOD is known, this could also be used to assess the accuracy of POD-Galerkin models at different timescales. 

  Here, we considered generating multiple trajectories of data from ROMs by starting at different initial conditions, which may be obtained from snapshots of the original data. An extension of this idea could involve combining time-resolved ROM predictions with non-time-resolved data through the use of a Kalman filter or smoother at times when original data is available.

We have formulated the GSE method using standard versions of both Galerkin projection and SPOD. It would be relatively straightforward to incorporate additional variants of both methods to potentially improve results for certain applications. For example, closure models can be used to improve the accuracy of POD-Galerkin models \cite{wang2012podclosure,osth2014podvisc,cordier2013identification,callaham2022multiscale} for cases where it is difficult to capture all dissipative effects. Similarly, recent work using 
multi-taper estimates \cite{schmidt2022spectral} and other aliasing-mitigation strategies \cite{karban2022solutions} could be incorporated to improve the convergence of the SPOD computation. 
 Another variant of the proposed methodology could involve performing a projection onto the linearized (rather than nonlinear) governing equations, which would enable prediction of the spectral content of the system through direct analysis of the pseudospecra 
of the underlying operator, rather than through applying SPOD on data generated by the ROM.

 The GSE methodology could potentially be useful for experimental data collected through techniques such as particle image velocimetry, where the high sampling rate requirements often constrict the spatial window that can be captured. While it is typically difficult to acquire accurate POD-Galerkin projection models from experimental data, the fact that long term stability is not required for GSE could make such analysis feasible. Future work will look to apply it to a broader range of both numerical and experimental data.

\backmatter


\section*{Declarations}

The authors are not aware of any biases or conflicts of interest that might be interpreted as affecting the objectivity of this work. 

\bmhead{Ethical Approval}

Not applicable.

\bmhead{Competing interests}

Note applicable

\bmhead{Authors' contributions}

K.J.A. and S.T.M.D. provided the original conceptualization for the method presented in this work; however, the methodology was finalized equally among all authors. K.J.A. and S.T.M.D. contributed equally to the example problem formulation and results for section~\ref{results_toy}. A.A. performed the simulations and analysis for the results shown in section~\ref{results_twoplates}. All authors contributed equally to editing and reviewing the manuscript.

\bmhead{Funding}

The authors gratefully acknowledge the support for this work from the the Achievement Rewards for College Scientists Foundation, Inc.'s Scholar Illinois Chapter, and the Illinois Space Grant Consortium.

\bmhead{Availability of data and materials}

Access to datasets presented in this work may be available upon request.

\bibliography{spectral_bib}


\begin{thebibliography}{79}
\ifx \bisbn   \undefined \def \bisbn  #1{ISBN #1}\fi
\ifx \binits  \undefined \def \binits#1{#1}\fi
\ifx \bauthor  \undefined \def \bauthor#1{#1}\fi
\ifx \batitle  \undefined \def \batitle#1{#1}\fi
\ifx \bjtitle  \undefined \def \bjtitle#1{#1}\fi
\ifx \bvolume  \undefined \def \bvolume#1{\textbf{#1}}\fi
\ifx \byear  \undefined \def \byear#1{#1}\fi
\ifx \bissue  \undefined \def \bissue#1{#1}\fi
\ifx \bfpage  \undefined \def \bfpage#1{#1}\fi
\ifx \blpage  \undefined \def \blpage #1{#1}\fi
\ifx \burl  \undefined \def \burl#1{\textsf{#1}}\fi
\ifx \doiurl  \undefined \def \doiurl#1{\url{https://doi.org/#1}}\fi
\ifx \betal  \undefined \def \betal{\textit{et al.}}\fi
\ifx \binstitute  \undefined \def \binstitute#1{#1}\fi
\ifx \binstitutionaled  \undefined \def \binstitutionaled#1{#1}\fi
\ifx \bctitle  \undefined \def \bctitle#1{#1}\fi
\ifx \beditor  \undefined \def \beditor#1{#1}\fi
\ifx \bpublisher  \undefined \def \bpublisher#1{#1}\fi
\ifx \bbtitle  \undefined \def \bbtitle#1{#1}\fi
\ifx \bedition  \undefined \def \bedition#1{#1}\fi
\ifx \bseriesno  \undefined \def \bseriesno#1{#1}\fi
\ifx \blocation  \undefined \def \blocation#1{#1}\fi
\ifx \bsertitle  \undefined \def \bsertitle#1{#1}\fi
\ifx \bsnm \undefined \def \bsnm#1{#1}\fi
\ifx \bsuffix \undefined \def \bsuffix#1{#1}\fi
\ifx \bparticle \undefined \def \bparticle#1{#1}\fi
\ifx \barticle \undefined \def \barticle#1{#1}\fi
\bibcommenthead
\ifx \bconfdate \undefined \def \bconfdate #1{#1}\fi
\ifx \botherref \undefined \def \botherref #1{#1}\fi
\ifx \url \undefined \def \url#1{\textsf{#1}}\fi
\ifx \bchapter \undefined \def \bchapter#1{#1}\fi
\ifx \bbook \undefined \def \bbook#1{#1}\fi
\ifx \bcomment \undefined \def \bcomment#1{#1}\fi
\ifx \oauthor \undefined \def \oauthor#1{#1}\fi
\ifx \citeauthoryear \undefined \def \citeauthoryear#1{#1}\fi
\ifx \endbibitem  \undefined \def \endbibitem {}\fi
\ifx \bconflocation  \undefined \def \bconflocation#1{#1}\fi
\ifx \arxivurl  \undefined \def \arxivurl#1{\textsf{#1}}\fi
\csname PreBibitemsHook\endcsname

\bibitem{Lumley1967}
\begin{bchapter}
\bauthor{\bsnm{Lumley}, \binits{J.L.}}:
\bctitle{The Structure of Inhomogeneous Turbulent Flows}.
In: \beditor{\bsnm{Yaglam}, \binits{A.M.}},
\beditor{\bsnm{Tatarsky}, \binits{V.I.}} (eds.)
\bbtitle{Proceedings of the International Colloquium on the Fine Scale
  Structure of the Atmosphere and Its Influence on Radio Wave Propagation}.
\bpublisher{Doklady Akademii Nauk},
\blocation{SSSR, Moscow, Nauka}
(\byear{1967})
\end{bchapter}
\endbibitem

\bibitem{lumley1970stochastic}
\begin{botherref}
\oauthor{\bsnm{Lumley}, \binits{J.L.}}:
Stochastic tools in turbulence. volume 12. applied mathematics and mechanics.
Technical report,
Pennsylvania State University Park Dept. of Aerospace Engineering
(1970)
\end{botherref}
\endbibitem

\bibitem{sirovich1987turbulence}
\begin{barticle}
\bauthor{\bsnm{Sirovich}, \binits{L.}}:
\batitle{Turbulence and the dynamics of coherent structures, parts i-iii}.
\bjtitle{Quarterly of Applied Mathematics}
\bvolume{45}(\bissue{3}),
\bfpage{561}--\blpage{582}
(\byear{1987})
\end{barticle}
\endbibitem

\bibitem{holmes2012pod}
\begin{bbook}
\bauthor{\bsnm{Holmes}, \binits{P.}},
\bauthor{\bsnm{Lumley}, \binits{J.L.}},
\bauthor{\bsnm{Berkooz}, \binits{G.}},
\bauthor{\bsnm{Rowley}, \binits{C.W.}}:
\bbtitle{Turbulence, Coherent Structures, Dynamical Systems and Symmetry}.
\bpublisher{Cambridge University Press},
\blocation{Cambridge, England}
(\byear{2012})
\end{bbook}
\endbibitem

\bibitem{rempfer2000low}
\begin{barticle}
\bauthor{\bsnm{Rempfer}, \binits{D.}}:
\batitle{On low-dimensional galerkin models for fluid flow}.
\bjtitle{Theoretical and Computational Fluid Dynamics}
\bvolume{14}(\bissue{2}),
\bfpage{75}--\blpage{88}
(\byear{2000})
\end{barticle}
\endbibitem

\bibitem{aubry1988turbbl}
\begin{barticle}
\bauthor{\bsnm{Aubry}, \binits{N.}},
\bauthor{\bsnm{Holmes}, \binits{P.}},
\bauthor{\bsnm{Lumley}, \binits{J.L.}},
\bauthor{\bsnm{Stone}, \binits{E.}}:
\batitle{The dynamics of coherent structures in the wall region of a turbulent
  boundary layer}.
\bjtitle{Journal of Fluid Mechanics}
\bvolume{192},
\bfpage{115}--\blpage{173}
(\byear{1988})
\end{barticle}
\endbibitem

\bibitem{rempfer1994bl}
\begin{barticle}
\bauthor{\bsnm{Rempfer}, \binits{D.}},
\bauthor{\bsnm{Fasel}, \binits{H.F.}}:
\batitle{Dynamics of three-dimensional coherent structures in a flat-plate
  boundary layer}.
\bjtitle{Journal of Fluid Mechanics}
\bvolume{275},
\bfpage{257}--\blpage{283}
(\byear{1994})
\end{barticle}
\endbibitem

\bibitem{moehlis2002couette}
\begin{barticle}
\bauthor{\bsnm{Moehlis}, \binits{J.}},
\bauthor{\bsnm{Smith}, \binits{T.}},
\bauthor{\bsnm{Holmes}, \binits{P.}},
\bauthor{\bsnm{Faisst}, \binits{H.}}:
\batitle{Models for turbulent plane couette flow using the proper orthogonal
  decomposition}.
\bjtitle{Physics of Fluids (1994-present)}
\bvolume{14}(\bissue{7}),
\bfpage{2493}--\blpage{2507}
(\byear{2002})
\end{barticle}
\endbibitem

\bibitem{smith2005pod}
\begin{barticle}
\bauthor{\bsnm{Smith}, \binits{T.R.}},
\bauthor{\bsnm{Moehlis}, \binits{J.}},
\bauthor{\bsnm{Holmes}, \binits{P.}}:
\batitle{Low-dimensional modelling of turbulence using the proper orthogonal
  decomposition: a tutorial}.
\bjtitle{Nonlinear Dynamics}
\bvolume{41}(\bissue{1-3}),
\bfpage{275}--\blpage{307}
(\byear{2005})
\end{barticle}
\endbibitem

\bibitem{borggaard2008galerkinpipe}
\begin{bchapter}
\bauthor{\bsnm{Borggaard}, \binits{J.}},
\bauthor{\bsnm{Duggleby}, \binits{A.}},
\bauthor{\bsnm{Hay}, \binits{A.}},
\bauthor{\bsnm{Iliescu}, \binits{T.}},
\bauthor{\bsnm{Wang}, \binits{Z.}}:
\bctitle{Reduced-order modeling of turbulent flows}.
In: \bbtitle{Proceedings of MTNS}
(\byear{2008})
\end{bchapter}
\endbibitem

\bibitem{podvin2009galerkin}
\begin{barticle}
\bauthor{\bsnm{Podvin}, \binits{B.}}:
\batitle{A proper-orthogonal-decomposition--based model for the wall layer of a
  turbulent channel flow}.
\bjtitle{Physics of Fluids (1994-present)}
\bvolume{21}(\bissue{1}),
\bfpage{015111}
(\byear{2009})
\end{barticle}
\endbibitem

\bibitem{deane1991galerkin}
\begin{barticle}
\bauthor{\bsnm{Deane}, \binits{A.}},
\bauthor{\bsnm{Kevrekidis}, \binits{I.}},
\bauthor{\bsnm{Karniadakis}, \binits{G.E.}},
\bauthor{\bsnm{Orszag}, \binits{S.}}:
\batitle{Low-dimensional models for complex geometry flows: Application to
  grooved channels and circular cylinders}.
\bjtitle{Physics of Fluids A: Fluid Dynamics (1989-1993)}
\bvolume{3}(\bissue{10}),
\bfpage{2337}--\blpage{2354}
(\byear{1991})
\end{barticle}
\endbibitem

\bibitem{noack1994galerkin}
\begin{barticle}
\bauthor{\bsnm{Noack}, \binits{B.R.}},
\bauthor{\bsnm{Eckelmann}, \binits{H.}}:
\batitle{A global stability analysis of the steady and periodic cylinder wake}.
\bjtitle{Journal of Fluid Mechanics}
\bvolume{270},
\bfpage{297}--\blpage{330}
(\byear{1994})
\end{barticle}
\endbibitem

\bibitem{noack:03cyl}
\begin{barticle}
\bauthor{\bsnm{Noack}, \binits{B.R.}},
\bauthor{\bsnm{Afanasiev}, \binits{K.}},
\bauthor{\bsnm{Morzynski}, \binits{M.}},
\bauthor{\bsnm{Tadmor}, \binits{G.}},
\bauthor{\bsnm{Thiele}, \binits{F.}}:
\batitle{A hierarchy of low-dimensional models for the transient and
  post-transient cylinder wake}.
\bjtitle{Journal of Fluid Mechanics}
\bvolume{497},
\bfpage{335}--\blpage{363}
(\byear{2003})
\end{barticle}
\endbibitem

\bibitem{rowley2001cavity}
\begin{barticle}
\bauthor{\bsnm{Rowley}, \binits{C.W.}},
\bauthor{\bsnm{Colonius}, \binits{T.}},
\bauthor{\bsnm{Murray}, \binits{R.M.}}:
\batitle{Dynamical models for control of cavity oscillations}.
\bjtitle{AIAA paper}
\bvolume{2126}(\bissue{2001}),
\bfpage{2126}--\blpage{34}
(\byear{2001})
\end{barticle}
\endbibitem

\bibitem{rowley2006cavity}
\begin{barticle}
\bauthor{\bsnm{Rowley}, \binits{C.W.}},
\bauthor{\bsnm{Williams}, \binits{D.R.}}:
\batitle{Dynamics and control of high-reynolds-number flow over open cavities}.
\bjtitle{Annu. Rev. Fluid Mech.}
\bvolume{38},
\bfpage{251}--\blpage{276}
(\byear{2006})
\end{barticle}
\endbibitem

\bibitem{rajaee1994mixing}
\begin{barticle}
\bauthor{\bsnm{Rajaee}, \binits{M.}},
\bauthor{\bsnm{Karlsson}, \binits{S.K.}},
\bauthor{\bsnm{Sirovich}, \binits{L.}}:
\batitle{Low-dimensional description of free-shear-flow coherent structures and
  their dynamical behaviour}.
\bjtitle{Journal of Fluid Mechanics}
\bvolume{258},
\bfpage{1}--\blpage{29}
(\byear{1994})
\end{barticle}
\endbibitem

\bibitem{ukeiley2001mixing}
\begin{barticle}
\bauthor{\bsnm{Ukeiley}, \binits{L.}},
\bauthor{\bsnm{Cordier}, \binits{L.}},
\bauthor{\bsnm{Manceau}, \binits{R.}},
\bauthor{\bsnm{Delville}, \binits{J.}},
\bauthor{\bsnm{Glauser}, \binits{M.}},
\bauthor{\bsnm{Bonnet}, \binits{J.}}:
\batitle{Examination of large-scale structures in a turbulent plane mixing
  layer. part 2. dynamical systems model}.
\bjtitle{Journal of Fluid Mechanics}
\bvolume{441},
\bfpage{67}--\blpage{108}
(\byear{2001})
\end{barticle}
\endbibitem

\bibitem{balajewicz2013jfm}
\begin{barticle}
\bauthor{\bsnm{Balajewicz}, \binits{M.J.}},
\bauthor{\bsnm{Dowell}, \binits{E.H.}},
\bauthor{\bsnm{Noack}, \binits{B.R.}}:
\batitle{Low-dimensional modelling of high-reynolds-number shear flows
  incorporating constraints from the {Navier--Stokes} equation}.
\bjtitle{Journal of Fluid Mechanics}
\bvolume{729},
\bfpage{285}--\blpage{308}
(\byear{2013})
\end{barticle}
\endbibitem

\bibitem{osth2014podvisc}
\begin{barticle}
\bauthor{\bsnm{{\"O}sth}, \binits{J.}},
\bauthor{\bsnm{Noack}, \binits{B.R.}},
\bauthor{\bsnm{Krajnovi{\'c}}, \binits{S.}},
\bauthor{\bsnm{Barros}, \binits{D.}},
\bauthor{\bsnm{Bor{\'e}e}, \binits{J.}}:
\batitle{On the need for a nonlinear subscale turbulence term in {POD} models
  as exemplified for a high-reynolds-number flow over an ahmed body}.
\bjtitle{Journal of Fluid Mechanics}
\bvolume{747},
\bfpage{518}--\blpage{544}
(\byear{2014})
\end{barticle}
\endbibitem

\bibitem{wang2012podclosure}
\begin{barticle}
\bauthor{\bsnm{Wang}, \binits{Z.}},
\bauthor{\bsnm{Akhtar}, \binits{I.}},
\bauthor{\bsnm{Borggaard}, \binits{J.}},
\bauthor{\bsnm{Iliescu}, \binits{T.}}:
\batitle{Proper orthogonal decomposition closure models for turbulent flows: a
  numerical comparison}.
\bjtitle{Computer Methods in Applied Mechanics and Engineering}
\bvolume{237},
\bfpage{10}--\blpage{26}
(\byear{2012})
\end{barticle}
\endbibitem

\bibitem{noack2008finite}
\begin{barticle}
\bauthor{\bsnm{Noack}, \binits{B.R.}},
\bauthor{\bsnm{Schlegel}, \binits{M.}},
\bauthor{\bsnm{Ahlborn}, \binits{B.}},
\bauthor{\bsnm{Mutschke}, \binits{G.}},
\bauthor{\bsnm{Morzy{\'n}ski}, \binits{M.}},
\bauthor{\bsnm{Comte}, \binits{P.}},
\bauthor{\bsnm{Tadmor}, \binits{G.}}:
\batitle{A finite-time thermodynamics of unsteady fluid flows}.
\bjtitle{Journal of Non Equilibrium Thermodynamics}
\bvolume{33}(\bissue{2}),
\bfpage{103}--\blpage{148}
(\byear{2008})
\end{barticle}
\endbibitem

\bibitem{callaham2022multiscale}
\begin{botherref}
\oauthor{\bsnm{Callaham}, \binits{J.L.}},
\oauthor{\bsnm{Loiseau}, \binits{J.-C.}},
\oauthor{\bsnm{Brunton}, \binits{S.L.}}:
Multiscale model reduction for incompressible flows.
arXiv preprint arXiv:2206.13205
(2022)
\end{botherref}
\endbibitem

\bibitem{balajewicz2015minimal}
\begin{botherref}
\oauthor{\bsnm{Balajewicz}, \binits{M.}},
\oauthor{\bsnm{Tezaur}, \binits{I.}},
\oauthor{\bsnm{Dowell}, \binits{E.}}:
Minimal subspace rotation on the {Stiefel} manifold for stabilization and
  enhancement of projection-based reduced order models for the compressible
  {N}avier-{S}tokes equations.
arXiv preprint arXiv:1504.06661
(2015)
\end{botherref}
\endbibitem

\bibitem{cordier2013identification}
\begin{barticle}
\bauthor{\bsnm{Cordier}, \binits{L.}},
\bauthor{\bsnm{Noack}, \binits{B.R.}},
\bauthor{\bsnm{Tissot}, \binits{G.}},
\bauthor{\bsnm{Lehnasch}, \binits{G.}},
\bauthor{\bsnm{Delville}, \binits{J.}},
\bauthor{\bsnm{Balajewicz}, \binits{M.}},
\bauthor{\bsnm{Daviller}, \binits{G.}},
\bauthor{\bsnm{Niven}, \binits{R.K.}}:
\batitle{Identification strategies for model-based control}.
\bjtitle{Experiments in fluids}
\bvolume{54}(\bissue{8}),
\bfpage{1}--\blpage{21}
(\byear{2013})
\end{barticle}
\endbibitem

\bibitem{carlberg2011efficient}
\begin{barticle}
\bauthor{\bsnm{Carlberg}, \binits{K.}},
\bauthor{\bsnm{Bou-Mosleh}, \binits{C.}},
\bauthor{\bsnm{Farhat}, \binits{C.}}:
\batitle{Efficient non-linear model reduction via a least-squares
  petrov--galerkin projection and compressive tensor approximations}.
\bjtitle{International Journal for numerical methods in engineering}
\bvolume{86}(\bissue{2}),
\bfpage{155}--\blpage{181}
(\byear{2011})
\end{barticle}
\endbibitem

\bibitem{carlberg2013gnat}
\begin{barticle}
\bauthor{\bsnm{Carlberg}, \binits{K.}},
\bauthor{\bsnm{Farhat}, \binits{C.}},
\bauthor{\bsnm{Cortial}, \binits{J.}},
\bauthor{\bsnm{Amsallem}, \binits{D.}}:
\batitle{The gnat method for nonlinear model reduction: effective
  implementation and application to computational fluid dynamics and turbulent
  flows}.
\bjtitle{Journal of Computational Physics}
\bvolume{242},
\bfpage{623}--\blpage{647}
(\byear{2013})
\end{barticle}
\endbibitem

\bibitem{willcox:2002}
\begin{barticle}
\bauthor{\bsnm{Willcox}, \binits{K.}},
\bauthor{\bsnm{Peraire}, \binits{J.}}:
\batitle{Balanced model reduction via the proper orthogonal decomposition}.
\bjtitle{AIAA Journal}
\bvolume{40}(\bissue{11}),
\bfpage{2323}--\blpage{2330}
(\byear{2002})
\end{barticle}
\endbibitem

\bibitem{rowley:05pod}
\begin{barticle}
\bauthor{\bsnm{Rowley}, \binits{C.W.}}:
\batitle{Model reduction for fluids using balanced proper orthogonal
  decomposition.}
\bjtitle{International Journal of Bifurcation and Chaos}
\bvolume{15}(\bissue{3}),
\bfpage{997}--\blpage{1013}
(\byear{2005})
\end{barticle}
\endbibitem

\bibitem{glauser1987coherent}
\begin{bchapter}
\bauthor{\bsnm{Glauser}, \binits{M.N.}},
\bauthor{\bsnm{Leib}, \binits{S.J.}},
\bauthor{\bsnm{George}, \binits{W.K.}}:
\bctitle{Coherent structures in the axisymmetric turbulent jet mixing layer}.
In: \bbtitle{Turbulent Shear Flows 5},
pp. \bfpage{134}--\blpage{145}.
\bpublisher{Springer},
\blocation{Berlin, Heidelberg}
(\byear{1987})
\end{bchapter}
\endbibitem

\bibitem{picard2000pressure}
\begin{barticle}
\bauthor{\bsnm{Picard}, \binits{C.}},
\bauthor{\bsnm{Delville}, \binits{J.}}:
\batitle{Pressure velocity coupling in a subsonic round jet}.
\bjtitle{International Journal of Heat and Fluid Flow}
\bvolume{21}(\bissue{3}),
\bfpage{359}--\blpage{364}
(\byear{2000})
\end{barticle}
\endbibitem

\bibitem{towne2018spectral}
\begin{barticle}
\bauthor{\bsnm{Towne}, \binits{A.}},
\bauthor{\bsnm{Schmidt}, \binits{O.T.}},
\bauthor{\bsnm{Colonius}, \binits{T.}}:
\batitle{Spectral proper orthogonal decomposition and its relationship to
  dynamic mode decomposition and resolvent analysis}.
\bjtitle{Journal of Fluid Mechanics}
\bvolume{847},
\bfpage{821}--\blpage{867}
(\byear{2018})
\end{barticle}
\endbibitem

\bibitem{schmidt2020guide}
\begin{barticle}
\bauthor{\bsnm{Schmidt}, \binits{O.T.}},
\bauthor{\bsnm{Colonius}, \binits{T.}}:
\batitle{Guide to spectral proper orthogonal decomposition}.
\bjtitle{AIAA Journal}
\bvolume{58}(\bissue{3}),
\bfpage{1023}--\blpage{1033}
(\byear{2020})
\end{barticle}
\endbibitem

\bibitem{schmid2008}
\begin{bchapter}
\bauthor{\bsnm{Schmid}, \binits{P.J.}},
\bauthor{\bsnm{Sesterhenn}, \binits{J.}}:
\bctitle{Dynamic Mode Decomposition of Numerical and Experimental Data}.
In: \bbtitle{61st Annual Meeting of the APS Division of Fluid Dynamics}.
\bpublisher{American Physical Society},
\blocation{San Antonio, TX}
(\byear{2008})
\end{bchapter}
\endbibitem

\bibitem{schmid2010dynamic}
\begin{barticle}
\bauthor{\bsnm{Schmid}, \binits{P.J.}}:
\batitle{Dynamic mode decomposition of numerical and experimental data}.
\bjtitle{Journal of Fluid Mechanics}
\bvolume{656},
\bfpage{5}--\blpage{28}
(\byear{2010})
\end{barticle}
\endbibitem

\bibitem{rowley2009spectral}
\begin{barticle}
\bauthor{\bsnm{Rowley}, \binits{C.W.}},
\bauthor{\bsnm{Mezi{\'c}}, \binits{I.}},
\bauthor{\bsnm{Bagheri}, \binits{S.}},
\bauthor{\bsnm{Schlatter}, \binits{P.}},
\bauthor{\bsnm{Henningson}, \binits{D.S.}}:
\batitle{Spectral analysis of nonlinear flows}.
\bjtitle{Journal of Fluid Mechanics}
\bvolume{641}(\bissue{1}),
\bfpage{115}--\blpage{127}
(\byear{2009})
\end{barticle}
\endbibitem

\bibitem{chen2012variants}
\begin{barticle}
\bauthor{\bsnm{Chen}, \binits{K.K.}},
\bauthor{\bsnm{Tu}, \binits{J.H.}},
\bauthor{\bsnm{Rowley}, \binits{C.W.}}:
\batitle{Variants of dynamic mode decomposition: boundary condition, {Koopman},
  and {Fourier} analyses}.
\bjtitle{Journal of Nonlinear Science}
\bvolume{22}(\bissue{6}),
\bfpage{887}--\blpage{915}
(\byear{2012})
\end{barticle}
\endbibitem

\bibitem{gueniat2015dynamic}
\begin{barticle}
\bauthor{\bsnm{Gu{\'e}niat}, \binits{F.}},
\bauthor{\bsnm{Mathelin}, \binits{L.}},
\bauthor{\bsnm{Pastur}, \binits{L.R.}}:
\batitle{A dynamic mode decomposition approach for large and arbitrarily
  sampled systems}.
\bjtitle{Physics of Fluids}
\bvolume{27}(\bissue{2}),
\bfpage{025113}
(\byear{2015})
\end{barticle}
\endbibitem

\bibitem{leroux2016dynamic}
\begin{barticle}
\bauthor{\bsnm{Leroux}, \binits{R.}},
\bauthor{\bsnm{Cordier}, \binits{L.}}:
\batitle{Dynamic mode decomposition for non-uniformly sampled data}.
\bjtitle{Experiments in Fluids}
\bvolume{57}(\bissue{5}),
\bfpage{94}
(\byear{2016})
\end{barticle}
\endbibitem

\bibitem{askham2018variable}
\begin{barticle}
\bauthor{\bsnm{Askham}, \binits{T.}},
\bauthor{\bsnm{Kutz}, \binits{J.N.}}:
\batitle{Variable projection methods for an optimized dynamic mode
  decomposition}.
\bjtitle{SIAM Journal on Applied Dynamical Systems}
\bvolume{17}(\bissue{1}),
\bfpage{380}--\blpage{416}
(\byear{2018})
\end{barticle}
\endbibitem

\bibitem{donoho2006compressed}
\begin{barticle}
\bauthor{\bsnm{Donoho}, \binits{D.L.}}:
\batitle{Compressed sensing}.
\bjtitle{IEEE Transactions on Information Theory}
\bvolume{52}(\bissue{4}),
\bfpage{1289}--\blpage{1306}
(\byear{2006})
\end{barticle}
\endbibitem

\bibitem{tu2014spectral}
\begin{barticle}
\bauthor{\bsnm{Tu}, \binits{J.H.}},
\bauthor{\bsnm{Rowley}, \binits{C.W.}},
\bauthor{\bsnm{Kutz}, \binits{J.N.}},
\bauthor{\bsnm{Shang}, \binits{J.K.}}:
\batitle{Spectral analysis of fluid flows using sub-{N}yquist-rate {PIV} data}.
\bjtitle{Experiments in Fluids}
\bvolume{55}(\bissue{9}),
\bfpage{1}--\blpage{13}
(\byear{2014})
\end{barticle}
\endbibitem

\bibitem{nyquist1928certain}
\begin{barticle}
\bauthor{\bsnm{Nyquist}, \binits{H.}}:
\batitle{Certain topics in telegraph transmission theory}.
\bjtitle{Transactions of the American Institute of Electrical Engineers}
\bvolume{47}(\bissue{2}),
\bfpage{617}--\blpage{644}
(\byear{1928})
\end{barticle}
\endbibitem

\bibitem{shannon1949communication}
\begin{barticle}
\bauthor{\bsnm{Shannon}, \binits{C.E.}}:
\batitle{Communication in the presence of noise}.
\bjtitle{Proceedings of the IRE}
\bvolume{37}(\bissue{1}),
\bfpage{10}--\blpage{21}
(\byear{1949})
\end{barticle}
\endbibitem

\bibitem{tu2013integration}
\begin{barticle}
\bauthor{\bsnm{Tu}, \binits{J.H.}},
\bauthor{\bsnm{Griffin}, \binits{J.}},
\bauthor{\bsnm{Hart}, \binits{A.}},
\bauthor{\bsnm{Rowley}, \binits{C.W.}},
\bauthor{\bsnm{Cattafesta~III}, \binits{L.N.}},
\bauthor{\bsnm{Ukeiley}, \binits{L.S.}}:
\batitle{Integration of non-time-resolved piv and time-resolved velocity point
  sensors for dynamic estimation of velocity fields}.
\bjtitle{Experiments in Fluids}
\bvolume{54}(\bissue{2}),
\bfpage{1}--\blpage{20}
(\byear{2013})
\end{barticle}
\endbibitem

\bibitem{zhang2020spectral}
\begin{barticle}
\bauthor{\bsnm{Zhang}, \binits{Y.}},
\bauthor{\bsnm{Cattafesta}, \binits{L.N.}},
\bauthor{\bsnm{Ukeiley}, \binits{L.}}:
\batitle{Spectral analysis modal methods (samms) using non-time-resolved piv}.
\bjtitle{Experiments in Fluids}
\bvolume{61}(\bissue{11}),
\bfpage{1}--\blpage{12}
(\byear{2020})
\end{barticle}
\endbibitem

\bibitem{tinney2006spectral}
\begin{barticle}
\bauthor{\bsnm{Tinney}, \binits{C.}},
\bauthor{\bsnm{Coiffet}, \binits{F.}},
\bauthor{\bsnm{Delville}, \binits{J.}},
\bauthor{\bsnm{Hall}, \binits{A.}},
\bauthor{\bsnm{Jordan}, \binits{P.}},
\bauthor{\bsnm{Glauser}, \binits{M.}}:
\batitle{On spectral linear stochastic estimation}.
\bjtitle{Experiments in Fluids}
\bvolume{41}(\bissue{5}),
\bfpage{763}--\blpage{775}
(\byear{2006})
\end{barticle}
\endbibitem

\bibitem{krishna2020reconstructing}
\begin{barticle}
\bauthor{\bsnm{Krishna}, \binits{C.V.}},
\bauthor{\bsnm{Wang}, \binits{M.}},
\bauthor{\bsnm{Hemati}, \binits{M.S.}},
\bauthor{\bsnm{Luhar}, \binits{M.}}:
\batitle{Reconstructing the time evolution of wall-bounded turbulent flows from
  non-time-resolved {PIV} measurements}.
\bjtitle{Physical Review Fluids}
\bvolume{5}(\bissue{5}),
\bfpage{054604}
(\byear{2020})
\end{barticle}
\endbibitem

\bibitem{wang2021model}
\begin{barticle}
\bauthor{\bsnm{Wang}, \binits{M.}},
\bauthor{\bsnm{Krishna}, \binits{C.V.}},
\bauthor{\bsnm{Luhar}, \binits{M.}},
\bauthor{\bsnm{Hemati}, \binits{M.S.}}:
\batitle{Model-based multi-sensor fusion for reconstructing wall-bounded
  turbulence}.
\bjtitle{Theoretical and Computational Fluid Dynamics}
\bvolume{35}(\bissue{5}),
\bfpage{683}--\blpage{707}
(\byear{2021})
\end{barticle}
\endbibitem

\bibitem{chu2021stochastic}
\begin{barticle}
\bauthor{\bsnm{Chu}, \binits{T.}},
\bauthor{\bsnm{Schmidt}, \binits{O.T.}}:
\batitle{A stochastic spod-galerkin model for broadband turbulent flows}.
\bjtitle{Theoretical and Computational Fluid Dynamics}
\bvolume{35}(\bissue{6}),
\bfpage{759}--\blpage{782}
(\byear{2021})
\end{barticle}
\endbibitem

\bibitem{towne2021space}
\begin{bchapter}
\bauthor{\bsnm{Towne}, \binits{A.}}:
\bctitle{Space-time galerkin projection via spectral proper orthogonal
  decomposition and resolvent modes}.
In: \bbtitle{AIAA Scitech 2021 Forum},
p. \bfpage{1676}
(\byear{2021})
\end{bchapter}
\endbibitem

\bibitem{sohankar2006flow}
\begin{barticle}
\bauthor{\bsnm{Sohankar}, \binits{A.}}:
\batitle{Flow over a bluff body from moderate to high {R}eynolds numbers using
  large eddy simulation}.
\bjtitle{Computers \& Fluids}
\bvolume{35}(\bissue{10}),
\bfpage{1154}--\blpage{1168}
(\byear{2006})
\end{barticle}
\endbibitem

\bibitem{gao2008airborne}
\begin{barticle}
\bauthor{\bsnm{Gao}, \binits{N.}},
\bauthor{\bsnm{Niu}, \binits{J.}},
\bauthor{\bsnm{Perino}, \binits{M.}},
\bauthor{\bsnm{Heiselberg}, \binits{P.}}:
\batitle{The airborne transmission of infection between flats in high-rise
  residential buildings: tracer gas simulation}.
\bjtitle{Building and Environment}
\bvolume{43}(\bissue{11}),
\bfpage{1805}--\blpage{1817}
(\byear{2008})
\end{barticle}
\endbibitem

\bibitem{williamson1985evolution}
\begin{barticle}
\bauthor{\bsnm{Williamson}, \binits{C.}}:
\batitle{Evolution of a single wake behind a pair of bluff bodies}.
\bjtitle{Journal of Fluid Mechanics}
\bvolume{159},
\bfpage{1}--\blpage{18}
(\byear{1985})
\end{barticle}
\endbibitem

\bibitem{supradeepananalysis}
\begin{botherref}
\oauthor{\bsnm{Supradeepan}, \binits{C.}},
\oauthor{\bsnm{Roy}, \binits{A.}}:
Analysis of flow over two side by side cylinders for different gaps at low
  {R}eynolds number: a numerical approach.
Physics of Fluids
\textbf{26}(6)
(2014)
\end{botherref}
\endbibitem

\bibitem{bai2016flip}
\begin{barticle}
\bauthor{\bsnm{Bai}, \binits{X.-D.}},
\bauthor{\bsnm{Zhang}, \binits{W.}},
\bauthor{\bsnm{Guo}, \binits{A.-X.}},
\bauthor{\bsnm{Wang}, \binits{Y.}}:
\batitle{The flip-flopping wake pattern behind two side-by-side circular
  cylinders: A global stability analysis}.
\bjtitle{Physics of Fluids}
\bvolume{28}(\bissue{4}),
\bfpage{044102}
(\byear{2016})
\end{barticle}
\endbibitem

\bibitem{alam2003aerodynamic}
\begin{barticle}
\bauthor{\bsnm{Alam}, \binits{M.M.}},
\bauthor{\bsnm{Moriya}, \binits{M.}},
\bauthor{\bsnm{Sakamoto}, \binits{H.}}:
\batitle{Aerodynamic characteristics of two side-by-side circular cylinders and
  application of wavelet analysis on the switching phenomenon}.
\bjtitle{Journal of Fluids and Structures}
\bvolume{18}(\bissue{3-4}),
\bfpage{325}--\blpage{346}
(\byear{2003})
\end{barticle}
\endbibitem

\bibitem{kang2003characteristics}
\begin{barticle}
\bauthor{\bsnm{Kang}, \binits{S.}}:
\batitle{Characteristics of flow over two circular cylinders in a side-by-side
  arrangement at low {R}eynolds numbers}.
\bjtitle{Physics of Fluids}
\bvolume{15}(\bissue{9}),
\bfpage{2486}--\blpage{2498}
(\byear{2003})
\end{barticle}
\endbibitem

\bibitem{zhou2016wake}
\begin{barticle}
\bauthor{\bsnm{Zhou}, \binits{Y.}},
\bauthor{\bsnm{Alam}, \binits{M.M.}}:
\batitle{Wake of two interacting circular cylinders: a review}.
\bjtitle{International Journal of Heat and Fluid Flow}
\bvolume{62},
\bfpage{510}--\blpage{537}
(\byear{2016})
\end{barticle}
\endbibitem

\bibitem{guillaume2000investigation}
\begin{barticle}
\bauthor{\bsnm{Guillaume}, \binits{D.}},
\bauthor{\bsnm{LaRue}, \binits{J.}}:
\batitle{Investigation of the flopping regime of two-, three-, and four-plate
  arrays}.
\bjtitle{Journal of Fluids Engineering}
\bvolume{122}(\bissue{4}),
\bfpage{677}--\blpage{682}
(\byear{2000})
\end{barticle}
\endbibitem

\bibitem{miau1996flopping}
\begin{barticle}
\bauthor{\bsnm{Miau}, \binits{J.-J.}},
\bauthor{\bsnm{Wang}, \binits{H.}},
\bauthor{\bsnm{Chou}, \binits{J.}}:
\batitle{Flopping phenomenon of flow behind two plates placed side-by-side
  normal to the flow direction}.
\bjtitle{Fluid dynamics research}
\bvolume{17}(\bissue{6}),
\bfpage{311}
(\byear{1996})
\end{barticle}
\endbibitem

\bibitem{deng2020low}
\begin{botherref}
\oauthor{\bsnm{Deng}, \binits{N.}},
\oauthor{\bsnm{Noack}, \binits{B.R.}},
\oauthor{\bsnm{Morzy{\'n}ski}, \binits{M.}},
\oauthor{\bsnm{Pastur}, \binits{L.R.}}:
Low-order model for successive bifurcations of the fluidic pinball.
Journal of fluid mechanics
\textbf{884}
(2020)
\end{botherref}
\endbibitem

\bibitem{deng2022cluster}
\begin{botherref}
\oauthor{\bsnm{Deng}, \binits{N.}},
\oauthor{\bsnm{Noack}, \binits{B.R.}},
\oauthor{\bsnm{Morzy{\'n}ski}, \binits{M.}},
\oauthor{\bsnm{Pastur}, \binits{L.R.}}:
Cluster-based hierarchical network model of the fluidic pinball--cartographing
  transient and post-transient, multi-frequency, multi-attractor behaviour.
Journal of Fluid Mechanics
\textbf{934}
(2022)
\end{botherref}
\endbibitem

\bibitem{maceda2021stabilization}
\begin{botherref}
\oauthor{\bsnm{Maceda}, \binits{G.Y.C.}},
\oauthor{\bsnm{Li}, \binits{Y.}},
\oauthor{\bsnm{Lusseyran}, \binits{F.}},
\oauthor{\bsnm{Morzy{\'n}ski}, \binits{M.}},
\oauthor{\bsnm{Noack}, \binits{B.R.}}:
Stabilization of the fluidic pinball with gradient-enriched machine learning
  control.
Journal of Fluid Mechanics
\textbf{917}
(2021)
\end{botherref}
\endbibitem

\bibitem{schmidt2019efficient}
\begin{barticle}
\bauthor{\bsnm{Schmidt}, \binits{O.T.}},
\bauthor{\bsnm{Towne}, \binits{A.}}:
\batitle{An efficient streaming algorithm for spectral proper orthogonal
  decomposition}.
\bjtitle{Computer Physics Communications}
\bvolume{237},
\bfpage{98}--\blpage{109}
(\byear{2019})
\end{barticle}
\endbibitem

\bibitem{welch1967use}
\begin{barticle}
\bauthor{\bsnm{Welch}, \binits{P.}}:
\batitle{The use of fast fourier transform for the estimation of power spectra:
  a method based on time averaging over short, modified periodograms}.
\bjtitle{IEEE Transactions on audio and electroacoustics}
\bvolume{15}(\bissue{2}),
\bfpage{70}--\blpage{73}
(\byear{1967})
\end{barticle}
\endbibitem

\bibitem{berkooz1993pod}
\begin{barticle}
\bauthor{\bsnm{Berkooz}, \binits{G.}},
\bauthor{\bsnm{Holmes}, \binits{P.}},
\bauthor{\bsnm{Lumley}, \binits{J.L.}}:
\batitle{The proper orthogonal decomposition in the analysis of turbulent
  flows}.
\bjtitle{Annual Review of Fluid Mechanics}
\bvolume{25}(\bissue{1}),
\bfpage{539}--\blpage{575}
(\byear{1993})
\end{barticle}
\endbibitem

\bibitem{moore1981ieeetac}
\begin{barticle}
\bauthor{\bsnm{Moore}, \binits{B.}}:
\batitle{Principal component analysis in linear systems: Controllability,
  observability, and model reduction}.
\bjtitle{IEEE transactions on automatic control}
\bvolume{26}(\bissue{1}),
\bfpage{17}--\blpage{32}
(\byear{1981})
\end{barticle}
\endbibitem

\bibitem{taira2007immersed}
\begin{barticle}
\bauthor{\bsnm{Taira}, \binits{K.}},
\bauthor{\bsnm{Colonius}, \binits{T.}}:
\batitle{The immersed boundary method: a projection approach}.
\bjtitle{Journal of Computational Physics}
\bvolume{225}(\bissue{2}),
\bfpage{2118}--\blpage{2137}
(\byear{2007})
\end{barticle}
\endbibitem

\bibitem{colonius2008fast}
\begin{barticle}
\bauthor{\bsnm{Colonius}, \binits{T.}},
\bauthor{\bsnm{Taira}, \binits{K.}}:
\batitle{A fast immersed boundary method using a nullspace approach and
  multi-domain far-field boundary conditions}.
\bjtitle{Computer Methods in Applied Mechanics and Engineering}
\bvolume{197}(\bissue{25-28}),
\bfpage{2131}--\blpage{2146}
(\byear{2008})
\end{barticle}
\endbibitem

\bibitem{chorin1968numerical}
\begin{barticle}
\bauthor{\bsnm{Chorin}, \binits{A.J.}}:
\batitle{Numerical solution of the navier-stokes equations}.
\bjtitle{Mathematics of computation}
\bvolume{22}(\bissue{104}),
\bfpage{745}--\blpage{762}
(\byear{1968})
\end{barticle}
\endbibitem

\bibitem{temam1969approximation}
\begin{barticle}
\bauthor{\bsnm{Temam}, \binits{R.}}:
\batitle{Sur l'approximation de la solution des {\'e}quations de navier-stokes
  par la m{\'e}thode des pas fractionnaires (ii)}.
\bjtitle{Archive for rational mechanics and analysis}
\bvolume{33},
\bfpage{377}--\blpage{385}
(\byear{1969})
\end{barticle}
\endbibitem

\bibitem{brunton2013reduced}
\begin{barticle}
\bauthor{\bsnm{Brunton}, \binits{S.L.}},
\bauthor{\bsnm{Rowley}, \binits{C.W.}},
\bauthor{\bsnm{Williams}, \binits{D.R.}}:
\batitle{Reduced-order unsteady aerodynamic models at low reynolds numbers}.
\bjtitle{Journal of Fluid Mechanics}
\bvolume{724},
\bfpage{203}--\blpage{233}
(\byear{2013})
\end{barticle}
\endbibitem

\bibitem{brunton2014state}
\begin{barticle}
\bauthor{\bsnm{Brunton}, \binits{S.L.}},
\bauthor{\bsnm{Dawson}, \binits{S.T.M.}},
\bauthor{\bsnm{Rowley}, \binits{C.W.}}:
\batitle{State-space model identification and feedback control of unsteady
  aerodynamic forces}.
\bjtitle{Journal of Fluids and Structures}
\bvolume{50},
\bfpage{253}--\blpage{270}
(\byear{2014})
\end{barticle}
\endbibitem

\bibitem{dawson2016lift}
\begin{bchapter}
\bauthor{\bsnm{Dawson}, \binits{S.T.M.}},
\bauthor{\bsnm{Hemati}, \binits{M.}},
\bauthor{\bsnm{Floryan}, \binits{D.C.}},
\bauthor{\bsnm{Rowley}, \binits{C.W.}}:
\bctitle{Lift enhancement of high angle of attack airfoils using periodic
  pitching}.
In: \bbtitle{54th AIAA Aerospace Sciences Meeting},
p. \bfpage{2069}
(\byear{2016})
\end{bchapter}
\endbibitem

\bibitem{dawson2017reduced}
\begin{botherref}
\oauthor{\bsnm{Dawson}, \binits{S.T.M.}}:
Reduced-order modeling of fluids systems, with applications in unsteady
  aerodynamics.
PhD thesis,
Princeton University
(2017)
\end{botherref}
\endbibitem

\bibitem{almashjary2021reduced}
\begin{botherref}
\oauthor{\bsnm{Almashjary}, \binits{A.N.}}:
Reduced-order modeling of unsteady flow over two collinear plates at low
  {R}eynolds numbers.
Master's thesis,
Illinois Institute of Technology
(2021)
\end{botherref}
\endbibitem

\bibitem{schmidt2022spectral}
\begin{barticle}
\bauthor{\bsnm{Schmidt}, \binits{O.T.}}:
\batitle{Spectral proper orthogonal decomposition using multitaper estimates}.
\bjtitle{Theoretical and Computational Fluid Dynamics}
\bvolume{36}(\bissue{5}),
\bfpage{741}--\blpage{754}
(\byear{2022})
\end{barticle}
\endbibitem

\bibitem{karban2022solutions}
\begin{barticle}
\bauthor{\bsnm{Karban}, \binits{U.}},
\bauthor{\bsnm{Martini}, \binits{E.}},
\bauthor{\bsnm{Jordan}, \binits{P.}},
\bauthor{\bsnm{Br{\`e}s}, \binits{G.A.}},
\bauthor{\bsnm{Towne}, \binits{A.}}:
\batitle{Solutions to aliasing in time-resolved flow data}.
\bjtitle{Theoretical and Computational Fluid Dynamics}
\bvolume{36}(\bissue{6}),
\bfpage{887}--\blpage{914}
(\byear{2022})
\end{barticle}
\endbibitem

\end{thebibliography}



\end{document}